\documentclass[aps,prd,reprint,groupedaddress,amsfonts,amssymb,amsmath,showkeys]{revtex4-2}
\usepackage{graphicx}
\usepackage{hyperref}
\hypersetup{colorlinks,linkcolor=blue,citecolor=blue,urlcolor=blue}

\begin{document}

\title{Symmetry analysis of compact tetraquark states and implications for the fully charmed candidates $X(6600)$, $X(6900)$, and $X(7100)$}
\thanks{This work was supported by the National Natural Science Foundation of China under Grant No. 12205075 and by the startup research funds of Henan University.}

\author{Shuai Yin}
\author{Ti Gong}
\email[E-mail: ]{ttigong@henu.edu.cn}
\author{Jingya Zhu}
\email[E-mail: ]{zhujy@henu.edu.cn}
\affiliation{School of Physics and Electronics, Henan University, Kaifeng 475004, China}

\date{\today}

\begin{abstract}
Motivated by recent experimental observations, we investigate the $J^P$ distribution of low-energy compact tetraquark states using symmetry analysis based on inherent nodal structures. Assuming tetrahedral and square configurations for the $qq\bar q\bar q$ system, we derive the allowed orbital structures from the restricted representations of $S_4$ onto $S_2\times S_2$ for $L\leq3$. The accessible-state distribution is particularly prominent in the $J^P=2^+$, $2^-$, and $3^-$ sectors, with the $2^+$ sector showing the strongest low-energy preference. We further find that the symmetry-driven distribution is qualitatively similar to that of the three-flavor four-quark system, and that the dominant $J^P=2^+$ pattern persists under phenomenological weightings inspired by chromomagnetic interaction (CMI) considerations. These results suggest that the low-lying compact tetraquark spectrum is primarily constrained by symmetry, while the detailed distribution exhibits sensitivity to dynamical weightings. Applying this framework to the fully charmed candidates $X(6600)$, $X(6900)$, and $X(7100)$, we find that their observed $J^{PC}=2^{++}$ quantum numbers are consistent with a low-lying compact tetraquark interpretation. The present analysis identifies the relative ordering of the $1^-$ and $1^+$ states as a sensitive channel, suggesting a direction for future non-perturbative investigations.
\end{abstract}


\maketitle

\section{\label{sec1}Introduction}
Conventional quark models have successfully described the spectra of mesons ($q\bar q$) and baryons ($qqq$ or $\bar q\bar q\bar q$)~\cite{ParticleDataGroup:2026aaa,Godfrey:1998pd,Hey:1982aj,Capstick:2000qj,Klempt:2009pi,Crede:2013kia}. However, since the discovery of the $X(3872)$ in 2003~\cite{Belle:2003nnu}, numerous exotic hadrons beyond the conventional meson--baryon framework have been observed~\cite{Belle:2007hrb,BESIII:2013ris,Belle:2013yex,BESIII:2015cld,LHCb:2014zfx,WASA-at-COSY:2014lmt,LHCb:2015yax,LHCb:2020jpq,BESIII:2020qkh,LHCb:2020bwg,LHCb:2021vvq}. Understanding their internal structures and formation mechanisms has therefore become a central problem in hadron spectroscopy (for reviews, see Refs.~\cite{Amsler:2004ps,Zhu:2004xa,Jaffe:2004ph,Swanson:2006st,Zhu:2007wz,Godfrey:2008nc,Drenska:2010kg,Esposito:2014rxa,Chen:2016qju,Richard:2016eis,Lebed:2016hpi,Ali:2017jda,Olsen:2017bmm,Karliner:2017qhf,Liu:2019zoy,Brambilla:2019esw,Guo:2019twa,Barabanov:2020jvn,Chen:2022asf,Mai:2022eur}).

Although multiquark states have been studied for more than half a century~\cite{Gell-Mann:1964ewy,Zweig:1964ruk}, the nature of many exotic hadrons remains unsettled. Recently, CMS reported three resonance structures near 6.6, 6.9, and 7.1~GeV in the $J/\psi J/\psi$ invariant-mass spectrum~\cite{CMS:2023owd}, identified as $X(6600)$, $X(6900)$, and $X(7100)$, respectively. These states are widely considered candidates for fully charmed tetraquarks~\cite{Liu:2024biq,Zhu:2024swp}, with the existence of $X(6900)$ independently confirmed by LHCb and ATLAS~\cite{LHCb:2020bwg,ATLAS:2023bft}. More recently, CMS determined their quantum numbers as $J^{PC}=2^{++}$~\cite{CMS:2025fpt}, excluding $J=0$ and $J=1$ assignments and providing important constraints on their internal structures.

Theoretical investigations of fully charmed tetraquarks have employed various approaches~\cite{Iwasaki:1975pv,Iwasaki:1976cn,Chao:1980dv,Ader:1981db,Heller:1985cb,Badalian:1985es,Lloyd:2003yc,Berezhnoy:2011xy,Berezhnoy:2011xn,Heupel:2012ua,Wu:2016vtq,Chen:2016jxd,Karliner:2016zzc,Wang:2017jtz,Debastiani:2017msn,Wang:2018poa,Liu:2019zuc,Wang:2019rdo,Bedolla:2019zwg,Liu:2020eha,Wang:2020ols,Jin:2020jfc,Becchi:2020uvq,Lu:2020cns,Chen:2020xwe,Wang:2020gmd,Albuquerque:2020hio,Giron:2020wpx,Maiani:2020pur,Richard:2020hdw,Wang:2020wrp,Chao:2020dml,Maciula:2020wri,Karliner:2020dta,Wang:2020dlo,Dong:2020nwy,Zhang:2020hoh,Feng:2020riv,Zhao:2020nwy,Faustov:2020qfm,Weng:2020jao,Zhang:2020xtb,Zhu:2020xni,Guo:2020pvt,Cao:2020gul,Gong:2020bmg,Wan:2020fsk,Yang:2020wkh,Zhao:2020jvl,Ke:2021iyh,Mutuk:2021hmi,Wang:2021kfv,Tiwari:2021tmz,Wang:2021mma,Liu:2021rtn,Kuang:2022vdy,Yang:2022bfu,Gong:2022hgd,Zhou:2022xpd,Wang:2022xja,Chen:2022sbf,Biloshytskyi:2022dmo,Wang:2022yes,Niu:2022cug,Yu:2022lak,Kuang:2023vac,Lu:2023ccs,Agaev:2023wua,Agaev:2023gaq,Agaev:2023rpj,Wang:2023kir,Anwar:2023fbp,Tang:2024zvf,Yang:2024guo,Wu:2024ocq,Chen:2024orv,Kalamidas:2025gen,Aydin:2025lbl,Silva:2025bdg,Wang:2025hex,Lu:2025lyu,Liu:2025mxv,Feng:2026orq,Celiberto:2026kks,Wang:2026kcw}. However, several assumptions adopted in previous studies require reassessment in light of the experimentally established $J^{PC}=2^{++}$ quantum numbers. As the first confirmed exotic hadrons with spin $J=2$, the $X(6600)$, $X(6900)$, and $X(7100)$ states provide a unique opportunity to explore compact multiquark dynamics.

\begin{figure}[htbp!]
\includegraphics[width=0.5\linewidth]{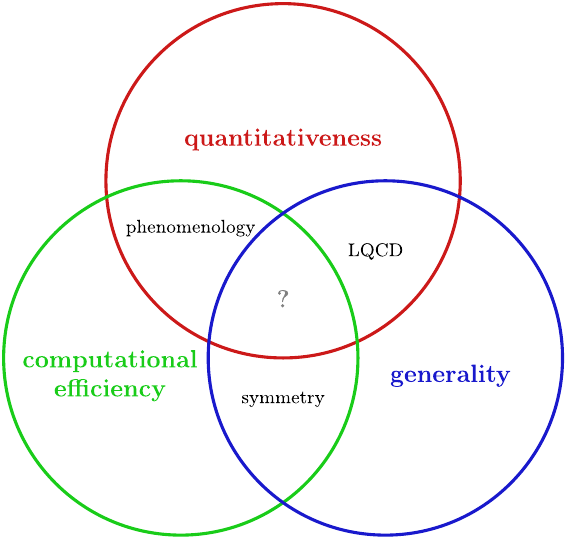}
\caption{\label{imptri}Schematic illustration of the positions of LQCD, phenomenological, and symmetry approaches in the ``impossible trinity'' defined by quantitativeness, generality, and computational efficiency.}
\end{figure}

Since perturbative quantum chromodynamics (QCD) is not applicable in the low-energy regime, hadron spectroscopy relies on non-perturbative methods, including lattice QCD (LQCD)~\cite{MILC:1997usn,Sasaki:2003gi,Okiharu:2004ve,Prelovsek:2008rf,Dudek:2009kk,Prelovsek:2010kg,NPLQCD:2011naw,HadronSpectrum:2012gic,Bicudo:2012qt,Ikeda:2013vwa,Bicudo:2015vta,HALQCD:2016ofq,Francis:2016hui,Cheung:2016bym,Bicudo:2016ooe,Cheung:2017tnt,Hughes:2017xie,Francis:2018jyb,Junnarkar:2018twb,Leskovec:2019ioa,CLQCD:2019npr,Hudspith:2020tdf,Ryan:2020iog,Padmanath:2022cvl,Bicudo:2022cqi,Lyu:2023xro,Wilson:2023anv,Meng:2024czd,Guo:2025imr}, phenomenological models, such as diquark models~\cite{Maiani:2004vq,Maiani:2014aja,Lebed:2015tna,Anwar:2023svj}, chiral effective field theory~\cite{AlFiky:2005jd,Haidenbauer:2011za,Guo:2017vcf,Meng:2022ozq}, QCD sum rules~\cite{Nielsen:2009uh,Albuquerque:2018jkn,Wang:2025sic}, hadronic molecular models~\cite{Dong:2017gaw,Guo:2017jvc,Chen:2019asm,Guo:2019fdo,Yamaguchi:2019vea,Dong:2021juy,Dong:2021bvy,Wu:2022ftm,Su:2025toa}, etc., as well as symmetry-based analyses~\cite{Chen:2019thk}. LQCD provides a systematic first-principles description but remains computationally demanding~\cite{Edwards:2004sx,Clark:2009wm}, while phenomenological approaches often depend on model assumptions~\cite{Chen:2022asf}. Symmetry-based methods offer a complementary perspective by determining allowed quantum numbers from group-theoretical constraints without detailed dynamical input. Fig.~\ref{imptri} illustrates the relative positioning of LQCD, phenomenological approaches, and symmetry-based methods within the ``impossible trinity'' of quantitativeness, generality, and computational efficiency.

Symmetry analyses of compact multiquark systems originated from the MIT bag model and chromomagnetic interaction (CMI) framework~\cite{Jaffe:1976ig,Jaffe:1976ih,Jaffe:1976yi}. Subsequent studies established the connection between spatial symmetries, permutation groups, and accessible quantum numbers in few-body systems~\cite{PhysRevA.52.3586,Bao:1997zz,Bao:1998an,Bao:1999,Bao:1998kj,Liu:2002qm,Liu:2002vi,Liu:2003gi,Liu:2004xyi}. In particular, the inherent nodal surface (INS) framework provides a model-independent classification of low-energy states according to spatial symmetry constraints.

Applying INS analysis to identical-particle four-body systems, Ref.~\cite{Chen:2019thk} showed that low-lying pentaquark states $\text{P}_c^+$ may favor $J^P=\frac{3}{2}^{\pm}$ and $\frac{5}{2}^{\pm}$ states, demonstrating the predictive capability of symmetry-based methods. Motivated by this approach, we investigate the low-energy state space of compact tetraquark systems using group-theoretical methods within the INS framework.

The remainder of this paper is organized as follows. In Sec.~\ref{sec2}, we derive the allowed tetraquark states under spin-statistics and color-singlet constraints. In Sec.~\ref{sec3}, we analyze the INS structure and establish the qualitative ordering of low-lying states for $L\leq3$. A phenomenological weighting scheme inspired by CMI is then introduced to test the robustness of the symmetry-based ordering, and the results are compared with QCD sum-rule predictions. Finally, we summarize our conclusions.

\section{\label{sec2}Solution to spin-statistics and color-singlet constraint}
Compact QCD bound states include mesons, baryons, multiquark states, glueballs, and hybrid hadrons. Their internal structure is characterized by four degrees of freedom: orbital, spin, flavor, and color. Neglecting electromagnetic interactions and flavor-dependent effects, quarks and antiquarks can be treated as two types of identical fermions, while gluons behave as identical bosons.

The wave function of a compact QCD bound state is constrained by two fundamental principles. First, the spin-statistics theorem requires the total wave function to be antisymmetric under exchange of any two quarks or antiquarks and symmetric under exchange of identical bosons. Second, color confinement requires physical hadrons to be color singlets, meaning that the color wave function must transform according to the trivial representation of $SU(3)_{\text C}$. These constraints are independent of any dynamical model.

We denote the complete Hilbert space of compact tetraquark states by $\mathfrak{T}$. It can be decomposed into the product of the quark and antiquark sectors as
\begin{equation}
\mathfrak{T}=\mathfrak{q}^2\otimes\bar{\mathfrak{q}}^2.
\end{equation}
Expressing $\mathfrak{q}^2$ and ${\mathfrak{\bar q}}^2$ in terms of the Schur--Weyl duality expansion, we have
\begin{equation}
\left\{\begin{aligned}
\mathfrak{q}^2&=\bigoplus_{[f]}[f]\otimes\mathfrak{q}_{[f]}^2,\\
\mathfrak{\bar q}^2&=\bigoplus_{[\bar f]}[\bar f]\otimes\mathfrak{\bar q}_{[\bar f]}^2,
\end{aligned}\right.
\label{desp}
\end{equation}
where $[f]$ and $[\bar f]$ denote irreducible representations (IRREPs) of the permutation group $S_2$. Next, decompose $\mathfrak{q}^2$ and $\bar{\mathfrak {q}}^2$ into the direct product of subspaces corresponding to the four degrees of freedom orbit (O), spin (S), flavor (F), and color (C) as
\begin{equation}
\left\{\begin{aligned}
\mathfrak{q}^2&=\text O^2\otimes\text S^2\otimes\text F^2\otimes\text C^2,\\
\mathfrak{\bar q}^2&={\bar{\text O}}^2\otimes{\bar{\text S}}^2\otimes{\bar{\text F}}^2\otimes{\bar{\text C}}^2.
\end{aligned}\right.
\end{equation}
Then perform the Schur--Weyl duality expansion on each degree of freedom, yielding
\begin{equation}
\left\{\begin{aligned}
\mathfrak{q}^2&
=\bigoplus_{[f]_R}\left[\left(\bigotimes_R[f]_R\right)\otimes\left(\bigotimes_RR_{[f]_R}^2\right)\right],\\
\mathfrak{\bar q}^2&
=\bigoplus_{[\bar f]_R}\left[\left(\bigotimes_R[\bar f]_R\right)\otimes\left(\bigotimes_R\bar R_{[\bar f]_R}^2\right)\right],
\end{aligned}\right.
\label{desp-DOF}
\end{equation}
where $R\in\{\text O,\text S,\text F,\text C\}$. Comparing Eqs.~(\ref{desp}) and ~(\ref{desp-DOF}), we have
\begin{equation}
\left\{\begin{aligned}
[f]&=\bigotimes_R[f]_R,\\
[\bar f]&=\bigotimes_R[\bar f]_R.
\end{aligned}\right.
\end{equation}
In the decomposition of Eq.~(\ref{desp}), the physically allowed terms must satisfy the spin-statistics constraint for fermions, which requires
\begin{equation}
[f]=[\bar f]=[11].
\label{q2qbar2-s-s-constr}
\end{equation}

If compact tetraquark states are viewed as composed of a diquark and an antidiquark, the color-singlet constraint can be formulated as follows: the symmetry properties of the color wave function under the exchange of the two quarks must match those under the exchange of the two antiquarks. In other words, the diquark and antidiquark sectors must transform consistently so as to yield an overall color-singlet state, which means
\begin{equation}
[f]_{\text C}=[\bar f]_{\text C}.
\label{q2qbar2-c-s-constr}
\end{equation}

By combining Eqs.~(\ref{q2qbar2-s-s-constr}) and~(\ref{q2qbar2-c-s-constr}), the full set of allowed IRREP combinations is determined by solving the associated Kronecker coefficients, as listed in Table~\ref{solu-2constr}.

\begin{table}[htbp!]
\caption{\label{solu-2constr}Allowed IRREP combinations of compact tetraquark wave functions under the spin-statistics and color-singlet constraints.}
\begin{ruledtabular}
\begin{tabular}{cccccccc}
$[f]_{\text O}$ & $[f]_{\text S}$ & $[f]_{\text F}$ & $[f]_{\text C}$ & $[\bar f]_{\text O}$ & $[\bar f]_{\text S}$ & $[\bar f]_{\text F}$ & $[\bar f]_{\text C}$\\
\hline
$[2]$  & $[2]$  & $[2]$  & $[11]$ & $[2]$  & $[2]$  & $[2]$  & $[11]$\\
$[2]$  & $[2]$  & $[2]$  & $[11]$ & $[2]$  & $[11]$ & $[11]$ & $[11]$\\
$[2]$  & $[2]$  & $[2]$  & $[11]$ & $[11]$ & $[2]$  & $[11]$ & $[11]$\\
$[2]$  & $[2]$  & $[2]$  & $[11]$ & $[11]$ & $[11]$ & $[2]$  & $[11]$\\
$[2]$  & $[2]$  & $[11]$ & $[2]$  & $[2]$  & $[2]$  & $[11]$  & $[2]$\\
$[2]$  & $[2]$  & $[11]$ & $[2]$  & $[2]$  & $[11]$ & $[2]$   & $[2]$\\
$[2]$  & $[2]$  & $[11]$ & $[2]$  & $[11]$ & $[2]$  & $[2]$   & $[2]$\\
$[2]$  & $[2]$  & $[11]$ & $[2]$  & $[11]$ & $[11]$ & $[11]$  & $[2]$\\
\hline
$[2]$  & $[11]$ & $[2]$  & $[2]$  & $[2]$  & $[2]$  & $[11]$  & $[2]$\\
$[2]$  & $[11]$ & $[2]$  & $[2]$  & $[2]$  & $[11]$ & $[2]$   & $[2]$\\
$[2]$  & $[11]$ & $[2]$  & $[2]$  & $[11]$ & $[2]$  & $[2]$   & $[2]$\\
$[2]$  & $[11]$ & $[2]$  & $[2]$  & $[11]$ & $[11]$ & $[11]$  & $[2]$\\
$[2]$  & $[11]$ & $[11]$ & $[11]$ & $[2]$  & $[2]$  & $[2]$  & $[11]$\\
$[2]$  & $[11]$ & $[11]$ & $[11]$ & $[2]$  & $[11]$ & $[11]$ & $[11]$\\
$[2]$  & $[11]$ & $[11]$ & $[11]$ & $[11]$ & $[2]$  & $[11]$ & $[11]$\\
$[2]$  & $[11]$ & $[11]$ & $[11]$ & $[11]$ & $[11]$ & $[2]$  & $[11]$\\
\hline
$[11]$ & $[2]$  & $[2]$  & $[2]$  & $[2]$  & $[2]$  & $[11]$  & $[2]$\\
$[11]$ & $[2]$  & $[2]$  & $[2]$  & $[2]$  & $[11]$ & $[2]$   & $[2]$\\
$[11]$ & $[2]$  & $[2]$  & $[2]$  & $[11]$ & $[2]$  & $[2]$   & $[2]$\\
$[11]$ & $[2]$  & $[2]$  & $[2]$  & $[11]$ & $[11]$ & $[11]$  & $[2]$\\
$[11]$ & $[2]$  & $[11]$ & $[11]$ & $[2]$  & $[2]$  & $[2]$  & $[11]$\\
$[11]$ & $[2]$  & $[11]$ & $[11]$ & $[2]$  & $[11]$ & $[11]$ & $[11]$\\
$[11]$ & $[2]$  & $[11]$ & $[11]$ & $[11]$ & $[2]$  & $[11]$ & $[11]$\\
$[11]$ & $[2]$  & $[11]$ & $[11]$ & $[11]$ & $[11]$ & $[2]$  & $[11]$\\
\hline
$[11]$ & $[11]$ & $[2]$  & $[11]$ & $[2]$  & $[2]$  & $[2]$  & $[11]$\\
$[11]$ & $[11]$ & $[2]$  & $[11]$ & $[2]$  & $[11]$ & $[11]$ & $[11]$\\
$[11]$ & $[11]$ & $[2]$  & $[11]$ & $[11]$ & $[2]$  & $[11]$ & $[11]$\\
$[11]$ & $[11]$ & $[2]$  & $[11]$ & $[11]$ & $[11]$ & $[2]$  & $[11]$\\
$[11]$ & $[11]$ & $[11]$ & $[2]$  & $[2]$  & $[2]$  & $[11]$  & $[2]$\\
$[11]$ & $[11]$ & $[11]$ & $[2]$  & $[2]$  & $[11]$ & $[2]$   & $[2]$\\
$[11]$ & $[11]$ & $[11]$ & $[2]$  & $[11]$ & $[2]$  & $[2]$   & $[2]$\\
$[11]$ & $[11]$ & $[11]$ & $[2]$  & $[11]$ & $[11]$ & $[11]$  & $[2]$\\
\end{tabular}
\end{ruledtabular}
\end{table}

\section{\label{sec3}INS structure of orbital wave functions}
A nodal surface is a hypersurface in configuration space where the orbital wave function vanishes. When such a surface is determined solely by geometric symmetry and independent of dynamics, it is called an invariant nodal surface (INS). The INS framework determines accessible quantum states by solving the constraints imposed by symmetry transformations, without requiring a specific interaction model. The general formalism for identical-particle few-body systems has been developed in Refs.~\cite{PhysRevA.52.3586,Bao:1997zz,Bao:1998an,Bao:1999,Bao:1998kj,Liu:2002qm,Liu:2002vi,Liu:2003gi,Liu:2004xyi}. Combining the INS constraints with the spin-statistics and color-singlet conditions introduced above allows the low-energy structure of compact tetraquark systems to be analyzed systematically.

\subsection{Basic conclusions of INS analysis}
We consider two representative spatial configurations: the equilateral tetrahedron (ETH), where the four particles occupy the vertices of a regular tetrahedron, and the square (Sqr), where they occupy the vertices of a square. These configurations characterize the dominant spatial correlations of low-energy four-body systems. Following Ref.~\cite{Chen:2019thk}, we also introduce partially symmetric configurations ETH$_3$ and Sqr$_3$, where only a subset of the full symmetry constraints is imposed.

In the Sqr configuration, two possible arrangements of quark--antiquark pairs are considered: parallel (Sqr$^{\parallel}$) and perpendicular (Sqr$^{\perp}$). The corresponding orbital permutation symmetries of the quark and antiquark pairs are denoted by $\lambda^2$ and $\bar\lambda^2$, respectively.

\begin{figure}[htbp!]
\includegraphics[width=0.32\linewidth]{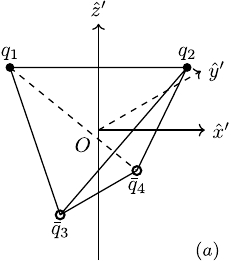}
\includegraphics[width=0.32\linewidth]{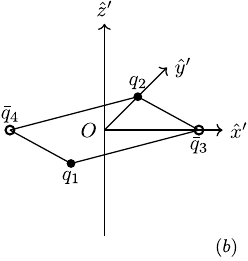}
\includegraphics[width=0.32\linewidth]{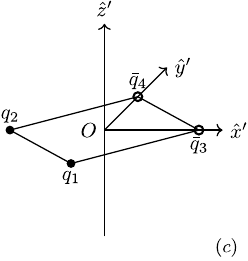}
\caption{\label{configs}Schematic construction of the body-fixed coordinate systems $Ox'y'z'$ for compact tetraquark states in the (a) ETH, (b) Sqr$^\perp$, and (c) Sqr$^\parallel$ configurations. The coordinate systems are defined following the convention adopted in Fig.~1 of Ref.~\cite{Chen:2019thk}.}
\end{figure}

Introducing body-fixed coordinates $Ox'y'z'$ as shown in Fig.~\ref{configs}, and restricting the $S_4$ IRREPs onto the subgroup $S_2\times S_2$, one obtains
\begin{equation}\left\{
\begin{aligned}
[4]&\downarrow[2]\otimes[2],\\
[31]&\downarrow\left([2]\otimes[2]\right)\oplus\left([2]\otimes[11]\right)\oplus\left([11]\otimes[2]\right),\\
[22]&\downarrow\left([2]\otimes[2]\right)\oplus\left([11]\otimes[11]\right),\\
[211]&\downarrow\left([2]\otimes[11]\right)\oplus\left([11]\otimes[2]\right)\oplus\left([11]\otimes[11]\right),\\
[1^4]&\downarrow[11]\otimes[11].
\end{aligned}\right.
\label{restr}
\end{equation}
Combining these branching relations with the accessibility results listed in Table~6 of Ref.~\cite{Chen:2019thk}, all allowed $L^\pi\lambda^2\bar\lambda^2$ components for the ETH, ETH$_3$, Sqr, and Sqr$_3$ configurations can be determined. The results are summarized in Table~\ref{ThisINS}.

\begin{table}[htbp!]
\caption{\label{ThisINS}Accessibility of the $L^\pi\lambda^2\bar{\lambda}^2$ IRREP components of the orbital wave functions for compact tetraquark states obtained from the restriction of $S_4$ to $S_2\times S_2$, with orbital angular momentum $L\leq3$, in the ETH and Sqr configurations as well as the ETH$_3$ and Sqr$_3$ configurations. The capital letters A, B, and C denote cases in which the total multiplicity of the corresponding $S_2\times S_2$ representation in the restriction of an accessible $S_4$ IRREP is 1, 2, and 3, respectively. The symbol ``--'' indicates that the corresponding IRREP is inaccessible in the given configuration.}
\begin{ruledtabular}
\begin{tabular}{lccccc}
& $L^\pi$ & $[2]\otimes[2]$ & $[2]\otimes[11]$ & $[11]\otimes[2]$ & $[11]\otimes[11]$\\
\hline
ETH & $0^+$ & A & A & A & A\\
& $1^-$ & B & B & B & C\\
& $2^+$ & C & B & B & C\\
& $3^-$ & C & B & B & C\\
\hline
Sqr & $0^+$ & B & -- & -- & A\\
& $1^-$ & A & B & B & A\\
& $2^+$ & C & B & B & C\\
& $3^-$ & A & B & B & A\\
\hline
ETH$_3$ & $0^+$ & B & A & A & B\\
& $1^-$ & B & B & B & C\\
& $2^+$ & C & B & B & C\\
& $3^-$ & C & B & B & C\\
\hline
Sqr$_3$ & $0^+$ & C & B & B & B\\
& $1^-$ & A & B & B & A\\
& $2^+$ & C & B & B & C\\
& $3^-$ & A & B & B & A\\
\end{tabular}
\end{ruledtabular}
\end{table}

Let $\mathcal{A}$ denote the spatial configuration and $Z$ the IRREP of the color wavefunction. The full state space of compact tetraquark states can be decomposed into a direct sum according to $L^\pi$, $\lambda^2\bar{\lambda}^2$, $\mathcal{A}$, $s^2\bar s^2$, and $Z$. We denote the subspaces arising from this decomposition by $\mathfrak{T}_{L^\pi,\lambda^2\bar{\lambda}^2}^{s^2\bar s^2,Z}[\mathcal{A}]$.

For compact tetraquark states in the ETH and Sqr configurations, as well as the partially symmetric ETH$_3$ and Sqr$_3$ configurations, the distribution of accessible states as a function of $J^P$ for $L \leq 3$ is obtained by combining the results of Tables~\ref{solu-2constr} and~\ref{ThisINS} with the standard angular-momentum coupling relations. The resulting distributions are summarized in Table~\ref{Na-this}. For comparison, Table~7 of Ref.~\cite{Chen:2019thk} presents the corresponding statistical results for the number of accessible states in the three-flavor four-quark system.

\begin{table*}[htbp!]
\caption{\label{Na-this}Distribution of the number of accessible states as a function of $J^P$ for compact tetraquark states in the ETH and Sqr configurations, as well as the ETH$_3$ and Sqr$_3$ configurations, with orbital angular momentum restricted to $L\leq3$.}
\begin{ruledtabular}
\begin{tabular}{lcccccccccccc}
Configuration & $0^+$ & $0^-$ & $1^+$ & $1^-$ & $2^+$ & $2^-$ & $3^+$ & $3^-$ & $4^+$ & $4^-$ & $5^+$ & $5^-$\\
\hline
ETH & 36 & 54 & 104 & 128 & 128 & 152 & 80 & 138 & 20 & 80 & 0 & 20\\
Sqr & 32 & 36 & 98  & 84  & 126 & 96  & 80 & 84  & 20 & 48 & 0 & 12\\
ETH$_3$ & 44 & 54 & 116 & 128 & 132 & 152 & 80 & 138 & 20 & 80 & 0 & 20\\
Sqr$_3$ & 56 & 36 & 134 & 84  & 138 & 96  & 80 & 84  & 20 & 48 & 0 & 12\\
\end{tabular}
\end{ruledtabular}
\end{table*}

\begin{figure}[htbp!]
\includegraphics[width=0.7\linewidth]{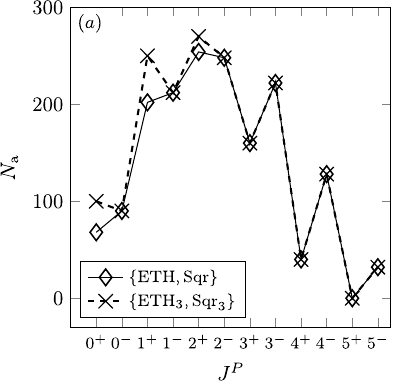}
\includegraphics[width=0.7\linewidth]{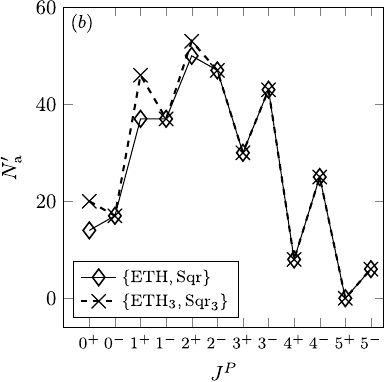}
\caption{\label{Na}Distributions of the number of accessible states as functions of $J^P$ for (a) compact tetraquark states, $N_{\text a}$, and (b) three-flavor four-quark systems, $N_{\text a}'$, obtained by restricting the configuration space to the sets \{ETH, Sqr\} and \{ETH$_3$, Sqr$_3$\}.}
\end{figure}

For each selected set of spatial configurations, such as \{ETH, Sqr\} or \{ETH$_3$, Sqr$_3$\}, the total number of accessible states associated with a given $J^P$ is defined as the sum over all allowed states with that quantum number and is referred to as the accessible-state number of $J^P$. Fig.~\ref{Na} illustrates the resulting distributions, where $N_{\text a}$ denotes the number of accessible states for compact tetraquark systems and $N_{\text a}'$ denotes the corresponding quantity for three-flavor four-quark systems, both shown as functions of $J^P$.

In this work, the number of accessible states associated with a given $J^P$ is defined as the total number of subspaces $\mathfrak{T}_{L^\pi,\lambda^2\bar{\lambda}^2}^{s^2\bar s^2,Z}[\mathcal{A}]$ corresponding to that set of quantum numbers. For a generic quantum bound system, states with fewer nodal surfaces are generally expected to lie lower in energy~\cite{Chen:2019thk}. Under this principle, the distribution of accessible states as a function of $J^P$ provides a qualitative ordering of the low-lying spectrum. For nodeless low-$L$ configurations, the number $N(J^P)$ of accessible independent states reflects the variational flexibility within that symmetry sector. A larger $N(J^P)$ allows for a smoother redistribution of the wavefunction density without introducing additional nodes, which generally leads to a lower kinetic energy expectation. We therefore adopt $N(J^P)$ as a heuristic indicator for the relative ordering among the competing $J^P$ channels. Restricting the analysis to the configuration set \{ETH, Sqr\}, we infer that the lowest-lying states are most likely characterized by $J^P=2^+$, followed successively by states with $J^P=2^-$, $3^-$, and $1^-$. Restricting the analysis to the partially symmetric configuration set \{ETH$_3$, Sqr$_3$\}, we infer that the lowest-lying states are most likely characterized by $J^P=2^+$, $1^+$, $2^-$, and $3^-$, in the order suggested by the accessible-state counting.

Further analysis shows that, when the investigation is restricted to the configuration sets \{ETH, Sqr\} and \{ETH$_3$, Sqr$_3$\}, the ratio $N_{\text a}/N_{\text a}'$ fluctuates about the mean values 5.24 and 5.25, with standard deviations of 0.23 and 0.21, respectively. This remarkable consistency indicates that, for $L \leq 3$, the accessible-state distributions of compact tetraquark systems and three-flavor four-quark systems exhibit nearly identical dependence on $J^P$. In other words, although the two systems differ in their color-singlet constraints and in the representation structures of their flavor and color degrees of freedom, these differences have only a negligible effect on the overall shape of the accessible-state distribution.

A caveat is in order regarding the scope of the INS-based ordering presented above. The accessible-state counting does not determine absolute masses or provide a rigorous dynamical ordering; it only suggests a qualitative tendency under the assumption that fewer nodal structures correlate with lower excitation. Actual spectra may be modified by color--spin interactions, correlations, and threshold couplings. The following sensitivity analysis examines the robustness of the qualitative ordering against one class of such effects.

\subsection{\label{sec3B}Sensitivity analysis with phenomenological weighting}
The CMI model describes the color--spin interaction between (anti)quarks and provides an important mechanism for understanding hadron spectroscopy~\cite{Jaffe:1976ig}. For a $qq$ (or $\bar q\bar q$) pair, antisymmetric color and spin configurations are generally more favorable than symmetric ones~\cite{Wilczek:2004im}. Accordingly, diquarks with antisymmetric color and spin structures are referred to as ``good diquarks'', while the corresponding symmetric configurations are ``bad diquarks''. Different color--spin symmetries lead to distinct chromomagnetic energies, resulting in the splitting and ordering of low-lying states.

To qualitatively assess whether such dynamical effects could alter the symmetry-based counting, we introduce a phenomenological weight factor for each color–spin configuration:
\begin{equation}
w\left(s^2,{\bar{s}}^2,Z\right)=\frac{1}{W_0}\exp\left[-\frac{\delta E\left(s^2,{\bar{s}}^2,Z\right)}{\Lambda}\right],
\label{weight-fun}
\end{equation}

\noindent where $\Lambda$ is a parameter with dimension of energy, and $\delta E$ denotes the phenomenological energy correction associated with spin symmetries and color representations. Retaining the leading-order contribution, we parameterize the energy correction as
\begin{equation}
\begin{aligned}
&\delta E\left(s^2,{\bar{s}}^2,Z\right)\\
=&\left(s^2-\frac{1}{2}\right)\cdot E_1+\left({\bar{s}}^2-\frac{1}{2}\right)\cdot E_2+\frac{\delta_{Z,S}-\delta_{Z,A}}{2}\cdot E_3,
\end{aligned}
\end{equation}
where $E_1$, $E_2$, and $E_3$ describe the relative shifts among different spin-color configurations. The normalization factor $W_0$ is determined by
\begin{equation}
\sum_{J^P}\rho_{\text a}=1,
\end{equation}
where $\rho_{\text a}$ denotes the normalized, weighted number of accessible states.

The physical interpretation of Eq.~(\ref{weight-fun}) is straightforward. The parameter $\Lambda$ controls the relative strength of dynamical effects in the state-counting statistics. As $\Lambda$ decreases, the distribution becomes more sensitive to the energy correction $\delta E$, whereas in the limit $\Lambda\rightarrow\infty$, the weighting becomes independent of $\delta E$ and $\rho_{\text a}$ approaches the symmetry-only distribution shown in Fig.~\ref{Na} (a).

As a baseline choice for the sensitivity scan, we adopt:
\begin{equation}
E_1=E_2=E_3.
\end{equation}
To characterize the strength of dynamical effects, we introduce a dimensionless parameter $\theta=\Lambda/E_1$, which is sampled over the discrete set $\left\{0.01,0.5,1,5,\infty\right\}$, where $\theta=\infty$ corresponds to the symmetry-dominated limit.

\begin{figure}[htbp!]
\includegraphics[width=0.7\linewidth]{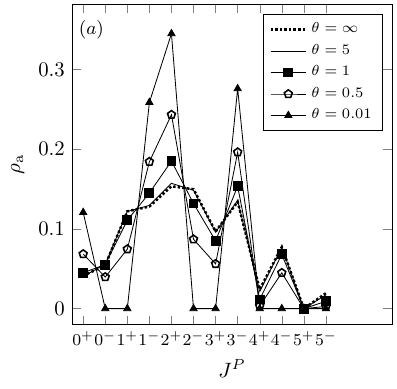} 
\includegraphics[width=0.7\linewidth]{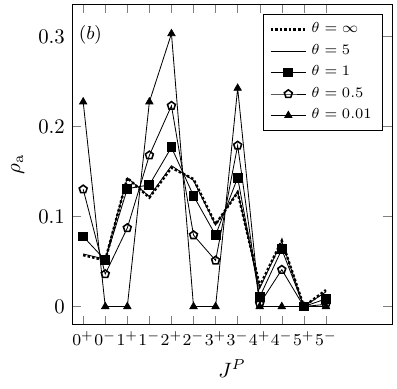}
\caption{\label{Na-CMI}Distributions of the weighted and normalized number of accessible states $\rho_{\text a}$ as functions of $J^P$ for compact tetraquark states, evaluated at different values of the parameter $\theta$ and with the configuration space restricted to the sets (a) \{ETH, Sqr\} and (b) \{ETH$_3$, Sqr$_3$\}.}
\end{figure}

The resulting distributions $\rho_{\text a}(J^P)$ are shown in Fig.~\ref{Na-CMI}. Several channels, including $0^-$, $1^+$, $2^-$, $3^+$, $4^+$, $4^-$, $5^+$, and $5^-$, are strongly suppressed when the dynamical weighting is enhanced (i.e. $\theta$ decreases), while the distribution becomes concentrated mainly in the $0^+$, $1^-$, $2^+$, and $3^-$ channels.

Despite these modifications, the $2^+$ component remains the dominant contribution throughout the explored parameter region. In fact, its relative weight increases further as the dynamical weighting becomes more pronounced, suggesting that the symmetry-based preference for $2^+$ states is not only robust but also reinforced under this phenomenological weighting scheme. For the partially symmetric configuration set \{ETH$_3$, Sqr$_3$\}, the $0^+$ contribution decreases and becomes comparable to the $1^-$ component.

Comparing the weighted distributions with the symmetry-only case at $\theta=\infty$, we find that the overall pattern remains qualitatively unchanged. Therefore, the INS-induced structure provides a robust constraint on the low-energy spectrum even after dynamical effects are included.

\subsection{Comparison with a result based on QCD sum rules}
Recently, a QCD sum-rule analysis investigated fully charmed compact tetraquark states with the color structure $8_{[c\bar c]}\otimes8_{[c\bar c]}$~\cite{Tang:2024zvf}. The study considered several $J^{PC}$ channels, including $0^{-+}$, $0^{--}$, $1^{-+}$, $1^{+-}$, $1^{--}$, and $2^{++}$. The predicted masses indicate that the $1^{+-}$ and $2^{++}$ states lie around 6.48--6.62~GeV, while the remaining channels appear at higher masses of 6.85--7.02~GeV~\cite{Tang:2024zvf}.

In the compact tetraquark picture developed here, the symmetry analysis suggests that $J^P=2^+$ states tend to appear at lower energies, whereas states with $J^P=3^-$, $1^-$, $0^+$, and other quantum numbers are expected to lie higher. 

For the fully charmed system, the flavor configuration is fixed as $cc\bar c\bar c$. Consequently, the flavor sector only contributes an overall multiplicative factor to the accessible-state number and does not change the relative distribution among different $J^P$ channels.

Comparing the ordering obtained here with the QCD sum-rule result, the common feature is that the $2^+$ state is favored to appear among the low-lying states. Therefore, if the experimentally observed states $X(6600)$, $X(6900)$, and $X(7100)$ are compact fully charmed tetraquarks, their experimentally determined quantum numbers $J^{PC}=2^{++}$ are consistent with their interpretation as low-lying compact tetraquark states.

\begin{figure}[htbp!]
\includegraphics[width=0.7\linewidth]{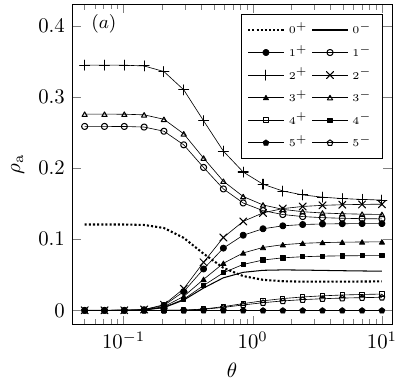}
\includegraphics[width=0.7\linewidth]{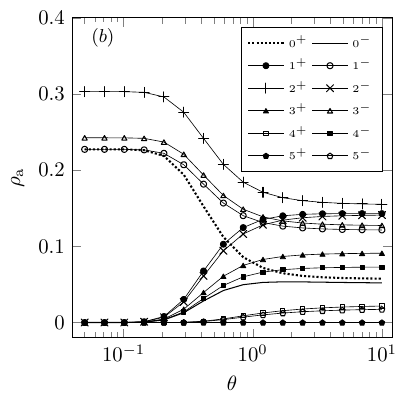}
\caption{\label{order}Dependence of the weighted and normalized number of accessible states $\rho_{\text a}$ on the parameter $\theta$ for each $J^P$ channel in compact tetraquark states, with $\theta\in[0.05,10]$ and the configuration space restricted to the sets (a) \{ETH, Sqr\} and (b) \{ETH$_3$, Sqr$_3$\}.}
\end{figure}

This consistency supports the interpretation that the $2^+$ state is naturally favored by the symmetry structure. However, for the relative ordering of the $1^-$ and $1^+$ states, the present analysis reveals a notable sensitivity to the phenomenological smoothing parameter, as illustrated in Fig.~\ref{order}. This figure shows the evolution of accessible-state numbers with the parameter $\theta$. The intersections between different curves correspond to changes in the relative ordering of the symmetry-based accessibility measure. For the partially symmetric configuration set \{ETH$_3$, Sqr$_3$\}, the $1^-$ and $1^+$ channels cross at $\theta\simeq1.0$. The crossing indicates that the $1^-/1^+$ ordering is sensitive to the phenomenological weighting scheme, highlighting a theoretical uncertainty that warrants further investigation with more rigorous methods.

Before concluding this section, we revisit the dependence of our results on the phenomenological parameters $E_1$, $E_2$, and $E_3$ introduced in Sec.~\ref{sec3B}. We have verified that varying the ratios $E_2/E_1$ and $E_3/E_1$ within a moderate range may affect the ordering of the curves at a fixed value of $\theta$, but does not alter the intersection pattern: whether any two curves intersect, and the sequence in which these intersections occur as $\theta$ increases, remain unchanged. This indicates that the qualitative evolution of the level ordering with $\theta$ is robust against the specific values of the energy parameters, and that the conclusions drawn from the sensitivity analysis are not artifacts of the adopted parametrization.

\section{Summary and remarks}
In this work, we investigate the low-energy structure of compact tetraquark states using symmetry constraints from the INS framework combined with spin-statistics and color-singlet conditions. We find that the accessible-state distribution is dominated by higher-angular-momentum components, particularly $J^P=2^+$, $2^-$, and $3^-$, while low-angular-momentum states such as $0^+$ and $0^-$ are comparatively suppressed.

After introducing dynamical weightings, the $0^+$ contribution increases significantly, but the dominance of the $2^+$ and $3^-$ components remains. This indicates that the spatial symmetry constraints encoded by INS play a leading role in determining the qualitative structure of the low-energy spectrum.

Applying these results to the fully charmed candidates $X(6600)$, $X(6900)$, and $X(7100)$, the present analysis suggests that $2^+$ configurations are natural candidates for these structures, which is noteworthy given their observed $J^{PC}=2^{++}$ assignment. Individual peak assignments, however, require dynamical calculations beyond the present symmetry analysis.

Furthermore, the comparison between compact tetraquark systems and three-flavor four-quark systems shows that their accessible-state distributions exhibit similar qualitative patterns despite differences in color-singlet constraints and representation structures. This suggests that, for multiquark systems with equal numbers of quarks and antiquarks, the INS-induced distribution may be largely insensitive to the detailed internal organization of quark and antiquark degrees of freedom.

Nevertheless, when dynamical effects are incorporated through phenomenological weightings, the distribution can shift in a system-dependent way. Thus, while symmetry provides a robust framework for classifying possible states, the actual ordering is likely influenced by dynamical details that require further investigation.

\bibliographystyle{apsrev4-2}
\bibliography{ref.bib}

\begin{thebibliography}{195}%
\makeatletter
\providecommand \@ifxundefined [1]{%
 \@ifx{#1\undefined}
}%
\providecommand \@ifnum [1]{%
 \ifnum #1\expandafter \@firstoftwo
 \else \expandafter \@secondoftwo
 \fi
}%
\providecommand \@ifx [1]{%
 \ifx #1\expandafter \@firstoftwo
 \else \expandafter \@secondoftwo
 \fi
}%
\providecommand \natexlab [1]{#1}%
\providecommand \enquote  [1]{``#1''}%
\providecommand \bibnamefont  [1]{#1}%
\providecommand \bibfnamefont [1]{#1}%
\providecommand \citenamefont [1]{#1}%
\providecommand \href@noop [0]{\@secondoftwo}%
\providecommand \href [0]{\begingroup \@sanitize@url \@href}%
\providecommand \@href[1]{\@@startlink{#1}\@@href}%
\providecommand \@@href[1]{\endgroup#1\@@endlink}%
\providecommand \@sanitize@url [0]{\catcode `\\12\catcode `\$12\catcode
  `\&12\catcode `\#12\catcode `\^12\catcode `\_12\catcode `\%12\relax}%
\providecommand \@@startlink[1]{}%
\providecommand \@@endlink[0]{}%
\providecommand \url  [0]{\begingroup\@sanitize@url \@url }%
\providecommand \@url [1]{\endgroup\@href {#1}{\urlprefix }}%
\providecommand \urlprefix  [0]{URL }%
\providecommand \Eprint [0]{\href }%
\providecommand \doibase [0]{https://doi.org/}%
\providecommand \selectlanguage [0]{\@gobble}%
\providecommand \bibinfo  [0]{\@secondoftwo}%
\providecommand \bibfield  [0]{\@secondoftwo}%
\providecommand \translation [1]{[#1]}%
\providecommand \BibitemOpen [0]{}%
\providecommand \bibitemStop [0]{}%
\providecommand \bibitemNoStop [0]{.\EOS\space}%
\providecommand \EOS [0]{\spacefactor3000\relax}%
\providecommand \BibitemShut  [1]{\csname bibitem#1\endcsname}%
\let\auto@bib@innerbib\@empty
\bibitem [{\citenamefont {Takahashi}\ \emph {et~al.}(2026)\citenamefont
  {Takahashi} \emph {et~al.}}]{ParticleDataGroup:2026aaa}%
  \BibitemOpen
  \bibfield  {author} {\bibinfo {author} {\bibfnamefont {F.}~\bibnamefont
  {Takahashi}} \emph {et~al.} (\bibinfo {collaboration} {Particle Data
  Group}),\ }\href {https://doi.org/10.1142/S0217751X26300115} {\bibfield
  {journal} {\bibinfo  {journal} {Int. J. Mod. Phys. A}\ }\textbf {\bibinfo
  {volume} {41}},\ \bibinfo {pages} {2630011} (\bibinfo {year}
  {2026})}\BibitemShut {NoStop}%
\bibitem [{\citenamefont {Godfrey}\ and\ \citenamefont
  {Napolitano}(1999)}]{Godfrey:1998pd}%
  \BibitemOpen
  \bibfield  {author} {\bibinfo {author} {\bibfnamefont {S.}~\bibnamefont
  {Godfrey}}\ and\ \bibinfo {author} {\bibfnamefont {J.}~\bibnamefont
  {Napolitano}},\ }\href {https://doi.org/10.1103/RevModPhys.71.1411}
  {\bibfield  {journal} {\bibinfo  {journal} {Rev. Mod. Phys.}\ }\textbf
  {\bibinfo {volume} {71}},\ \bibinfo {pages} {1411} (\bibinfo {year}
  {1999})},\ \Eprint {https://arxiv.org/abs/hep-ph/9811410}
  {arXiv:hep-ph/9811410} \BibitemShut {NoStop}%
\bibitem [{\citenamefont {Hey}\ and\ \citenamefont {Kelly}(1983)}]{Hey:1982aj}%
  \BibitemOpen
  \bibfield  {author} {\bibinfo {author} {\bibfnamefont {A.~J.~G.}\
  \bibnamefont {Hey}}\ and\ \bibinfo {author} {\bibfnamefont {R.~L.}\
  \bibnamefont {Kelly}},\ }\href {https://doi.org/10.1016/0370-1573(83)90114-X}
  {\bibfield  {journal} {\bibinfo  {journal} {Phys. Rept.}\ }\textbf {\bibinfo
  {volume} {96}},\ \bibinfo {pages} {71} (\bibinfo {year} {1983})}\BibitemShut
  {NoStop}%
\bibitem [{\citenamefont {Capstick}\ and\ \citenamefont
  {Roberts}(2000)}]{Capstick:2000qj}%
  \BibitemOpen
  \bibfield  {author} {\bibinfo {author} {\bibfnamefont {S.}~\bibnamefont
  {Capstick}}\ and\ \bibinfo {author} {\bibfnamefont {W.}~\bibnamefont
  {Roberts}},\ }\href {https://doi.org/10.1016/S0146-6410(00)00109-5}
  {\bibfield  {journal} {\bibinfo  {journal} {Prog. Part. Nucl. Phys.}\
  }\textbf {\bibinfo {volume} {45}},\ \bibinfo {pages} {S241} (\bibinfo {year}
  {2000})},\ \Eprint {https://arxiv.org/abs/nucl-th/0008028}
  {arXiv:nucl-th/0008028} \BibitemShut {NoStop}%
\bibitem [{\citenamefont {Klempt}\ and\ \citenamefont
  {Richard}(2010)}]{Klempt:2009pi}%
  \BibitemOpen
  \bibfield  {author} {\bibinfo {author} {\bibfnamefont {E.}~\bibnamefont
  {Klempt}}\ and\ \bibinfo {author} {\bibfnamefont {J.-M.}\ \bibnamefont
  {Richard}},\ }\href {https://doi.org/10.1103/RevModPhys.82.1095} {\bibfield
  {journal} {\bibinfo  {journal} {Rev. Mod. Phys.}\ }\textbf {\bibinfo {volume}
  {82}},\ \bibinfo {pages} {1095} (\bibinfo {year} {2010})},\ \Eprint
  {https://arxiv.org/abs/0901.2055} {arXiv:0901.2055 [hep-ph]} \BibitemShut
  {NoStop}%
\bibitem [{\citenamefont {Crede}\ and\ \citenamefont
  {Roberts}(2013)}]{Crede:2013kia}%
  \BibitemOpen
  \bibfield  {author} {\bibinfo {author} {\bibfnamefont {V.}~\bibnamefont
  {Crede}}\ and\ \bibinfo {author} {\bibfnamefont {W.}~\bibnamefont
  {Roberts}},\ }\href {https://doi.org/10.1088/0034-4885/76/7/076301}
  {\bibfield  {journal} {\bibinfo  {journal} {Rept. Prog. Phys.}\ }\textbf
  {\bibinfo {volume} {76}},\ \bibinfo {pages} {076301} (\bibinfo {year}
  {2013})},\ \Eprint {https://arxiv.org/abs/1302.7299} {arXiv:1302.7299
  [nucl-ex]} \BibitemShut {NoStop}%
\bibitem [{\citenamefont {Choi}\ \emph {et~al.}(2003)\citenamefont {Choi} \emph
  {et~al.}}]{Belle:2003nnu}%
  \BibitemOpen
  \bibfield  {author} {\bibinfo {author} {\bibfnamefont {S.~K.}\ \bibnamefont
  {Choi}} \emph {et~al.} (\bibinfo {collaboration} {Belle}),\ }\href
  {https://doi.org/10.1103/PhysRevLett.91.262001} {\bibfield  {journal}
  {\bibinfo  {journal} {Phys. Rev. Lett.}\ }\textbf {\bibinfo {volume} {91}},\
  \bibinfo {pages} {262001} (\bibinfo {year} {2003})},\ \Eprint
  {https://arxiv.org/abs/hep-ex/0309032} {arXiv:hep-ex/0309032} \BibitemShut
  {NoStop}%
\bibitem [{\citenamefont {Choi}\ \emph {et~al.}(2008)\citenamefont {Choi} \emph
  {et~al.}}]{Belle:2007hrb}%
  \BibitemOpen
  \bibfield  {author} {\bibinfo {author} {\bibfnamefont {S.~K.}\ \bibnamefont
  {Choi}} \emph {et~al.} (\bibinfo {collaboration} {Belle}),\ }\href
  {https://doi.org/10.1103/PhysRevLett.100.142001} {\bibfield  {journal}
  {\bibinfo  {journal} {Phys. Rev. Lett.}\ }\textbf {\bibinfo {volume} {100}},\
  \bibinfo {pages} {142001} (\bibinfo {year} {2008})},\ \Eprint
  {https://arxiv.org/abs/0708.1790} {arXiv:0708.1790 [hep-ex]} \BibitemShut
  {NoStop}%
\bibitem [{\citenamefont {Ablikim}\ \emph {et~al.}(2013)\citenamefont {Ablikim}
  \emph {et~al.}}]{BESIII:2013ris}%
  \BibitemOpen
  \bibfield  {author} {\bibinfo {author} {\bibfnamefont {M.}~\bibnamefont
  {Ablikim}} \emph {et~al.} (\bibinfo {collaboration} {BESIII}),\ }\href
  {https://doi.org/10.1103/PhysRevLett.110.252001} {\bibfield  {journal}
  {\bibinfo  {journal} {Phys. Rev. Lett.}\ }\textbf {\bibinfo {volume} {110}},\
  \bibinfo {pages} {252001} (\bibinfo {year} {2013})},\ \Eprint
  {https://arxiv.org/abs/1303.5949} {arXiv:1303.5949 [hep-ex]} \BibitemShut
  {NoStop}%
\bibitem [{\citenamefont {Liu}\ \emph {et~al.}(2013)\citenamefont {Liu} \emph
  {et~al.}}]{Belle:2013yex}%
  \BibitemOpen
  \bibfield  {author} {\bibinfo {author} {\bibfnamefont {Z.~Q.}\ \bibnamefont
  {Liu}} \emph {et~al.} (\bibinfo {collaboration} {Belle}),\ }\href
  {https://doi.org/10.1103/PhysRevLett.110.252002} {\bibfield  {journal}
  {\bibinfo  {journal} {Phys. Rev. Lett.}\ }\textbf {\bibinfo {volume} {110}},\
  \bibinfo {pages} {252002} (\bibinfo {year} {2013})},\ \bibinfo {note}
  {[Erratum: Phys.Rev.Lett. 111, 019901 (2013)]},\ \Eprint
  {https://arxiv.org/abs/1304.0121} {arXiv:1304.0121 [hep-ex]} \BibitemShut
  {NoStop}%
\bibitem [{\citenamefont {Ablikim}\ \emph {et~al.}(2015)\citenamefont {Ablikim}
  \emph {et~al.}}]{BESIII:2015cld}%
  \BibitemOpen
  \bibfield  {author} {\bibinfo {author} {\bibfnamefont {M.}~\bibnamefont
  {Ablikim}} \emph {et~al.} (\bibinfo {collaboration} {BESIII}),\ }\href
  {https://doi.org/10.1103/PhysRevLett.115.112003} {\bibfield  {journal}
  {\bibinfo  {journal} {Phys. Rev. Lett.}\ }\textbf {\bibinfo {volume} {115}},\
  \bibinfo {pages} {112003} (\bibinfo {year} {2015})},\ \Eprint
  {https://arxiv.org/abs/1506.06018} {arXiv:1506.06018 [hep-ex]} \BibitemShut
  {NoStop}%
\bibitem [{\citenamefont {Aaij}\ \emph {et~al.}(2014)\citenamefont {Aaij} \emph
  {et~al.}}]{LHCb:2014zfx}%
  \BibitemOpen
  \bibfield  {author} {\bibinfo {author} {\bibfnamefont {R.}~\bibnamefont
  {Aaij}} \emph {et~al.} (\bibinfo {collaboration} {LHCb}),\ }\href
  {https://doi.org/10.1103/PhysRevLett.112.222002} {\bibfield  {journal}
  {\bibinfo  {journal} {Phys. Rev. Lett.}\ }\textbf {\bibinfo {volume} {112}},\
  \bibinfo {pages} {222002} (\bibinfo {year} {2014})},\ \Eprint
  {https://arxiv.org/abs/1404.1903} {arXiv:1404.1903 [hep-ex]} \BibitemShut
  {NoStop}%
\bibitem [{\citenamefont {Adlarson}\ \emph {et~al.}(2014)\citenamefont
  {Adlarson} \emph {et~al.}}]{WASA-at-COSY:2014lmt}%
  \BibitemOpen
  \bibfield  {author} {\bibinfo {author} {\bibfnamefont {P.}~\bibnamefont
  {Adlarson}} \emph {et~al.} (\bibinfo {collaboration} {WASA-at-COSY}),\ }\href
  {https://doi.org/10.1103/PhysRevC.90.035204} {\bibfield  {journal} {\bibinfo
  {journal} {Phys. Rev. C}\ }\textbf {\bibinfo {volume} {90}},\ \bibinfo
  {pages} {035204} (\bibinfo {year} {2014})},\ \Eprint
  {https://arxiv.org/abs/1408.4928} {arXiv:1408.4928 [nucl-ex]} \BibitemShut
  {NoStop}%
\bibitem [{\citenamefont {Aaij}\ \emph {et~al.}(2015)\citenamefont {Aaij} \emph
  {et~al.}}]{LHCb:2015yax}%
  \BibitemOpen
  \bibfield  {author} {\bibinfo {author} {\bibfnamefont {R.}~\bibnamefont
  {Aaij}} \emph {et~al.} (\bibinfo {collaboration} {LHCb}),\ }\href
  {https://doi.org/10.1103/PhysRevLett.115.072001} {\bibfield  {journal}
  {\bibinfo  {journal} {Phys. Rev. Lett.}\ }\textbf {\bibinfo {volume} {115}},\
  \bibinfo {pages} {072001} (\bibinfo {year} {2015})},\ \Eprint
  {https://arxiv.org/abs/1507.03414} {arXiv:1507.03414 [hep-ex]} \BibitemShut
  {NoStop}%
\bibitem [{\citenamefont {Aaij}\ \emph {et~al.}(2021)\citenamefont {Aaij} \emph
  {et~al.}}]{LHCb:2020jpq}%
  \BibitemOpen
  \bibfield  {author} {\bibinfo {author} {\bibfnamefont {R.}~\bibnamefont
  {Aaij}} \emph {et~al.} (\bibinfo {collaboration} {LHCb}),\ }\href
  {https://doi.org/10.1016/j.scib.2021.02.030} {\bibfield  {journal} {\bibinfo
  {journal} {Sci. Bull.}\ }\textbf {\bibinfo {volume} {66}},\ \bibinfo {pages}
  {1278} (\bibinfo {year} {2021})},\ \Eprint {https://arxiv.org/abs/2012.10380}
  {arXiv:2012.10380 [hep-ex]} \BibitemShut {NoStop}%
\bibitem [{\citenamefont {Ablikim}\ \emph {et~al.}(2021)\citenamefont {Ablikim}
  \emph {et~al.}}]{BESIII:2020qkh}%
  \BibitemOpen
  \bibfield  {author} {\bibinfo {author} {\bibfnamefont {M.}~\bibnamefont
  {Ablikim}} \emph {et~al.} (\bibinfo {collaboration} {BESIII}),\ }\href
  {https://doi.org/10.1103/PhysRevLett.126.102001} {\bibfield  {journal}
  {\bibinfo  {journal} {Phys. Rev. Lett.}\ }\textbf {\bibinfo {volume} {126}},\
  \bibinfo {pages} {102001} (\bibinfo {year} {2021})},\ \Eprint
  {https://arxiv.org/abs/2011.07855} {arXiv:2011.07855 [hep-ex]} \BibitemShut
  {NoStop}%
\bibitem [{\citenamefont {Aaij}\ \emph {et~al.}(2020)\citenamefont {Aaij} \emph
  {et~al.}}]{LHCb:2020bwg}%
  \BibitemOpen
  \bibfield  {author} {\bibinfo {author} {\bibfnamefont {R.}~\bibnamefont
  {Aaij}} \emph {et~al.} (\bibinfo {collaboration} {LHCb}),\ }\href
  {https://doi.org/10.1016/j.scib.2020.08.032} {\bibfield  {journal} {\bibinfo
  {journal} {Sci. Bull.}\ }\textbf {\bibinfo {volume} {65}},\ \bibinfo {pages}
  {1983} (\bibinfo {year} {2020})},\ \Eprint {https://arxiv.org/abs/2006.16957}
  {arXiv:2006.16957 [hep-ex]} \BibitemShut {NoStop}%
\bibitem [{\citenamefont {Aaij}\ \emph {et~al.}(2022)\citenamefont {Aaij} \emph
  {et~al.}}]{LHCb:2021vvq}%
  \BibitemOpen
  \bibfield  {author} {\bibinfo {author} {\bibfnamefont {R.}~\bibnamefont
  {Aaij}} \emph {et~al.} (\bibinfo {collaboration} {LHCb}),\ }\href
  {https://doi.org/10.1038/s41567-022-01614-y} {\bibfield  {journal} {\bibinfo
  {journal} {Nature Phys.}\ }\textbf {\bibinfo {volume} {18}},\ \bibinfo
  {pages} {751} (\bibinfo {year} {2022})},\ \Eprint
  {https://arxiv.org/abs/2109.01038} {arXiv:2109.01038 [hep-ex]} \BibitemShut
  {NoStop}%
\bibitem [{\citenamefont {Amsler}\ and\ \citenamefont
  {Tornqvist}(2004)}]{Amsler:2004ps}%
  \BibitemOpen
  \bibfield  {author} {\bibinfo {author} {\bibfnamefont {C.}~\bibnamefont
  {Amsler}}\ and\ \bibinfo {author} {\bibfnamefont {N.~A.}\ \bibnamefont
  {Tornqvist}},\ }\href {https://doi.org/10.1016/j.physrep.2003.09.003}
  {\bibfield  {journal} {\bibinfo  {journal} {Phys. Rept.}\ }\textbf {\bibinfo
  {volume} {389}},\ \bibinfo {pages} {61} (\bibinfo {year} {2004})}\BibitemShut
  {NoStop}%
\bibitem [{\citenamefont {Zhu}(2004)}]{Zhu:2004xa}%
  \BibitemOpen
  \bibfield  {author} {\bibinfo {author} {\bibfnamefont {S.-L.}\ \bibnamefont
  {Zhu}},\ }\href {https://doi.org/10.1142/S0217751X04019676} {\bibfield
  {journal} {\bibinfo  {journal} {Int. J. Mod. Phys. A}\ }\textbf {\bibinfo
  {volume} {19}},\ \bibinfo {pages} {3439} (\bibinfo {year} {2004})},\ \Eprint
  {https://arxiv.org/abs/hep-ph/0406204} {arXiv:hep-ph/0406204} \BibitemShut
  {NoStop}%
\bibitem [{\citenamefont {Jaffe}(2005)}]{Jaffe:2004ph}%
  \BibitemOpen
  \bibfield  {author} {\bibinfo {author} {\bibfnamefont {R.~L.}\ \bibnamefont
  {Jaffe}},\ }\href {https://doi.org/10.1016/j.physrep.2004.11.005} {\bibfield
  {journal} {\bibinfo  {journal} {Phys. Rept.}\ }\textbf {\bibinfo {volume}
  {409}},\ \bibinfo {pages} {1} (\bibinfo {year} {2005})},\ \Eprint
  {https://arxiv.org/abs/hep-ph/0409065} {arXiv:hep-ph/0409065} \BibitemShut
  {NoStop}%
\bibitem [{\citenamefont {Swanson}(2006)}]{Swanson:2006st}%
  \BibitemOpen
  \bibfield  {author} {\bibinfo {author} {\bibfnamefont {E.~S.}\ \bibnamefont
  {Swanson}},\ }\href {https://doi.org/10.1016/j.physrep.2006.04.003}
  {\bibfield  {journal} {\bibinfo  {journal} {Phys. Rept.}\ }\textbf {\bibinfo
  {volume} {429}},\ \bibinfo {pages} {243} (\bibinfo {year} {2006})},\ \Eprint
  {https://arxiv.org/abs/hep-ph/0601110} {arXiv:hep-ph/0601110} \BibitemShut
  {NoStop}%
\bibitem [{\citenamefont {Zhu}(2008)}]{Zhu:2007wz}%
  \BibitemOpen
  \bibfield  {author} {\bibinfo {author} {\bibfnamefont {S.-L.}\ \bibnamefont
  {Zhu}},\ }\href {https://doi.org/10.1142/S0218301308009446} {\bibfield
  {journal} {\bibinfo  {journal} {Int. J. Mod. Phys. E}\ }\textbf {\bibinfo
  {volume} {17}},\ \bibinfo {pages} {283} (\bibinfo {year} {2008})},\ \Eprint
  {https://arxiv.org/abs/hep-ph/0703225} {arXiv:hep-ph/0703225} \BibitemShut
  {NoStop}%
\bibitem [{\citenamefont {Godfrey}\ and\ \citenamefont
  {Olsen}(2008)}]{Godfrey:2008nc}%
  \BibitemOpen
  \bibfield  {author} {\bibinfo {author} {\bibfnamefont {S.}~\bibnamefont
  {Godfrey}}\ and\ \bibinfo {author} {\bibfnamefont {S.~L.}\ \bibnamefont
  {Olsen}},\ }\href {https://doi.org/10.1146/annurev.nucl.58.110707.171145}
  {\bibfield  {journal} {\bibinfo  {journal} {Ann. Rev. Nucl. Part. Sci.}\
  }\textbf {\bibinfo {volume} {58}},\ \bibinfo {pages} {51} (\bibinfo {year}
  {2008})},\ \Eprint {https://arxiv.org/abs/0801.3867} {arXiv:0801.3867
  [hep-ph]} \BibitemShut {NoStop}%
\bibitem [{\citenamefont {Drenska}\ \emph {et~al.}(2010)\citenamefont
  {Drenska}, \citenamefont {Faccini}, \citenamefont {Piccinini}, \citenamefont
  {Polosa}, \citenamefont {Renga},\ and\ \citenamefont
  {Sabelli}}]{Drenska:2010kg}%
  \BibitemOpen
  \bibfield  {author} {\bibinfo {author} {\bibfnamefont {N.}~\bibnamefont
  {Drenska}}, \bibinfo {author} {\bibfnamefont {R.}~\bibnamefont {Faccini}},
  \bibinfo {author} {\bibfnamefont {F.}~\bibnamefont {Piccinini}}, \bibinfo
  {author} {\bibfnamefont {A.}~\bibnamefont {Polosa}}, \bibinfo {author}
  {\bibfnamefont {F.}~\bibnamefont {Renga}},\ and\ \bibinfo {author}
  {\bibfnamefont {C.}~\bibnamefont {Sabelli}},\ }\href
  {https://doi.org/10.1393/ncr/i2010-10059-8} {\bibfield  {journal} {\bibinfo
  {journal} {Riv. Nuovo Cim.}\ }\textbf {\bibinfo {volume} {33}},\ \bibinfo
  {pages} {633} (\bibinfo {year} {2010})},\ \Eprint
  {https://arxiv.org/abs/1006.2741} {arXiv:1006.2741 [hep-ph]} \BibitemShut
  {NoStop}%
\bibitem [{\citenamefont {Esposito}\ \emph {et~al.}(2015)\citenamefont
  {Esposito}, \citenamefont {Guerrieri}, \citenamefont {Piccinini},
  \citenamefont {Pilloni},\ and\ \citenamefont {Polosa}}]{Esposito:2014rxa}%
  \BibitemOpen
  \bibfield  {author} {\bibinfo {author} {\bibfnamefont {A.}~\bibnamefont
  {Esposito}}, \bibinfo {author} {\bibfnamefont {A.~L.}\ \bibnamefont
  {Guerrieri}}, \bibinfo {author} {\bibfnamefont {F.}~\bibnamefont
  {Piccinini}}, \bibinfo {author} {\bibfnamefont {A.}~\bibnamefont {Pilloni}},\
  and\ \bibinfo {author} {\bibfnamefont {A.~D.}\ \bibnamefont {Polosa}},\
  }\href {https://doi.org/10.1142/S0217751X15300021} {\bibfield  {journal}
  {\bibinfo  {journal} {Int. J. Mod. Phys. A}\ }\textbf {\bibinfo {volume}
  {30}},\ \bibinfo {pages} {1530002} (\bibinfo {year} {2015})},\ \Eprint
  {https://arxiv.org/abs/1411.5997} {arXiv:1411.5997 [hep-ph]} \BibitemShut
  {NoStop}%
\bibitem [{\citenamefont {Chen}\ \emph {et~al.}(2016)\citenamefont {Chen},
  \citenamefont {Chen}, \citenamefont {Liu},\ and\ \citenamefont
  {Zhu}}]{Chen:2016qju}%
  \BibitemOpen
  \bibfield  {author} {\bibinfo {author} {\bibfnamefont {H.-X.}\ \bibnamefont
  {Chen}}, \bibinfo {author} {\bibfnamefont {W.}~\bibnamefont {Chen}}, \bibinfo
  {author} {\bibfnamefont {X.}~\bibnamefont {Liu}},\ and\ \bibinfo {author}
  {\bibfnamefont {S.-L.}\ \bibnamefont {Zhu}},\ }\href
  {https://doi.org/10.1016/j.physrep.2016.05.004} {\bibfield  {journal}
  {\bibinfo  {journal} {Phys. Rept.}\ }\textbf {\bibinfo {volume} {639}},\
  \bibinfo {pages} {1} (\bibinfo {year} {2016})},\ \Eprint
  {https://arxiv.org/abs/1601.02092} {arXiv:1601.02092 [hep-ph]} \BibitemShut
  {NoStop}%
\bibitem [{\citenamefont {Richard}(2016)}]{Richard:2016eis}%
  \BibitemOpen
  \bibfield  {author} {\bibinfo {author} {\bibfnamefont {J.-M.}\ \bibnamefont
  {Richard}},\ }\href {https://doi.org/10.1007/s00601-016-1159-0} {\bibfield
  {journal} {\bibinfo  {journal} {Few Body Syst.}\ }\textbf {\bibinfo {volume}
  {57}},\ \bibinfo {pages} {1185} (\bibinfo {year} {2016})},\ \Eprint
  {https://arxiv.org/abs/1606.08593} {arXiv:1606.08593 [hep-ph]} \BibitemShut
  {NoStop}%
\bibitem [{\citenamefont {Lebed}\ \emph {et~al.}(2017)\citenamefont {Lebed},
  \citenamefont {Mitchell},\ and\ \citenamefont {Swanson}}]{Lebed:2016hpi}%
  \BibitemOpen
  \bibfield  {author} {\bibinfo {author} {\bibfnamefont {R.~F.}\ \bibnamefont
  {Lebed}}, \bibinfo {author} {\bibfnamefont {R.~E.}\ \bibnamefont
  {Mitchell}},\ and\ \bibinfo {author} {\bibfnamefont {E.~S.}\ \bibnamefont
  {Swanson}},\ }\href {https://doi.org/10.1016/j.ppnp.2016.11.003} {\bibfield
  {journal} {\bibinfo  {journal} {Prog. Part. Nucl. Phys.}\ }\textbf {\bibinfo
  {volume} {93}},\ \bibinfo {pages} {143} (\bibinfo {year} {2017})},\ \Eprint
  {https://arxiv.org/abs/1610.04528} {arXiv:1610.04528 [hep-ph]} \BibitemShut
  {NoStop}%
\bibitem [{\citenamefont {Ali}\ \emph {et~al.}(2017)\citenamefont {Ali},
  \citenamefont {Lange},\ and\ \citenamefont {Stone}}]{Ali:2017jda}%
  \BibitemOpen
  \bibfield  {author} {\bibinfo {author} {\bibfnamefont {A.}~\bibnamefont
  {Ali}}, \bibinfo {author} {\bibfnamefont {J.~S.}\ \bibnamefont {Lange}},\
  and\ \bibinfo {author} {\bibfnamefont {S.}~\bibnamefont {Stone}},\ }\href
  {https://doi.org/10.1016/j.ppnp.2017.08.003} {\bibfield  {journal} {\bibinfo
  {journal} {Prog. Part. Nucl. Phys.}\ }\textbf {\bibinfo {volume} {97}},\
  \bibinfo {pages} {123} (\bibinfo {year} {2017})},\ \Eprint
  {https://arxiv.org/abs/1706.00610} {arXiv:1706.00610 [hep-ph]} \BibitemShut
  {NoStop}%
\bibitem [{\citenamefont {Olsen}\ \emph {et~al.}(2018)\citenamefont {Olsen},
  \citenamefont {Skwarnicki},\ and\ \citenamefont {Zieminska}}]{Olsen:2017bmm}%
  \BibitemOpen
  \bibfield  {author} {\bibinfo {author} {\bibfnamefont {S.~L.}\ \bibnamefont
  {Olsen}}, \bibinfo {author} {\bibfnamefont {T.}~\bibnamefont {Skwarnicki}},\
  and\ \bibinfo {author} {\bibfnamefont {D.}~\bibnamefont {Zieminska}},\ }\href
  {https://doi.org/10.1103/RevModPhys.90.015003} {\bibfield  {journal}
  {\bibinfo  {journal} {Rev. Mod. Phys.}\ }\textbf {\bibinfo {volume} {90}},\
  \bibinfo {pages} {015003} (\bibinfo {year} {2018})},\ \Eprint
  {https://arxiv.org/abs/1708.04012} {arXiv:1708.04012 [hep-ph]} \BibitemShut
  {NoStop}%
\bibitem [{\citenamefont {Karliner}\ \emph {et~al.}(2018)\citenamefont
  {Karliner}, \citenamefont {Rosner},\ and\ \citenamefont
  {Skwarnicki}}]{Karliner:2017qhf}%
  \BibitemOpen
  \bibfield  {author} {\bibinfo {author} {\bibfnamefont {M.}~\bibnamefont
  {Karliner}}, \bibinfo {author} {\bibfnamefont {J.~L.}\ \bibnamefont
  {Rosner}},\ and\ \bibinfo {author} {\bibfnamefont {T.}~\bibnamefont
  {Skwarnicki}},\ }\href {https://doi.org/10.1146/annurev-nucl-101917-020902}
  {\bibfield  {journal} {\bibinfo  {journal} {Ann. Rev. Nucl. Part. Sci.}\
  }\textbf {\bibinfo {volume} {68}},\ \bibinfo {pages} {17} (\bibinfo {year}
  {2018})},\ \Eprint {https://arxiv.org/abs/1711.10626} {arXiv:1711.10626
  [hep-ph]} \BibitemShut {NoStop}%
\bibitem [{\citenamefont {Liu}\ \emph {et~al.}(2019{\natexlab{a}})\citenamefont
  {Liu}, \citenamefont {Chen}, \citenamefont {Chen}, \citenamefont {Liu},\ and\
  \citenamefont {Zhu}}]{Liu:2019zoy}%
  \BibitemOpen
  \bibfield  {author} {\bibinfo {author} {\bibfnamefont {Y.-R.}\ \bibnamefont
  {Liu}}, \bibinfo {author} {\bibfnamefont {H.-X.}\ \bibnamefont {Chen}},
  \bibinfo {author} {\bibfnamefont {W.}~\bibnamefont {Chen}}, \bibinfo {author}
  {\bibfnamefont {X.}~\bibnamefont {Liu}},\ and\ \bibinfo {author}
  {\bibfnamefont {S.-L.}\ \bibnamefont {Zhu}},\ }\href
  {https://doi.org/10.1016/j.ppnp.2019.04.003} {\bibfield  {journal} {\bibinfo
  {journal} {Prog. Part. Nucl. Phys.}\ }\textbf {\bibinfo {volume} {107}},\
  \bibinfo {pages} {237} (\bibinfo {year} {2019}{\natexlab{a}})},\ \Eprint
  {https://arxiv.org/abs/1903.11976} {arXiv:1903.11976 [hep-ph]} \BibitemShut
  {NoStop}%
\bibitem [{\citenamefont {Brambilla}\ \emph {et~al.}(2020)\citenamefont
  {Brambilla}, \citenamefont {Eidelman}, \citenamefont {Hanhart}, \citenamefont
  {Nefediev}, \citenamefont {Shen}, \citenamefont {Thomas}, \citenamefont
  {Vairo},\ and\ \citenamefont {Yuan}}]{Brambilla:2019esw}%
  \BibitemOpen
  \bibfield  {author} {\bibinfo {author} {\bibfnamefont {N.}~\bibnamefont
  {Brambilla}}, \bibinfo {author} {\bibfnamefont {S.}~\bibnamefont {Eidelman}},
  \bibinfo {author} {\bibfnamefont {C.}~\bibnamefont {Hanhart}}, \bibinfo
  {author} {\bibfnamefont {A.}~\bibnamefont {Nefediev}}, \bibinfo {author}
  {\bibfnamefont {C.-P.}\ \bibnamefont {Shen}}, \bibinfo {author}
  {\bibfnamefont {C.~E.}\ \bibnamefont {Thomas}}, \bibinfo {author}
  {\bibfnamefont {A.}~\bibnamefont {Vairo}},\ and\ \bibinfo {author}
  {\bibfnamefont {C.-Z.}\ \bibnamefont {Yuan}},\ }\href
  {https://doi.org/10.1016/j.physrep.2020.05.001} {\bibfield  {journal}
  {\bibinfo  {journal} {Phys. Rept.}\ }\textbf {\bibinfo {volume} {873}},\
  \bibinfo {pages} {1} (\bibinfo {year} {2020})},\ \Eprint
  {https://arxiv.org/abs/1907.07583} {arXiv:1907.07583 [hep-ex]} \BibitemShut
  {NoStop}%
\bibitem [{\citenamefont {Guo}\ \emph {et~al.}(2020)\citenamefont {Guo},
  \citenamefont {Liu},\ and\ \citenamefont {Sakai}}]{Guo:2019twa}%
  \BibitemOpen
  \bibfield  {author} {\bibinfo {author} {\bibfnamefont {F.-K.}\ \bibnamefont
  {Guo}}, \bibinfo {author} {\bibfnamefont {X.-H.}\ \bibnamefont {Liu}},\ and\
  \bibinfo {author} {\bibfnamefont {S.}~\bibnamefont {Sakai}},\ }\href
  {https://doi.org/10.1016/j.ppnp.2020.103757} {\bibfield  {journal} {\bibinfo
  {journal} {Prog. Part. Nucl. Phys.}\ }\textbf {\bibinfo {volume} {112}},\
  \bibinfo {pages} {103757} (\bibinfo {year} {2020})},\ \Eprint
  {https://arxiv.org/abs/1912.07030} {arXiv:1912.07030 [hep-ph]} \BibitemShut
  {NoStop}%
\bibitem [{\citenamefont {Barabanov}\ \emph {et~al.}(2021)\citenamefont
  {Barabanov} \emph {et~al.}}]{Barabanov:2020jvn}%
  \BibitemOpen
  \bibfield  {author} {\bibinfo {author} {\bibfnamefont {M.~Y.}\ \bibnamefont
  {Barabanov}} \emph {et~al.},\ }\href
  {https://doi.org/10.1016/j.ppnp.2020.103835} {\bibfield  {journal} {\bibinfo
  {journal} {Prog. Part. Nucl. Phys.}\ }\textbf {\bibinfo {volume} {116}},\
  \bibinfo {pages} {103835} (\bibinfo {year} {2021})},\ \Eprint
  {https://arxiv.org/abs/2008.07630} {arXiv:2008.07630 [hep-ph]} \BibitemShut
  {NoStop}%
\bibitem [{\citenamefont {Chen}\ \emph {et~al.}(2023)\citenamefont {Chen},
  \citenamefont {Chen}, \citenamefont {Liu}, \citenamefont {Liu},\ and\
  \citenamefont {Zhu}}]{Chen:2022asf}%
  \BibitemOpen
  \bibfield  {author} {\bibinfo {author} {\bibfnamefont {H.-X.}\ \bibnamefont
  {Chen}}, \bibinfo {author} {\bibfnamefont {W.}~\bibnamefont {Chen}}, \bibinfo
  {author} {\bibfnamefont {X.}~\bibnamefont {Liu}}, \bibinfo {author}
  {\bibfnamefont {Y.-R.}\ \bibnamefont {Liu}},\ and\ \bibinfo {author}
  {\bibfnamefont {S.-L.}\ \bibnamefont {Zhu}},\ }\href
  {https://doi.org/10.1088/1361-6633/aca3b6} {\bibfield  {journal} {\bibinfo
  {journal} {Rept. Prog. Phys.}\ }\textbf {\bibinfo {volume} {86}},\ \bibinfo
  {pages} {026201} (\bibinfo {year} {2023})},\ \Eprint
  {https://arxiv.org/abs/2204.02649} {arXiv:2204.02649 [hep-ph]} \BibitemShut
  {NoStop}%
\bibitem [{\citenamefont {Mai}\ \emph {et~al.}(2023)\citenamefont {Mai},
  \citenamefont {Mei{\ss}ner},\ and\ \citenamefont {Urbach}}]{Mai:2022eur}%
  \BibitemOpen
  \bibfield  {author} {\bibinfo {author} {\bibfnamefont {M.}~\bibnamefont
  {Mai}}, \bibinfo {author} {\bibfnamefont {U.-G.}\ \bibnamefont
  {Mei{\ss}ner}},\ and\ \bibinfo {author} {\bibfnamefont {C.}~\bibnamefont
  {Urbach}},\ }\href {https://doi.org/10.1016/j.physrep.2022.11.005} {\bibfield
   {journal} {\bibinfo  {journal} {Phys. Rept.}\ }\textbf {\bibinfo {volume}
  {1001}},\ \bibinfo {pages} {1} (\bibinfo {year} {2023})},\ \Eprint
  {https://arxiv.org/abs/2206.01477} {arXiv:2206.01477 [hep-ph]} \BibitemShut
  {NoStop}%
\bibitem [{\citenamefont {Gell-Mann}(1964)}]{Gell-Mann:1964ewy}%
  \BibitemOpen
  \bibfield  {author} {\bibinfo {author} {\bibfnamefont {M.}~\bibnamefont
  {Gell-Mann}},\ }\href {https://doi.org/10.1016/S0031-9163(64)92001-3}
  {\bibfield  {journal} {\bibinfo  {journal} {Phys. Lett.}\ }\textbf {\bibinfo
  {volume} {8}},\ \bibinfo {pages} {214} (\bibinfo {year} {1964})}\BibitemShut
  {NoStop}%
\bibitem [{\citenamefont {Zweig}(1964)}]{Zweig:1964ruk}%
  \BibitemOpen
  \bibfield  {author} {\bibinfo {author} {\bibfnamefont {G.}~\bibnamefont
  {Zweig}}\ }\href {https://doi.org/10.17181/CERN-TH-401}
  {10.17181/CERN-TH-401} (\bibinfo {year} {1964})\BibitemShut {NoStop}%
\bibitem [{\citenamefont {Hayrapetyan}\ \emph {et~al.}(2024)\citenamefont
  {Hayrapetyan} \emph {et~al.}}]{CMS:2023owd}%
  \BibitemOpen
  \bibfield  {author} {\bibinfo {author} {\bibfnamefont {A.}~\bibnamefont
  {Hayrapetyan}} \emph {et~al.} (\bibinfo {collaboration} {CMS}),\ }\href
  {https://doi.org/10.1103/PhysRevLett.132.111901} {\bibfield  {journal}
  {\bibinfo  {journal} {Phys. Rev. Lett.}\ }\textbf {\bibinfo {volume} {132}},\
  \bibinfo {pages} {111901} (\bibinfo {year} {2024})},\ \Eprint
  {https://arxiv.org/abs/2306.07164} {arXiv:2306.07164 [hep-ex]} \BibitemShut
  {NoStop}%
\bibitem [{\citenamefont {Liu}(2024)}]{Liu:2024biq}%
  \BibitemOpen
  \bibfield  {author} {\bibinfo {author} {\bibfnamefont {X.}~\bibnamefont
  {Liu}},\ }\href {https://doi.org/10.1016/j.scib.2024.07.033} {\bibfield
  {journal} {\bibinfo  {journal} {Sci. Bull.}\ }\textbf {\bibinfo {volume}
  {69}},\ \bibinfo {pages} {2802} (\bibinfo {year} {2024})},\ \Eprint
  {https://arxiv.org/abs/2407.13476} {arXiv:2407.13476 [hep-ph]} \BibitemShut
  {NoStop}%
\bibitem [{\citenamefont {Zhu}\ \emph {et~al.}(2024)\citenamefont {Zhu},
  \citenamefont {Bauer},\ and\ \citenamefont {Yi}}]{Zhu:2024swp}%
  \BibitemOpen
  \bibfield  {author} {\bibinfo {author} {\bibfnamefont {F.}~\bibnamefont
  {Zhu}}, \bibinfo {author} {\bibfnamefont {G.}~\bibnamefont {Bauer}},\ and\
  \bibinfo {author} {\bibfnamefont {K.}~\bibnamefont {Yi}},\ }\href
  {https://doi.org/10.1088/0256-307X/41/11/111201} {\bibfield  {journal}
  {\bibinfo  {journal} {Chin. Phys. Lett.}\ }\textbf {\bibinfo {volume} {41}},\
  \bibinfo {pages} {111201} (\bibinfo {year} {2024})},\ \Eprint
  {https://arxiv.org/abs/2410.11210} {arXiv:2410.11210 [hep-ph]} \BibitemShut
  {NoStop}%
\bibitem [{\citenamefont {Aad}\ \emph {et~al.}(2023)\citenamefont {Aad} \emph
  {et~al.}}]{ATLAS:2023bft}%
  \BibitemOpen
  \bibfield  {author} {\bibinfo {author} {\bibfnamefont {G.}~\bibnamefont
  {Aad}} \emph {et~al.} (\bibinfo {collaboration} {ATLAS}),\ }\href
  {https://doi.org/10.1103/PhysRevLett.131.151902} {\bibfield  {journal}
  {\bibinfo  {journal} {Phys. Rev. Lett.}\ }\textbf {\bibinfo {volume} {131}},\
  \bibinfo {pages} {151902} (\bibinfo {year} {2023})},\ \Eprint
  {https://arxiv.org/abs/2304.08962} {arXiv:2304.08962 [hep-ex]} \BibitemShut
  {NoStop}%
\bibitem [{\citenamefont {Hayrapetyan}\ \emph {et~al.}(2025)\citenamefont
  {Hayrapetyan} \emph {et~al.}}]{CMS:2025fpt}%
  \BibitemOpen
  \bibfield  {author} {\bibinfo {author} {\bibfnamefont {A.}~\bibnamefont
  {Hayrapetyan}} \emph {et~al.} (\bibinfo {collaboration} {CMS}),\ }\href
  {https://doi.org/10.1038/s41586-025-09711-7} {\bibfield  {journal} {\bibinfo
  {journal} {Nature}\ }\textbf {\bibinfo {volume} {648}},\ \bibinfo {pages}
  {58} (\bibinfo {year} {2025})},\ \Eprint {https://arxiv.org/abs/2506.07944}
  {arXiv:2506.07944 [hep-ex]} \BibitemShut {NoStop}%
\bibitem [{\citenamefont {Iwasaki}(1975)}]{Iwasaki:1975pv}%
  \BibitemOpen
  \bibfield  {author} {\bibinfo {author} {\bibfnamefont {Y.}~\bibnamefont
  {Iwasaki}},\ }\href {https://doi.org/10.1143/PTP.54.492} {\bibfield
  {journal} {\bibinfo  {journal} {Prog. Theor. Phys.}\ }\textbf {\bibinfo
  {volume} {54}},\ \bibinfo {pages} {492} (\bibinfo {year} {1975})}\BibitemShut
  {NoStop}%
\bibitem [{\citenamefont {Iwasaki}(1976)}]{Iwasaki:1976cn}%
  \BibitemOpen
  \bibfield  {author} {\bibinfo {author} {\bibfnamefont {Y.}~\bibnamefont
  {Iwasaki}},\ }\href {https://doi.org/10.1103/PhysRevLett.36.1266} {\bibfield
  {journal} {\bibinfo  {journal} {Phys. Rev. Lett.}\ }\textbf {\bibinfo
  {volume} {36}},\ \bibinfo {pages} {1266} (\bibinfo {year}
  {1976})}\BibitemShut {NoStop}%
\bibitem [{\citenamefont {Chao}(1981)}]{Chao:1980dv}%
  \BibitemOpen
  \bibfield  {author} {\bibinfo {author} {\bibfnamefont {K.-T.}\ \bibnamefont
  {Chao}},\ }\href {https://doi.org/10.1007/BF01431564} {\bibfield  {journal}
  {\bibinfo  {journal} {Z. Phys. C}\ }\textbf {\bibinfo {volume} {7}},\
  \bibinfo {pages} {317} (\bibinfo {year} {1981})}\BibitemShut {NoStop}%
\bibitem [{\citenamefont {Ader}\ \emph {et~al.}(1982)\citenamefont {Ader},
  \citenamefont {Richard},\ and\ \citenamefont {Taxil}}]{Ader:1981db}%
  \BibitemOpen
  \bibfield  {author} {\bibinfo {author} {\bibfnamefont {J.~P.}\ \bibnamefont
  {Ader}}, \bibinfo {author} {\bibfnamefont {J.~M.}\ \bibnamefont {Richard}},\
  and\ \bibinfo {author} {\bibfnamefont {P.}~\bibnamefont {Taxil}},\ }\href
  {https://doi.org/10.1103/PhysRevD.25.2370} {\bibfield  {journal} {\bibinfo
  {journal} {Phys. Rev. D}\ }\textbf {\bibinfo {volume} {25}},\ \bibinfo
  {pages} {2370} (\bibinfo {year} {1982})}\BibitemShut {NoStop}%
\bibitem [{\citenamefont {Heller}\ and\ \citenamefont
  {Tjon}(1985)}]{Heller:1985cb}%
  \BibitemOpen
  \bibfield  {author} {\bibinfo {author} {\bibfnamefont {L.}~\bibnamefont
  {Heller}}\ and\ \bibinfo {author} {\bibfnamefont {J.~A.}\ \bibnamefont
  {Tjon}},\ }\href {https://doi.org/10.1103/PhysRevD.32.755} {\bibfield
  {journal} {\bibinfo  {journal} {Phys. Rev. D}\ }\textbf {\bibinfo {volume}
  {32}},\ \bibinfo {pages} {755} (\bibinfo {year} {1985})}\BibitemShut
  {NoStop}%
\bibitem [{\citenamefont {Badalian}\ \emph {et~al.}(1987)\citenamefont
  {Badalian}, \citenamefont {Ioffe},\ and\ \citenamefont
  {Smilga}}]{Badalian:1985es}%
  \BibitemOpen
  \bibfield  {author} {\bibinfo {author} {\bibfnamefont {A.~M.}\ \bibnamefont
  {Badalian}}, \bibinfo {author} {\bibfnamefont {B.~L.}\ \bibnamefont
  {Ioffe}},\ and\ \bibinfo {author} {\bibfnamefont {A.~V.}\ \bibnamefont
  {Smilga}},\ }\href {https://doi.org/10.1016/0550-3213(87)90248-3} {\bibfield
  {journal} {\bibinfo  {journal} {Nucl. Phys. B}\ }\textbf {\bibinfo {volume}
  {281}},\ \bibinfo {pages} {85} (\bibinfo {year} {1987})}\BibitemShut
  {NoStop}%
\bibitem [{\citenamefont {Lloyd}\ and\ \citenamefont
  {Vary}(2004)}]{Lloyd:2003yc}%
  \BibitemOpen
  \bibfield  {author} {\bibinfo {author} {\bibfnamefont {R.~J.}\ \bibnamefont
  {Lloyd}}\ and\ \bibinfo {author} {\bibfnamefont {J.~P.}\ \bibnamefont
  {Vary}},\ }\href {https://doi.org/10.1103/PhysRevD.70.014009} {\bibfield
  {journal} {\bibinfo  {journal} {Phys. Rev. D}\ }\textbf {\bibinfo {volume}
  {70}},\ \bibinfo {pages} {014009} (\bibinfo {year} {2004})},\ \Eprint
  {https://arxiv.org/abs/hep-ph/0311179} {arXiv:hep-ph/0311179} \BibitemShut
  {NoStop}%
\bibitem [{\citenamefont {Berezhnoy}\ \emph {et~al.}(2011)\citenamefont
  {Berezhnoy}, \citenamefont {Likhoded}, \citenamefont {Luchinsky},\ and\
  \citenamefont {Novoselov}}]{Berezhnoy:2011xy}%
  \BibitemOpen
  \bibfield  {author} {\bibinfo {author} {\bibfnamefont {A.~V.}\ \bibnamefont
  {Berezhnoy}}, \bibinfo {author} {\bibfnamefont {A.~K.}\ \bibnamefont
  {Likhoded}}, \bibinfo {author} {\bibfnamefont {A.~V.}\ \bibnamefont
  {Luchinsky}},\ and\ \bibinfo {author} {\bibfnamefont {A.~A.}\ \bibnamefont
  {Novoselov}},\ }\href {https://doi.org/10.1103/PhysRevD.84.094023} {\bibfield
   {journal} {\bibinfo  {journal} {Phys. Rev. D}\ }\textbf {\bibinfo {volume}
  {84}},\ \bibinfo {pages} {094023} (\bibinfo {year} {2011})},\ \Eprint
  {https://arxiv.org/abs/1101.5881} {arXiv:1101.5881 [hep-ph]} \BibitemShut
  {NoStop}%
\bibitem [{\citenamefont {Berezhnoy}\ \emph {et~al.}(2012)\citenamefont
  {Berezhnoy}, \citenamefont {Luchinsky},\ and\ \citenamefont
  {Novoselov}}]{Berezhnoy:2011xn}%
  \BibitemOpen
  \bibfield  {author} {\bibinfo {author} {\bibfnamefont {A.~V.}\ \bibnamefont
  {Berezhnoy}}, \bibinfo {author} {\bibfnamefont {A.~V.}\ \bibnamefont
  {Luchinsky}},\ and\ \bibinfo {author} {\bibfnamefont {A.~A.}\ \bibnamefont
  {Novoselov}},\ }\href {https://doi.org/10.1103/PhysRevD.86.034004} {\bibfield
   {journal} {\bibinfo  {journal} {Phys. Rev. D}\ }\textbf {\bibinfo {volume}
  {86}},\ \bibinfo {pages} {034004} (\bibinfo {year} {2012})},\ \Eprint
  {https://arxiv.org/abs/1111.1867} {arXiv:1111.1867 [hep-ph]} \BibitemShut
  {NoStop}%
\bibitem [{\citenamefont {Heupel}\ \emph {et~al.}(2012)\citenamefont {Heupel},
  \citenamefont {Eichmann},\ and\ \citenamefont {Fischer}}]{Heupel:2012ua}%
  \BibitemOpen
  \bibfield  {author} {\bibinfo {author} {\bibfnamefont {W.}~\bibnamefont
  {Heupel}}, \bibinfo {author} {\bibfnamefont {G.}~\bibnamefont {Eichmann}},\
  and\ \bibinfo {author} {\bibfnamefont {C.~S.}\ \bibnamefont {Fischer}},\
  }\href {https://doi.org/10.1016/j.physletb.2012.11.009} {\bibfield  {journal}
  {\bibinfo  {journal} {Phys. Lett. B}\ }\textbf {\bibinfo {volume} {718}},\
  \bibinfo {pages} {545} (\bibinfo {year} {2012})},\ \Eprint
  {https://arxiv.org/abs/1206.5129} {arXiv:1206.5129 [hep-ph]} \BibitemShut
  {NoStop}%
\bibitem [{\citenamefont {Wu}\ \emph {et~al.}(2018)\citenamefont {Wu},
  \citenamefont {Liu}, \citenamefont {Chen}, \citenamefont {Liu},\ and\
  \citenamefont {Zhu}}]{Wu:2016vtq}%
  \BibitemOpen
  \bibfield  {author} {\bibinfo {author} {\bibfnamefont {J.}~\bibnamefont
  {Wu}}, \bibinfo {author} {\bibfnamefont {Y.-R.}\ \bibnamefont {Liu}},
  \bibinfo {author} {\bibfnamefont {K.}~\bibnamefont {Chen}}, \bibinfo {author}
  {\bibfnamefont {X.}~\bibnamefont {Liu}},\ and\ \bibinfo {author}
  {\bibfnamefont {S.-L.}\ \bibnamefont {Zhu}},\ }\href
  {https://doi.org/10.1103/PhysRevD.97.094015} {\bibfield  {journal} {\bibinfo
  {journal} {Phys. Rev. D}\ }\textbf {\bibinfo {volume} {97}},\ \bibinfo
  {pages} {094015} (\bibinfo {year} {2018})},\ \Eprint
  {https://arxiv.org/abs/1605.01134} {arXiv:1605.01134 [hep-ph]} \BibitemShut
  {NoStop}%
\bibitem [{\citenamefont {Chen}\ \emph {et~al.}(2017)\citenamefont {Chen},
  \citenamefont {Chen}, \citenamefont {Liu}, \citenamefont {Steele},\ and\
  \citenamefont {Zhu}}]{Chen:2016jxd}%
  \BibitemOpen
  \bibfield  {author} {\bibinfo {author} {\bibfnamefont {W.}~\bibnamefont
  {Chen}}, \bibinfo {author} {\bibfnamefont {H.-X.}\ \bibnamefont {Chen}},
  \bibinfo {author} {\bibfnamefont {X.}~\bibnamefont {Liu}}, \bibinfo {author}
  {\bibfnamefont {T.~G.}\ \bibnamefont {Steele}},\ and\ \bibinfo {author}
  {\bibfnamefont {S.-L.}\ \bibnamefont {Zhu}},\ }\href
  {https://doi.org/10.1016/j.physletb.2017.08.034} {\bibfield  {journal}
  {\bibinfo  {journal} {Phys. Lett. B}\ }\textbf {\bibinfo {volume} {773}},\
  \bibinfo {pages} {247} (\bibinfo {year} {2017})},\ \Eprint
  {https://arxiv.org/abs/1605.01647} {arXiv:1605.01647 [hep-ph]} \BibitemShut
  {NoStop}%
\bibitem [{\citenamefont {Karliner}\ \emph {et~al.}(2017)\citenamefont
  {Karliner}, \citenamefont {Nussinov},\ and\ \citenamefont
  {Rosner}}]{Karliner:2016zzc}%
  \BibitemOpen
  \bibfield  {author} {\bibinfo {author} {\bibfnamefont {M.}~\bibnamefont
  {Karliner}}, \bibinfo {author} {\bibfnamefont {S.}~\bibnamefont {Nussinov}},\
  and\ \bibinfo {author} {\bibfnamefont {J.~L.}\ \bibnamefont {Rosner}},\
  }\href {https://doi.org/10.1103/PhysRevD.95.034011} {\bibfield  {journal}
  {\bibinfo  {journal} {Phys. Rev. D}\ }\textbf {\bibinfo {volume} {95}},\
  \bibinfo {pages} {034011} (\bibinfo {year} {2017})},\ \Eprint
  {https://arxiv.org/abs/1611.00348} {arXiv:1611.00348 [hep-ph]} \BibitemShut
  {NoStop}%
\bibitem [{\citenamefont {Wang}(2017)}]{Wang:2017jtz}%
  \BibitemOpen
  \bibfield  {author} {\bibinfo {author} {\bibfnamefont {Z.-G.}\ \bibnamefont
  {Wang}},\ }\href {https://doi.org/10.1140/epjc/s10052-017-4997-0} {\bibfield
  {journal} {\bibinfo  {journal} {Eur. Phys. J. C}\ }\textbf {\bibinfo {volume}
  {77}},\ \bibinfo {pages} {432} (\bibinfo {year} {2017})},\ \Eprint
  {https://arxiv.org/abs/1701.04285} {arXiv:1701.04285 [hep-ph]} \BibitemShut
  {NoStop}%
\bibitem [{\citenamefont {Debastiani}\ and\ \citenamefont
  {Navarra}(2019)}]{Debastiani:2017msn}%
  \BibitemOpen
  \bibfield  {author} {\bibinfo {author} {\bibfnamefont {V.~R.}\ \bibnamefont
  {Debastiani}}\ and\ \bibinfo {author} {\bibfnamefont {F.~S.}\ \bibnamefont
  {Navarra}},\ }\href {https://doi.org/10.1088/1674-1137/43/1/013105}
  {\bibfield  {journal} {\bibinfo  {journal} {Chin. Phys. C}\ }\textbf
  {\bibinfo {volume} {43}},\ \bibinfo {pages} {013105} (\bibinfo {year}
  {2019})},\ \Eprint {https://arxiv.org/abs/1706.07553} {arXiv:1706.07553
  [hep-ph]} \BibitemShut {NoStop}%
\bibitem [{\citenamefont {Wang}\ and\ \citenamefont {Di}(2019)}]{Wang:2018poa}%
  \BibitemOpen
  \bibfield  {author} {\bibinfo {author} {\bibfnamefont {Z.-G.}\ \bibnamefont
  {Wang}}\ and\ \bibinfo {author} {\bibfnamefont {Z.-Y.}\ \bibnamefont {Di}},\
  }\href {https://doi.org/10.5506/APhysPolB.50.1335} {\bibfield  {journal}
  {\bibinfo  {journal} {Acta Phys. Polon. B}\ }\textbf {\bibinfo {volume}
  {50}},\ \bibinfo {pages} {1335} (\bibinfo {year} {2019})},\ \Eprint
  {https://arxiv.org/abs/1807.08520} {arXiv:1807.08520 [hep-ph]} \BibitemShut
  {NoStop}%
\bibitem [{\citenamefont {Liu}\ \emph {et~al.}(2019{\natexlab{b}})\citenamefont
  {Liu}, \citenamefont {L{\"u}}, \citenamefont {Zhong},\ and\ \citenamefont
  {Zhao}}]{Liu:2019zuc}%
  \BibitemOpen
  \bibfield  {author} {\bibinfo {author} {\bibfnamefont {M.-S.}\ \bibnamefont
  {Liu}}, \bibinfo {author} {\bibfnamefont {Q.-F.}\ \bibnamefont {L{\"u}}},
  \bibinfo {author} {\bibfnamefont {X.-H.}\ \bibnamefont {Zhong}},\ and\
  \bibinfo {author} {\bibfnamefont {Q.}~\bibnamefont {Zhao}},\ }\href
  {https://doi.org/10.1103/PhysRevD.100.016006} {\bibfield  {journal} {\bibinfo
   {journal} {Phys. Rev. D}\ }\textbf {\bibinfo {volume} {100}},\ \bibinfo
  {pages} {016006} (\bibinfo {year} {2019}{\natexlab{b}})},\ \Eprint
  {https://arxiv.org/abs/1901.02564} {arXiv:1901.02564 [hep-ph]} \BibitemShut
  {NoStop}%
\bibitem [{\citenamefont {Wang}\ \emph {et~al.}(2019)\citenamefont {Wang},
  \citenamefont {Meng},\ and\ \citenamefont {Zhu}}]{Wang:2019rdo}%
  \BibitemOpen
  \bibfield  {author} {\bibinfo {author} {\bibfnamefont {G.-J.}\ \bibnamefont
  {Wang}}, \bibinfo {author} {\bibfnamefont {L.}~\bibnamefont {Meng}},\ and\
  \bibinfo {author} {\bibfnamefont {S.-L.}\ \bibnamefont {Zhu}},\ }\href
  {https://doi.org/10.1103/PhysRevD.100.096013} {\bibfield  {journal} {\bibinfo
   {journal} {Phys. Rev. D}\ }\textbf {\bibinfo {volume} {100}},\ \bibinfo
  {pages} {096013} (\bibinfo {year} {2019})},\ \Eprint
  {https://arxiv.org/abs/1907.05177} {arXiv:1907.05177 [hep-ph]} \BibitemShut
  {NoStop}%
\bibitem [{\citenamefont {Bedolla}\ \emph {et~al.}(2020)\citenamefont
  {Bedolla}, \citenamefont {Ferretti}, \citenamefont {Roberts},\ and\
  \citenamefont {Santopinto}}]{Bedolla:2019zwg}%
  \BibitemOpen
  \bibfield  {author} {\bibinfo {author} {\bibfnamefont {M.~A.}\ \bibnamefont
  {Bedolla}}, \bibinfo {author} {\bibfnamefont {J.}~\bibnamefont {Ferretti}},
  \bibinfo {author} {\bibfnamefont {C.~D.}\ \bibnamefont {Roberts}},\ and\
  \bibinfo {author} {\bibfnamefont {E.}~\bibnamefont {Santopinto}},\ }\href
  {https://doi.org/10.1140/epjc/s10052-020-08579-3} {\bibfield  {journal}
  {\bibinfo  {journal} {Eur. Phys. J. C}\ }\textbf {\bibinfo {volume} {80}},\
  \bibinfo {pages} {1004} (\bibinfo {year} {2020})},\ \Eprint
  {https://arxiv.org/abs/1911.00960} {arXiv:1911.00960 [hep-ph]} \BibitemShut
  {NoStop}%
\bibitem [{\citenamefont {liu}\ \emph {et~al.}(2024)\citenamefont {liu},
  \citenamefont {Liu}, \citenamefont {Zhong},\ and\ \citenamefont
  {Zhao}}]{Liu:2020eha}%
  \BibitemOpen
  \bibfield  {author} {\bibinfo {author} {\bibfnamefont {M.-S.}\ \bibnamefont
  {liu}}, \bibinfo {author} {\bibfnamefont {F.-X.}\ \bibnamefont {Liu}},
  \bibinfo {author} {\bibfnamefont {X.-H.}\ \bibnamefont {Zhong}},\ and\
  \bibinfo {author} {\bibfnamefont {Q.}~\bibnamefont {Zhao}},\ }\href
  {https://doi.org/10.1103/PhysRevD.109.076017} {\bibfield  {journal} {\bibinfo
   {journal} {Phys. Rev. D}\ }\textbf {\bibinfo {volume} {109}},\ \bibinfo
  {pages} {076017} (\bibinfo {year} {2024})},\ \Eprint
  {https://arxiv.org/abs/2006.11952} {arXiv:2006.11952 [hep-ph]} \BibitemShut
  {NoStop}%
\bibitem [{\citenamefont {Wang}(2020)}]{Wang:2020ols}%
  \BibitemOpen
  \bibfield  {author} {\bibinfo {author} {\bibfnamefont {Z.-G.}\ \bibnamefont
  {Wang}},\ }\href {https://doi.org/10.1088/1674-1137/abb080} {\bibfield
  {journal} {\bibinfo  {journal} {Chin. Phys. C}\ }\textbf {\bibinfo {volume}
  {44}},\ \bibinfo {pages} {113106} (\bibinfo {year} {2020})},\ \Eprint
  {https://arxiv.org/abs/2006.13028} {arXiv:2006.13028 [hep-ph]} \BibitemShut
  {NoStop}%
\bibitem [{\citenamefont {Jin}\ \emph {et~al.}(2020)\citenamefont {Jin},
  \citenamefont {Xue}, \citenamefont {Huang},\ and\ \citenamefont
  {Ping}}]{Jin:2020jfc}%
  \BibitemOpen
  \bibfield  {author} {\bibinfo {author} {\bibfnamefont {X.}~\bibnamefont
  {Jin}}, \bibinfo {author} {\bibfnamefont {Y.}~\bibnamefont {Xue}}, \bibinfo
  {author} {\bibfnamefont {H.}~\bibnamefont {Huang}},\ and\ \bibinfo {author}
  {\bibfnamefont {J.}~\bibnamefont {Ping}},\ }\href
  {https://doi.org/10.1140/epjc/s10052-020-08650-z} {\bibfield  {journal}
  {\bibinfo  {journal} {Eur. Phys. J. C}\ }\textbf {\bibinfo {volume} {80}},\
  \bibinfo {pages} {1083} (\bibinfo {year} {2020})},\ \Eprint
  {https://arxiv.org/abs/2006.13745} {arXiv:2006.13745 [hep-ph]} \BibitemShut
  {NoStop}%
\bibitem [{\citenamefont {Becchi}\ \emph {et~al.}(2020)\citenamefont {Becchi},
  \citenamefont {Ferretti}, \citenamefont {Giachino}, \citenamefont {Maiani},\
  and\ \citenamefont {Santopinto}}]{Becchi:2020uvq}%
  \BibitemOpen
  \bibfield  {author} {\bibinfo {author} {\bibfnamefont {C.}~\bibnamefont
  {Becchi}}, \bibinfo {author} {\bibfnamefont {J.}~\bibnamefont {Ferretti}},
  \bibinfo {author} {\bibfnamefont {A.}~\bibnamefont {Giachino}}, \bibinfo
  {author} {\bibfnamefont {L.}~\bibnamefont {Maiani}},\ and\ \bibinfo {author}
  {\bibfnamefont {E.}~\bibnamefont {Santopinto}},\ }\href
  {https://doi.org/10.1016/j.physletb.2020.135952} {\bibfield  {journal}
  {\bibinfo  {journal} {Phys. Lett. B}\ }\textbf {\bibinfo {volume} {811}},\
  \bibinfo {pages} {135952} (\bibinfo {year} {2020})},\ \Eprint
  {https://arxiv.org/abs/2006.14388} {arXiv:2006.14388 [hep-ph]} \BibitemShut
  {NoStop}%
\bibitem [{\citenamefont {L{\"u}}\ \emph {et~al.}(2020)\citenamefont {L{\"u}},
  \citenamefont {Chen},\ and\ \citenamefont {Dong}}]{Lu:2020cns}%
  \BibitemOpen
  \bibfield  {author} {\bibinfo {author} {\bibfnamefont {Q.-F.}\ \bibnamefont
  {L{\"u}}}, \bibinfo {author} {\bibfnamefont {D.-Y.}\ \bibnamefont {Chen}},\
  and\ \bibinfo {author} {\bibfnamefont {Y.-B.}\ \bibnamefont {Dong}},\ }\href
  {https://doi.org/10.1140/epjc/s10052-020-08454-1} {\bibfield  {journal}
  {\bibinfo  {journal} {Eur. Phys. J. C}\ }\textbf {\bibinfo {volume} {80}},\
  \bibinfo {pages} {871} (\bibinfo {year} {2020})},\ \Eprint
  {https://arxiv.org/abs/2006.14445} {arXiv:2006.14445 [hep-ph]} \BibitemShut
  {NoStop}%
\bibitem [{\citenamefont {Chen}\ \emph
  {et~al.}(2020{\natexlab{a}})\citenamefont {Chen}, \citenamefont {Chen},
  \citenamefont {Liu},\ and\ \citenamefont {Zhu}}]{Chen:2020xwe}%
  \BibitemOpen
  \bibfield  {author} {\bibinfo {author} {\bibfnamefont {H.-X.}\ \bibnamefont
  {Chen}}, \bibinfo {author} {\bibfnamefont {W.}~\bibnamefont {Chen}}, \bibinfo
  {author} {\bibfnamefont {X.}~\bibnamefont {Liu}},\ and\ \bibinfo {author}
  {\bibfnamefont {S.-L.}\ \bibnamefont {Zhu}},\ }\href
  {https://doi.org/10.1016/j.scib.2020.08.038} {\bibfield  {journal} {\bibinfo
  {journal} {Sci. Bull.}\ }\textbf {\bibinfo {volume} {65}},\ \bibinfo {pages}
  {1994} (\bibinfo {year} {2020}{\natexlab{a}})},\ \Eprint
  {https://arxiv.org/abs/2006.16027} {arXiv:2006.16027 [hep-ph]} \BibitemShut
  {NoStop}%
\bibitem [{\citenamefont {Wang}\ \emph {et~al.}(2020)\citenamefont {Wang},
  \citenamefont {Lin}, \citenamefont {Xu}, \citenamefont {Xie}, \citenamefont
  {Huang},\ and\ \citenamefont {Chen}}]{Wang:2020gmd}%
  \BibitemOpen
  \bibfield  {author} {\bibinfo {author} {\bibfnamefont {X.-Y.}\ \bibnamefont
  {Wang}}, \bibinfo {author} {\bibfnamefont {Q.-Y.}\ \bibnamefont {Lin}},
  \bibinfo {author} {\bibfnamefont {H.}~\bibnamefont {Xu}}, \bibinfo {author}
  {\bibfnamefont {Y.-P.}\ \bibnamefont {Xie}}, \bibinfo {author} {\bibfnamefont
  {Y.}~\bibnamefont {Huang}},\ and\ \bibinfo {author} {\bibfnamefont
  {X.}~\bibnamefont {Chen}},\ }\href
  {https://doi.org/10.1103/PhysRevD.102.116014} {\bibfield  {journal} {\bibinfo
   {journal} {Phys. Rev. D}\ }\textbf {\bibinfo {volume} {102}},\ \bibinfo
  {pages} {116014} (\bibinfo {year} {2020})},\ \Eprint
  {https://arxiv.org/abs/2007.09697} {arXiv:2007.09697 [hep-ph]} \BibitemShut
  {NoStop}%
\bibitem [{\citenamefont {Albuquerque}\ \emph {et~al.}(2020)\citenamefont
  {Albuquerque}, \citenamefont {Narison}, \citenamefont {Rabemananjara},
  \citenamefont {Rabetiarivony},\ and\ \citenamefont
  {Randriamanatrika}}]{Albuquerque:2020hio}%
  \BibitemOpen
  \bibfield  {author} {\bibinfo {author} {\bibfnamefont {R.~M.}\ \bibnamefont
  {Albuquerque}}, \bibinfo {author} {\bibfnamefont {S.}~\bibnamefont
  {Narison}}, \bibinfo {author} {\bibfnamefont {A.}~\bibnamefont
  {Rabemananjara}}, \bibinfo {author} {\bibfnamefont {D.}~\bibnamefont
  {Rabetiarivony}},\ and\ \bibinfo {author} {\bibfnamefont {G.}~\bibnamefont
  {Randriamanatrika}},\ }\href {https://doi.org/10.1103/PhysRevD.102.094001}
  {\bibfield  {journal} {\bibinfo  {journal} {Phys. Rev. D}\ }\textbf {\bibinfo
  {volume} {102}},\ \bibinfo {pages} {094001} (\bibinfo {year} {2020})},\
  \Eprint {https://arxiv.org/abs/2008.01569} {arXiv:2008.01569 [hep-ph]}
  \BibitemShut {NoStop}%
\bibitem [{\citenamefont {Giron}\ and\ \citenamefont
  {Lebed}(2020)}]{Giron:2020wpx}%
  \BibitemOpen
  \bibfield  {author} {\bibinfo {author} {\bibfnamefont {J.~F.}\ \bibnamefont
  {Giron}}\ and\ \bibinfo {author} {\bibfnamefont {R.~F.}\ \bibnamefont
  {Lebed}},\ }\href {https://doi.org/10.1103/PhysRevD.102.074003} {\bibfield
  {journal} {\bibinfo  {journal} {Phys. Rev. D}\ }\textbf {\bibinfo {volume}
  {102}},\ \bibinfo {pages} {074003} (\bibinfo {year} {2020})},\ \Eprint
  {https://arxiv.org/abs/2008.01631} {arXiv:2008.01631 [hep-ph]} \BibitemShut
  {NoStop}%
\bibitem [{\citenamefont {Maiani}(2020)}]{Maiani:2020pur}%
  \BibitemOpen
  \bibfield  {author} {\bibinfo {author} {\bibfnamefont {L.}~\bibnamefont
  {Maiani}},\ }\href {https://doi.org/10.1016/j.scib.2020.08.019} {\bibfield
  {journal} {\bibinfo  {journal} {Sci. Bull.}\ }\textbf {\bibinfo {volume}
  {65}},\ \bibinfo {pages} {1949} (\bibinfo {year} {2020})},\ \Eprint
  {https://arxiv.org/abs/2008.01637} {arXiv:2008.01637 [hep-ph]} \BibitemShut
  {NoStop}%
\bibitem [{\citenamefont {Richard}(2020)}]{Richard:2020hdw}%
  \BibitemOpen
  \bibfield  {author} {\bibinfo {author} {\bibfnamefont {J.-M.}\ \bibnamefont
  {Richard}},\ }\href {https://doi.org/10.1016/j.scib.2020.08.020} {\bibfield
  {journal} {\bibinfo  {journal} {Sci. Bull.}\ }\textbf {\bibinfo {volume}
  {65}},\ \bibinfo {pages} {1954} (\bibinfo {year} {2020})},\ \Eprint
  {https://arxiv.org/abs/2008.01962} {arXiv:2008.01962 [hep-ph]} \BibitemShut
  {NoStop}%
\bibitem [{\citenamefont {Wang}\ \emph
  {et~al.}(2021{\natexlab{a}})\citenamefont {Wang}, \citenamefont {Chen},
  \citenamefont {Liu},\ and\ \citenamefont {Matsuki}}]{Wang:2020wrp}%
  \BibitemOpen
  \bibfield  {author} {\bibinfo {author} {\bibfnamefont {J.-Z.}\ \bibnamefont
  {Wang}}, \bibinfo {author} {\bibfnamefont {D.-Y.}\ \bibnamefont {Chen}},
  \bibinfo {author} {\bibfnamefont {X.}~\bibnamefont {Liu}},\ and\ \bibinfo
  {author} {\bibfnamefont {T.}~\bibnamefont {Matsuki}},\ }\href
  {https://doi.org/10.1103/PhysRevD.103.L071503} {\bibfield  {journal}
  {\bibinfo  {journal} {Phys. Rev. D}\ }\textbf {\bibinfo {volume} {103}},\
  \bibinfo {pages} {071503} (\bibinfo {year} {2021}{\natexlab{a}})},\ \Eprint
  {https://arxiv.org/abs/2008.07430} {arXiv:2008.07430 [hep-ph]} \BibitemShut
  {NoStop}%
\bibitem [{\citenamefont {Chao}\ and\ \citenamefont
  {Zhu}(2020)}]{Chao:2020dml}%
  \BibitemOpen
  \bibfield  {author} {\bibinfo {author} {\bibfnamefont {K.-T.}\ \bibnamefont
  {Chao}}\ and\ \bibinfo {author} {\bibfnamefont {S.-L.}\ \bibnamefont {Zhu}},\
  }\href {https://doi.org/10.1016/j.scib.2020.08.031} {\bibfield  {journal}
  {\bibinfo  {journal} {Sci. Bull.}\ }\textbf {\bibinfo {volume} {65}},\
  \bibinfo {pages} {1952} (\bibinfo {year} {2020})},\ \Eprint
  {https://arxiv.org/abs/2008.07670} {arXiv:2008.07670 [hep-ph]} \BibitemShut
  {NoStop}%
\bibitem [{\citenamefont {Maciu{\l}a}\ \emph {et~al.}(2021)\citenamefont
  {Maciu{\l}a}, \citenamefont {Sch{\"a}fer},\ and\ \citenamefont
  {Szczurek}}]{Maciula:2020wri}%
  \BibitemOpen
  \bibfield  {author} {\bibinfo {author} {\bibfnamefont {R.}~\bibnamefont
  {Maciu{\l}a}}, \bibinfo {author} {\bibfnamefont {W.}~\bibnamefont
  {Sch{\"a}fer}},\ and\ \bibinfo {author} {\bibfnamefont {A.}~\bibnamefont
  {Szczurek}},\ }\href {https://doi.org/10.1016/j.physletb.2020.136010}
  {\bibfield  {journal} {\bibinfo  {journal} {Phys. Lett. B}\ }\textbf
  {\bibinfo {volume} {812}},\ \bibinfo {pages} {136010} (\bibinfo {year}
  {2021})},\ \Eprint {https://arxiv.org/abs/2009.02100} {arXiv:2009.02100
  [hep-ph]} \BibitemShut {NoStop}%
\bibitem [{\citenamefont {Karliner}\ and\ \citenamefont
  {Rosner}(2020)}]{Karliner:2020dta}%
  \BibitemOpen
  \bibfield  {author} {\bibinfo {author} {\bibfnamefont {M.}~\bibnamefont
  {Karliner}}\ and\ \bibinfo {author} {\bibfnamefont {J.~L.}\ \bibnamefont
  {Rosner}},\ }\href {https://doi.org/10.1103/PhysRevD.102.114039} {\bibfield
  {journal} {\bibinfo  {journal} {Phys. Rev. D}\ }\textbf {\bibinfo {volume}
  {102}},\ \bibinfo {pages} {114039} (\bibinfo {year} {2020})},\ \Eprint
  {https://arxiv.org/abs/2009.04429} {arXiv:2009.04429 [hep-ph]} \BibitemShut
  {NoStop}%
\bibitem [{\citenamefont {Wang}(2021)}]{Wang:2020dlo}%
  \BibitemOpen
  \bibfield  {author} {\bibinfo {author} {\bibfnamefont {Z.-G.}\ \bibnamefont
  {Wang}},\ }\href {https://doi.org/10.1142/S0217751X21500147} {\bibfield
  {journal} {\bibinfo  {journal} {Int. J. Mod. Phys. A}\ }\textbf {\bibinfo
  {volume} {36}},\ \bibinfo {pages} {2150014} (\bibinfo {year} {2021})},\
  \Eprint {https://arxiv.org/abs/2009.05371} {arXiv:2009.05371 [hep-ph]}
  \BibitemShut {NoStop}%
\bibitem [{\citenamefont {Dong}\ \emph
  {et~al.}(2021{\natexlab{a}})\citenamefont {Dong}, \citenamefont {Baru},
  \citenamefont {Guo}, \citenamefont {Hanhart},\ and\ \citenamefont
  {Nefediev}}]{Dong:2020nwy}%
  \BibitemOpen
  \bibfield  {author} {\bibinfo {author} {\bibfnamefont {X.-K.}\ \bibnamefont
  {Dong}}, \bibinfo {author} {\bibfnamefont {V.}~\bibnamefont {Baru}}, \bibinfo
  {author} {\bibfnamefont {F.-K.}\ \bibnamefont {Guo}}, \bibinfo {author}
  {\bibfnamefont {C.}~\bibnamefont {Hanhart}},\ and\ \bibinfo {author}
  {\bibfnamefont {A.}~\bibnamefont {Nefediev}},\ }\href
  {https://doi.org/10.1103/PhysRevLett.127.119901} {\bibfield  {journal}
  {\bibinfo  {journal} {Phys. Rev. Lett.}\ }\textbf {\bibinfo {volume} {126}},\
  \bibinfo {pages} {132001} (\bibinfo {year} {2021}{\natexlab{a}})},\ \bibinfo
  {note} {[Erratum: Phys.Rev.Lett. 127, 119901 (2021)]},\ \Eprint
  {https://arxiv.org/abs/2009.07795} {arXiv:2009.07795 [hep-ph]} \BibitemShut
  {NoStop}%
\bibitem [{\citenamefont {Zhang}\ \emph {et~al.}(2025)\citenamefont {Zhang},
  \citenamefont {Ma},\ and\ \citenamefont {Sang}}]{Zhang:2020hoh}%
  \BibitemOpen
  \bibfield  {author} {\bibinfo {author} {\bibfnamefont {H.-F.}\ \bibnamefont
  {Zhang}}, \bibinfo {author} {\bibfnamefont {Y.-Q.}\ \bibnamefont {Ma}},\ and\
  \bibinfo {author} {\bibfnamefont {W.-L.}\ \bibnamefont {Sang}},\ }\href
  {https://doi.org/10.1016/j.scib.2025.04.035} {\bibfield  {journal} {\bibinfo
  {journal} {Sci. Bull.}\ }\textbf {\bibinfo {volume} {70}},\ \bibinfo {pages}
  {1915} (\bibinfo {year} {2025})},\ \Eprint {https://arxiv.org/abs/2009.08376}
  {arXiv:2009.08376 [hep-ph]} \BibitemShut {NoStop}%
\bibitem [{\citenamefont {Feng}\ \emph {et~al.}(2022)\citenamefont {Feng},
  \citenamefont {Huang}, \citenamefont {Jia}, \citenamefont {Sang},
  \citenamefont {Xiong},\ and\ \citenamefont {Zhang}}]{Feng:2020riv}%
  \BibitemOpen
  \bibfield  {author} {\bibinfo {author} {\bibfnamefont {F.}~\bibnamefont
  {Feng}}, \bibinfo {author} {\bibfnamefont {Y.}~\bibnamefont {Huang}},
  \bibinfo {author} {\bibfnamefont {Y.}~\bibnamefont {Jia}}, \bibinfo {author}
  {\bibfnamefont {W.-L.}\ \bibnamefont {Sang}}, \bibinfo {author}
  {\bibfnamefont {X.}~\bibnamefont {Xiong}},\ and\ \bibinfo {author}
  {\bibfnamefont {J.-Y.}\ \bibnamefont {Zhang}},\ }\href
  {https://doi.org/10.1103/PhysRevD.106.114029} {\bibfield  {journal} {\bibinfo
   {journal} {Phys. Rev. D}\ }\textbf {\bibinfo {volume} {106}},\ \bibinfo
  {pages} {114029} (\bibinfo {year} {2022})},\ \Eprint
  {https://arxiv.org/abs/2009.08450} {arXiv:2009.08450 [hep-ph]} \BibitemShut
  {NoStop}%
\bibitem [{\citenamefont {Zhao}\ \emph {et~al.}(2020)\citenamefont {Zhao},
  \citenamefont {Shi},\ and\ \citenamefont {Zhuang}}]{Zhao:2020nwy}%
  \BibitemOpen
  \bibfield  {author} {\bibinfo {author} {\bibfnamefont {J.}~\bibnamefont
  {Zhao}}, \bibinfo {author} {\bibfnamefont {S.}~\bibnamefont {Shi}},\ and\
  \bibinfo {author} {\bibfnamefont {P.}~\bibnamefont {Zhuang}},\ }\href
  {https://doi.org/10.1103/PhysRevD.102.114001} {\bibfield  {journal} {\bibinfo
   {journal} {Phys. Rev. D}\ }\textbf {\bibinfo {volume} {102}},\ \bibinfo
  {pages} {114001} (\bibinfo {year} {2020})},\ \Eprint
  {https://arxiv.org/abs/2009.10319} {arXiv:2009.10319 [hep-ph]} \BibitemShut
  {NoStop}%
\bibitem [{\citenamefont {Faustov}\ \emph {et~al.}(2020)\citenamefont
  {Faustov}, \citenamefont {Galkin},\ and\ \citenamefont
  {Savchenko}}]{Faustov:2020qfm}%
  \BibitemOpen
  \bibfield  {author} {\bibinfo {author} {\bibfnamefont {R.~N.}\ \bibnamefont
  {Faustov}}, \bibinfo {author} {\bibfnamefont {V.~O.}\ \bibnamefont
  {Galkin}},\ and\ \bibinfo {author} {\bibfnamefont {E.~M.}\ \bibnamefont
  {Savchenko}},\ }\href {https://doi.org/10.1103/PhysRevD.102.114030}
  {\bibfield  {journal} {\bibinfo  {journal} {Phys. Rev. D}\ }\textbf {\bibinfo
  {volume} {102}},\ \bibinfo {pages} {114030} (\bibinfo {year} {2020})},\
  \Eprint {https://arxiv.org/abs/2009.13237} {arXiv:2009.13237 [hep-ph]}
  \BibitemShut {NoStop}%
\bibitem [{\citenamefont {Weng}\ \emph {et~al.}(2021)\citenamefont {Weng},
  \citenamefont {Chen}, \citenamefont {Deng},\ and\ \citenamefont
  {Zhu}}]{Weng:2020jao}%
  \BibitemOpen
  \bibfield  {author} {\bibinfo {author} {\bibfnamefont {X.-Z.}\ \bibnamefont
  {Weng}}, \bibinfo {author} {\bibfnamefont {X.-L.}\ \bibnamefont {Chen}},
  \bibinfo {author} {\bibfnamefont {W.-Z.}\ \bibnamefont {Deng}},\ and\
  \bibinfo {author} {\bibfnamefont {S.-L.}\ \bibnamefont {Zhu}},\ }\href
  {https://doi.org/10.1103/PhysRevD.103.034001} {\bibfield  {journal} {\bibinfo
   {journal} {Phys. Rev. D}\ }\textbf {\bibinfo {volume} {103}},\ \bibinfo
  {pages} {034001} (\bibinfo {year} {2021})},\ \Eprint
  {https://arxiv.org/abs/2010.05163} {arXiv:2010.05163 [hep-ph]} \BibitemShut
  {NoStop}%
\bibitem [{\citenamefont {Zhang}(2021)}]{Zhang:2020xtb}%
  \BibitemOpen
  \bibfield  {author} {\bibinfo {author} {\bibfnamefont {J.-R.}\ \bibnamefont
  {Zhang}},\ }\href {https://doi.org/10.1103/PhysRevD.103.014018} {\bibfield
  {journal} {\bibinfo  {journal} {Phys. Rev. D}\ }\textbf {\bibinfo {volume}
  {103}},\ \bibinfo {pages} {014018} (\bibinfo {year} {2021})},\ \Eprint
  {https://arxiv.org/abs/2010.07719} {arXiv:2010.07719 [hep-ph]} \BibitemShut
  {NoStop}%
\bibitem [{\citenamefont {Zhu}(2021)}]{Zhu:2020xni}%
  \BibitemOpen
  \bibfield  {author} {\bibinfo {author} {\bibfnamefont {R.}~\bibnamefont
  {Zhu}},\ }\href {https://doi.org/10.1016/j.nuclphysb.2021.115393} {\bibfield
  {journal} {\bibinfo  {journal} {Nucl. Phys. B}\ }\textbf {\bibinfo {volume}
  {966}},\ \bibinfo {pages} {115393} (\bibinfo {year} {2021})},\ \Eprint
  {https://arxiv.org/abs/2010.09082} {arXiv:2010.09082 [hep-ph]} \BibitemShut
  {NoStop}%
\bibitem [{\citenamefont {Guo}\ and\ \citenamefont
  {Oller}(2021)}]{Guo:2020pvt}%
  \BibitemOpen
  \bibfield  {author} {\bibinfo {author} {\bibfnamefont {Z.-H.}\ \bibnamefont
  {Guo}}\ and\ \bibinfo {author} {\bibfnamefont {J.~A.}\ \bibnamefont
  {Oller}},\ }\href {https://doi.org/10.1103/PhysRevD.103.034024} {\bibfield
  {journal} {\bibinfo  {journal} {Phys. Rev. D}\ }\textbf {\bibinfo {volume}
  {103}},\ \bibinfo {pages} {034024} (\bibinfo {year} {2021})},\ \Eprint
  {https://arxiv.org/abs/2011.00978} {arXiv:2011.00978 [hep-ph]} \BibitemShut
  {NoStop}%
\bibitem [{\citenamefont {Cao}\ \emph {et~al.}(2021)\citenamefont {Cao},
  \citenamefont {Chen}, \citenamefont {Qi},\ and\ \citenamefont
  {Zheng}}]{Cao:2020gul}%
  \BibitemOpen
  \bibfield  {author} {\bibinfo {author} {\bibfnamefont {Q.-F.}\ \bibnamefont
  {Cao}}, \bibinfo {author} {\bibfnamefont {H.}~\bibnamefont {Chen}}, \bibinfo
  {author} {\bibfnamefont {H.-R.}\ \bibnamefont {Qi}},\ and\ \bibinfo {author}
  {\bibfnamefont {H.-Q.}\ \bibnamefont {Zheng}},\ }\href
  {https://doi.org/10.1088/1674-1137/ac0ee5} {\bibfield  {journal} {\bibinfo
  {journal} {Chin. Phys. C}\ }\textbf {\bibinfo {volume} {45}},\ \bibinfo
  {pages} {103102} (\bibinfo {year} {2021})},\ \Eprint
  {https://arxiv.org/abs/2011.04347} {arXiv:2011.04347 [hep-ph]} \BibitemShut
  {NoStop}%
\bibitem [{\citenamefont {Gong}\ \emph
  {et~al.}(2022{\natexlab{a}})\citenamefont {Gong}, \citenamefont {Du},
  \citenamefont {Zhao}, \citenamefont {Zhong},\ and\ \citenamefont
  {Zhou}}]{Gong:2020bmg}%
  \BibitemOpen
  \bibfield  {author} {\bibinfo {author} {\bibfnamefont {C.}~\bibnamefont
  {Gong}}, \bibinfo {author} {\bibfnamefont {M.-C.}\ \bibnamefont {Du}},
  \bibinfo {author} {\bibfnamefont {Q.}~\bibnamefont {Zhao}}, \bibinfo {author}
  {\bibfnamefont {X.-H.}\ \bibnamefont {Zhong}},\ and\ \bibinfo {author}
  {\bibfnamefont {B.}~\bibnamefont {Zhou}},\ }\href
  {https://doi.org/10.1016/j.physletb.2021.136794} {\bibfield  {journal}
  {\bibinfo  {journal} {Phys. Lett. B}\ }\textbf {\bibinfo {volume} {824}},\
  \bibinfo {pages} {136794} (\bibinfo {year} {2022}{\natexlab{a}})},\ \Eprint
  {https://arxiv.org/abs/2011.11374} {arXiv:2011.11374 [hep-ph]} \BibitemShut
  {NoStop}%
\bibitem [{\citenamefont {Wan}\ and\ \citenamefont {Qiao}(2021)}]{Wan:2020fsk}%
  \BibitemOpen
  \bibfield  {author} {\bibinfo {author} {\bibfnamefont {B.-D.}\ \bibnamefont
  {Wan}}\ and\ \bibinfo {author} {\bibfnamefont {C.-F.}\ \bibnamefont {Qiao}},\
  }\href {https://doi.org/10.1016/j.physletb.2021.136339} {\bibfield  {journal}
  {\bibinfo  {journal} {Phys. Lett. B}\ }\textbf {\bibinfo {volume} {817}},\
  \bibinfo {pages} {136339} (\bibinfo {year} {2021})},\ \Eprint
  {https://arxiv.org/abs/2012.00454} {arXiv:2012.00454 [hep-ph]} \BibitemShut
  {NoStop}%
\bibitem [{\citenamefont {Yang}\ \emph {et~al.}(2021)\citenamefont {Yang},
  \citenamefont {Tang},\ and\ \citenamefont {Qiao}}]{Yang:2020wkh}%
  \BibitemOpen
  \bibfield  {author} {\bibinfo {author} {\bibfnamefont {B.-C.}\ \bibnamefont
  {Yang}}, \bibinfo {author} {\bibfnamefont {L.}~\bibnamefont {Tang}},\ and\
  \bibinfo {author} {\bibfnamefont {C.-F.}\ \bibnamefont {Qiao}},\ }\href
  {https://doi.org/10.1140/epjc/s10052-021-09096-7} {\bibfield  {journal}
  {\bibinfo  {journal} {Eur. Phys. J. C}\ }\textbf {\bibinfo {volume} {81}},\
  \bibinfo {pages} {324} (\bibinfo {year} {2021})},\ \Eprint
  {https://arxiv.org/abs/2012.04463} {arXiv:2012.04463 [hep-ph]} \BibitemShut
  {NoStop}%
\bibitem [{\citenamefont {Zhao}\ \emph {et~al.}(2021)\citenamefont {Zhao},
  \citenamefont {Xu}, \citenamefont {Kaewsnod}, \citenamefont {Liu},
  \citenamefont {Limphirat},\ and\ \citenamefont {Yan}}]{Zhao:2020jvl}%
  \BibitemOpen
  \bibfield  {author} {\bibinfo {author} {\bibfnamefont {Z.}~\bibnamefont
  {Zhao}}, \bibinfo {author} {\bibfnamefont {K.}~\bibnamefont {Xu}}, \bibinfo
  {author} {\bibfnamefont {A.}~\bibnamefont {Kaewsnod}}, \bibinfo {author}
  {\bibfnamefont {X.}~\bibnamefont {Liu}}, \bibinfo {author} {\bibfnamefont
  {A.}~\bibnamefont {Limphirat}},\ and\ \bibinfo {author} {\bibfnamefont
  {Y.}~\bibnamefont {Yan}},\ }\href
  {https://doi.org/10.1103/PhysRevD.103.116027} {\bibfield  {journal} {\bibinfo
   {journal} {Phys. Rev. D}\ }\textbf {\bibinfo {volume} {103}},\ \bibinfo
  {pages} {116027} (\bibinfo {year} {2021})},\ \Eprint
  {https://arxiv.org/abs/2012.15554} {arXiv:2012.15554 [hep-ph]} \BibitemShut
  {NoStop}%
\bibitem [{\citenamefont {Ke}\ \emph {et~al.}(2021)\citenamefont {Ke},
  \citenamefont {Han}, \citenamefont {Liu},\ and\ \citenamefont
  {Shi}}]{Ke:2021iyh}%
  \BibitemOpen
  \bibfield  {author} {\bibinfo {author} {\bibfnamefont {H.-W.}\ \bibnamefont
  {Ke}}, \bibinfo {author} {\bibfnamefont {X.}~\bibnamefont {Han}}, \bibinfo
  {author} {\bibfnamefont {X.-H.}\ \bibnamefont {Liu}},\ and\ \bibinfo {author}
  {\bibfnamefont {Y.-L.}\ \bibnamefont {Shi}},\ }\href
  {https://doi.org/10.1140/epjc/s10052-021-09229-y} {\bibfield  {journal}
  {\bibinfo  {journal} {Eur. Phys. J. C}\ }\textbf {\bibinfo {volume} {81}},\
  \bibinfo {pages} {427} (\bibinfo {year} {2021})},\ \Eprint
  {https://arxiv.org/abs/2103.13140} {arXiv:2103.13140 [hep-ph]} \BibitemShut
  {NoStop}%
\bibitem [{\citenamefont {Mutuk}(2021)}]{Mutuk:2021hmi}%
  \BibitemOpen
  \bibfield  {author} {\bibinfo {author} {\bibfnamefont {H.}~\bibnamefont
  {Mutuk}},\ }\href {https://doi.org/10.1140/epjc/s10052-021-09176-8}
  {\bibfield  {journal} {\bibinfo  {journal} {Eur. Phys. J. C}\ }\textbf
  {\bibinfo {volume} {81}},\ \bibinfo {pages} {367} (\bibinfo {year} {2021})},\
  \Eprint {https://arxiv.org/abs/2104.11823} {arXiv:2104.11823 [hep-ph]}
  \BibitemShut {NoStop}%
\bibitem [{\citenamefont {Wang}\ \emph
  {et~al.}(2021{\natexlab{b}})\citenamefont {Wang}, \citenamefont {Meng},
  \citenamefont {Oka},\ and\ \citenamefont {Zhu}}]{Wang:2021kfv}%
  \BibitemOpen
  \bibfield  {author} {\bibinfo {author} {\bibfnamefont {G.-J.}\ \bibnamefont
  {Wang}}, \bibinfo {author} {\bibfnamefont {L.}~\bibnamefont {Meng}}, \bibinfo
  {author} {\bibfnamefont {M.}~\bibnamefont {Oka}},\ and\ \bibinfo {author}
  {\bibfnamefont {S.-L.}\ \bibnamefont {Zhu}},\ }\href
  {https://doi.org/10.1103/PhysRevD.104.036016} {\bibfield  {journal} {\bibinfo
   {journal} {Phys. Rev. D}\ }\textbf {\bibinfo {volume} {104}},\ \bibinfo
  {pages} {036016} (\bibinfo {year} {2021}{\natexlab{b}})},\ \Eprint
  {https://arxiv.org/abs/2105.13109} {arXiv:2105.13109 [hep-ph]} \BibitemShut
  {NoStop}%
\bibitem [{\citenamefont {Tiwari}\ \emph {et~al.}(2023)\citenamefont {Tiwari},
  \citenamefont {Rathaud},\ and\ \citenamefont {Rai}}]{Tiwari:2021tmz}%
  \BibitemOpen
  \bibfield  {author} {\bibinfo {author} {\bibfnamefont {R.}~\bibnamefont
  {Tiwari}}, \bibinfo {author} {\bibfnamefont {D.~P.}\ \bibnamefont
  {Rathaud}},\ and\ \bibinfo {author} {\bibfnamefont {A.~K.}\ \bibnamefont
  {Rai}},\ }\href {https://doi.org/10.1007/s12648-022-02427-8} {\bibfield
  {journal} {\bibinfo  {journal} {Indian J. Phys.}\ }\textbf {\bibinfo {volume}
  {97}},\ \bibinfo {pages} {943} (\bibinfo {year} {2023})},\ \Eprint
  {https://arxiv.org/abs/2108.04017} {arXiv:2108.04017 [hep-ph]} \BibitemShut
  {NoStop}%
\bibitem [{\citenamefont {Wang}\ \emph
  {et~al.}(2021{\natexlab{c}})\citenamefont {Wang}, \citenamefont {Yang},\ and\
  \citenamefont {Chen}}]{Wang:2021mma}%
  \BibitemOpen
  \bibfield  {author} {\bibinfo {author} {\bibfnamefont {Q.-N.}\ \bibnamefont
  {Wang}}, \bibinfo {author} {\bibfnamefont {Z.-Y.}\ \bibnamefont {Yang}},\
  and\ \bibinfo {author} {\bibfnamefont {W.}~\bibnamefont {Chen}},\ }\href
  {https://doi.org/10.1103/PhysRevD.104.114037} {\bibfield  {journal} {\bibinfo
   {journal} {Phys. Rev. D}\ }\textbf {\bibinfo {volume} {104}},\ \bibinfo
  {pages} {114037} (\bibinfo {year} {2021}{\natexlab{c}})},\ \Eprint
  {https://arxiv.org/abs/2109.08091} {arXiv:2109.08091 [hep-ph]} \BibitemShut
  {NoStop}%
\bibitem [{\citenamefont {Liu}\ \emph {et~al.}(2021)\citenamefont {Liu},
  \citenamefont {Liu}, \citenamefont {Zhong},\ and\ \citenamefont
  {Zhao}}]{Liu:2021rtn}%
  \BibitemOpen
  \bibfield  {author} {\bibinfo {author} {\bibfnamefont {F.-X.}\ \bibnamefont
  {Liu}}, \bibinfo {author} {\bibfnamefont {M.-S.}\ \bibnamefont {Liu}},
  \bibinfo {author} {\bibfnamefont {X.-H.}\ \bibnamefont {Zhong}},\ and\
  \bibinfo {author} {\bibfnamefont {Q.}~\bibnamefont {Zhao}},\ }\href
  {https://doi.org/10.1103/PhysRevD.104.116029} {\bibfield  {journal} {\bibinfo
   {journal} {Phys. Rev. D}\ }\textbf {\bibinfo {volume} {104}},\ \bibinfo
  {pages} {116029} (\bibinfo {year} {2021})},\ \Eprint
  {https://arxiv.org/abs/2110.09052} {arXiv:2110.09052 [hep-ph]} \BibitemShut
  {NoStop}%
\bibitem [{\citenamefont {Kuang}\ \emph {et~al.}(2022)\citenamefont {Kuang},
  \citenamefont {Serafin}, \citenamefont {Zhao},\ and\ \citenamefont
  {Vary}}]{Kuang:2022vdy}%
  \BibitemOpen
  \bibfield  {author} {\bibinfo {author} {\bibfnamefont {Z.}~\bibnamefont
  {Kuang}}, \bibinfo {author} {\bibfnamefont {K.}~\bibnamefont {Serafin}},
  \bibinfo {author} {\bibfnamefont {X.}~\bibnamefont {Zhao}},\ and\ \bibinfo
  {author} {\bibfnamefont {J.~P.}\ \bibnamefont {Vary}} (\bibinfo
  {collaboration} {BLFQ}),\ }\href
  {https://doi.org/10.1103/PhysRevD.105.094028} {\bibfield  {journal} {\bibinfo
   {journal} {Phys. Rev. D}\ }\textbf {\bibinfo {volume} {105}},\ \bibinfo
  {pages} {094028} (\bibinfo {year} {2022})},\ \Eprint
  {https://arxiv.org/abs/2201.06428} {arXiv:2201.06428 [hep-ph]} \BibitemShut
  {NoStop}%
\bibitem [{\citenamefont {Yang}\ \emph {et~al.}(2022)\citenamefont {Yang},
  \citenamefont {Ping},\ and\ \citenamefont {Segovia}}]{Yang:2022bfu}%
  \BibitemOpen
  \bibfield  {author} {\bibinfo {author} {\bibfnamefont {G.}~\bibnamefont
  {Yang}}, \bibinfo {author} {\bibfnamefont {J.}~\bibnamefont {Ping}},\ and\
  \bibinfo {author} {\bibfnamefont {J.}~\bibnamefont {Segovia}},\ }\href
  {https://doi.org/10.1103/PhysRevD.106.014005} {\bibfield  {journal} {\bibinfo
   {journal} {Phys. Rev. D}\ }\textbf {\bibinfo {volume} {106}},\ \bibinfo
  {pages} {014005} (\bibinfo {year} {2022})},\ \Eprint
  {https://arxiv.org/abs/2205.11548} {arXiv:2205.11548 [hep-ph]} \BibitemShut
  {NoStop}%
\bibitem [{\citenamefont {Gong}\ \emph
  {et~al.}(2022{\natexlab{b}})\citenamefont {Gong}, \citenamefont {Du},\ and\
  \citenamefont {Zhao}}]{Gong:2022hgd}%
  \BibitemOpen
  \bibfield  {author} {\bibinfo {author} {\bibfnamefont {C.}~\bibnamefont
  {Gong}}, \bibinfo {author} {\bibfnamefont {M.-C.}\ \bibnamefont {Du}},\ and\
  \bibinfo {author} {\bibfnamefont {Q.}~\bibnamefont {Zhao}},\ }\href
  {https://doi.org/10.1103/PhysRevD.106.054011} {\bibfield  {journal} {\bibinfo
   {journal} {Phys. Rev. D}\ }\textbf {\bibinfo {volume} {106}},\ \bibinfo
  {pages} {054011} (\bibinfo {year} {2022}{\natexlab{b}})},\ \Eprint
  {https://arxiv.org/abs/2206.13867} {arXiv:2206.13867 [hep-ph]} \BibitemShut
  {NoStop}%
\bibitem [{\citenamefont {Zhou}\ \emph {et~al.}(2022)\citenamefont {Zhou},
  \citenamefont {Guo}, \citenamefont {Kuang}, \citenamefont {Yang},\ and\
  \citenamefont {Dai}}]{Zhou:2022xpd}%
  \BibitemOpen
  \bibfield  {author} {\bibinfo {author} {\bibfnamefont {Q.}~\bibnamefont
  {Zhou}}, \bibinfo {author} {\bibfnamefont {D.}~\bibnamefont {Guo}}, \bibinfo
  {author} {\bibfnamefont {S.-Q.}\ \bibnamefont {Kuang}}, \bibinfo {author}
  {\bibfnamefont {Q.-H.}\ \bibnamefont {Yang}},\ and\ \bibinfo {author}
  {\bibfnamefont {L.-Y.}\ \bibnamefont {Dai}},\ }\href
  {https://doi.org/10.1103/PhysRevD.106.L111502} {\bibfield  {journal}
  {\bibinfo  {journal} {Phys. Rev. D}\ }\textbf {\bibinfo {volume} {106}},\
  \bibinfo {pages} {L111502} (\bibinfo {year} {2022})},\ \Eprint
  {https://arxiv.org/abs/2207.07537} {arXiv:2207.07537 [hep-ph]} \BibitemShut
  {NoStop}%
\bibitem [{\citenamefont {Wang}(2022)}]{Wang:2022xja}%
  \BibitemOpen
  \bibfield  {author} {\bibinfo {author} {\bibfnamefont {Z.-G.}\ \bibnamefont
  {Wang}},\ }\href {https://doi.org/10.1016/j.nuclphysb.2022.115983} {\bibfield
   {journal} {\bibinfo  {journal} {Nucl. Phys. B}\ }\textbf {\bibinfo {volume}
  {985}},\ \bibinfo {pages} {115983} (\bibinfo {year} {2022})},\ \Eprint
  {https://arxiv.org/abs/2207.08059} {arXiv:2207.08059 [hep-ph]} \BibitemShut
  {NoStop}%
\bibitem [{\citenamefont {Chen}\ \emph {et~al.}(2022)\citenamefont {Chen},
  \citenamefont {Yan},\ and\ \citenamefont {Chen}}]{Chen:2022sbf}%
  \BibitemOpen
  \bibfield  {author} {\bibinfo {author} {\bibfnamefont {H.-X.}\ \bibnamefont
  {Chen}}, \bibinfo {author} {\bibfnamefont {Y.-X.}\ \bibnamefont {Yan}},\ and\
  \bibinfo {author} {\bibfnamefont {W.}~\bibnamefont {Chen}},\ }\href
  {https://doi.org/10.1103/PhysRevD.106.094019} {\bibfield  {journal} {\bibinfo
   {journal} {Phys. Rev. D}\ }\textbf {\bibinfo {volume} {106}},\ \bibinfo
  {pages} {094019} (\bibinfo {year} {2022})},\ \Eprint
  {https://arxiv.org/abs/2207.08593} {arXiv:2207.08593 [hep-ph]} \BibitemShut
  {NoStop}%
\bibitem [{\citenamefont {Biloshytskyi}\ \emph {et~al.}(2022)\citenamefont
  {Biloshytskyi}, \citenamefont {Pascalutsa}, \citenamefont {Harland-Lang},
  \citenamefont {Malaescu}, \citenamefont {Schmieden},\ and\ \citenamefont
  {Schott}}]{Biloshytskyi:2022dmo}%
  \BibitemOpen
  \bibfield  {author} {\bibinfo {author} {\bibfnamefont {V.}~\bibnamefont
  {Biloshytskyi}}, \bibinfo {author} {\bibfnamefont {V.}~\bibnamefont
  {Pascalutsa}}, \bibinfo {author} {\bibfnamefont {L.}~\bibnamefont
  {Harland-Lang}}, \bibinfo {author} {\bibfnamefont {B.}~\bibnamefont
  {Malaescu}}, \bibinfo {author} {\bibfnamefont {K.}~\bibnamefont
  {Schmieden}},\ and\ \bibinfo {author} {\bibfnamefont {M.}~\bibnamefont
  {Schott}},\ }\href {https://doi.org/10.1103/PhysRevD.106.L111902} {\bibfield
  {journal} {\bibinfo  {journal} {Phys. Rev. D}\ }\textbf {\bibinfo {volume}
  {106}},\ \bibinfo {pages} {L111902} (\bibinfo {year} {2022})},\ \Eprint
  {https://arxiv.org/abs/2207.13623} {arXiv:2207.13623 [hep-ph]} \BibitemShut
  {NoStop}%
\bibitem [{\citenamefont {Wang}\ \emph {et~al.}(2022)\citenamefont {Wang},
  \citenamefont {Meng},\ and\ \citenamefont {Oka}}]{Wang:2022yes}%
  \BibitemOpen
  \bibfield  {author} {\bibinfo {author} {\bibfnamefont {G.-J.}\ \bibnamefont
  {Wang}}, \bibinfo {author} {\bibfnamefont {Q.}~\bibnamefont {Meng}},\ and\
  \bibinfo {author} {\bibfnamefont {M.}~\bibnamefont {Oka}},\ }\href
  {https://doi.org/10.1103/PhysRevD.106.096005} {\bibfield  {journal} {\bibinfo
   {journal} {Phys. Rev. D}\ }\textbf {\bibinfo {volume} {106}},\ \bibinfo
  {pages} {096005} (\bibinfo {year} {2022})},\ \Eprint
  {https://arxiv.org/abs/2208.07292} {arXiv:2208.07292 [hep-ph]} \BibitemShut
  {NoStop}%
\bibitem [{\citenamefont {Niu}\ \emph {et~al.}(2025)\citenamefont {Niu},
  \citenamefont {Ding}, \citenamefont {Wang},\ and\ \citenamefont
  {Yang}}]{Niu:2022cug}%
  \BibitemOpen
  \bibfield  {author} {\bibinfo {author} {\bibfnamefont {P.-Y.}\ \bibnamefont
  {Niu}}, \bibinfo {author} {\bibfnamefont {Y.-C.}\ \bibnamefont {Ding}},
  \bibinfo {author} {\bibfnamefont {Q.}~\bibnamefont {Wang}},\ and\ \bibinfo
  {author} {\bibfnamefont {S.}~\bibnamefont {Yang}},\ }\href
  {https://doi.org/10.1088/0256-307X/42/10/100202} {\bibfield  {journal}
  {\bibinfo  {journal} {Chin. Phys. Lett.}\ }\textbf {\bibinfo {volume} {42}},\
  \bibinfo {pages} {100202} (\bibinfo {year} {2025})},\ \Eprint
  {https://arxiv.org/abs/2209.01924} {arXiv:2209.01924 [hep-ph]} \BibitemShut
  {NoStop}%
\bibitem [{\citenamefont {Yu}\ \emph {et~al.}(2023)\citenamefont {Yu},
  \citenamefont {Li}, \citenamefont {Wang}, \citenamefont {Lu},\ and\
  \citenamefont {Yan}}]{Yu:2022lak}%
  \BibitemOpen
  \bibfield  {author} {\bibinfo {author} {\bibfnamefont {G.-L.}\ \bibnamefont
  {Yu}}, \bibinfo {author} {\bibfnamefont {Z.-Y.}\ \bibnamefont {Li}}, \bibinfo
  {author} {\bibfnamefont {Z.-G.}\ \bibnamefont {Wang}}, \bibinfo {author}
  {\bibfnamefont {J.}~\bibnamefont {Lu}},\ and\ \bibinfo {author}
  {\bibfnamefont {M.}~\bibnamefont {Yan}},\ }\href
  {https://doi.org/10.1140/epjc/s10052-023-11445-7} {\bibfield  {journal}
  {\bibinfo  {journal} {Eur. Phys. J. C}\ }\textbf {\bibinfo {volume} {83}},\
  \bibinfo {pages} {416} (\bibinfo {year} {2023})},\ \Eprint
  {https://arxiv.org/abs/2212.14339} {arXiv:2212.14339 [hep-ph]} \BibitemShut
  {NoStop}%
\bibitem [{\citenamefont {Kuang}\ \emph {et~al.}(2023)\citenamefont {Kuang},
  \citenamefont {Zhou}, \citenamefont {Guo}, \citenamefont {Yang},\ and\
  \citenamefont {Dai}}]{Kuang:2023vac}%
  \BibitemOpen
  \bibfield  {author} {\bibinfo {author} {\bibfnamefont {S.-Q.}\ \bibnamefont
  {Kuang}}, \bibinfo {author} {\bibfnamefont {Q.}~\bibnamefont {Zhou}},
  \bibinfo {author} {\bibfnamefont {D.}~\bibnamefont {Guo}}, \bibinfo {author}
  {\bibfnamefont {Q.-H.}\ \bibnamefont {Yang}},\ and\ \bibinfo {author}
  {\bibfnamefont {L.-Y.}\ \bibnamefont {Dai}},\ }\href
  {https://doi.org/10.1140/epjc/s10052-023-11473-3} {\bibfield  {journal}
  {\bibinfo  {journal} {Eur. Phys. J. C}\ }\textbf {\bibinfo {volume} {83}},\
  \bibinfo {pages} {383} (\bibinfo {year} {2023})},\ \Eprint
  {https://arxiv.org/abs/2302.03968} {arXiv:2302.03968 [hep-ph]} \BibitemShut
  {NoStop}%
\bibitem [{\citenamefont {Lu}\ \emph {et~al.}(2023)\citenamefont {Lu},
  \citenamefont {Chen}, \citenamefont {Kang}, \citenamefont {Qin},\ and\
  \citenamefont {Zheng}}]{Lu:2023ccs}%
  \BibitemOpen
  \bibfield  {author} {\bibinfo {author} {\bibfnamefont {Y.}~\bibnamefont
  {Lu}}, \bibinfo {author} {\bibfnamefont {C.}~\bibnamefont {Chen}}, \bibinfo
  {author} {\bibfnamefont {K.-G.}\ \bibnamefont {Kang}}, \bibinfo {author}
  {\bibfnamefont {G.-y.}\ \bibnamefont {Qin}},\ and\ \bibinfo {author}
  {\bibfnamefont {H.-Q.}\ \bibnamefont {Zheng}},\ }\href
  {https://doi.org/10.1103/PhysRevD.107.094006} {\bibfield  {journal} {\bibinfo
   {journal} {Phys. Rev. D}\ }\textbf {\bibinfo {volume} {107}},\ \bibinfo
  {pages} {094006} (\bibinfo {year} {2023})},\ \Eprint
  {https://arxiv.org/abs/2302.04150} {arXiv:2302.04150 [hep-ph]} \BibitemShut
  {NoStop}%
\bibitem [{\citenamefont {Agaev}\ \emph
  {et~al.}(2023{\natexlab{a}})\citenamefont {Agaev}, \citenamefont {Azizi},
  \citenamefont {Barsbay},\ and\ \citenamefont {Sundu}}]{Agaev:2023wua}%
  \BibitemOpen
  \bibfield  {author} {\bibinfo {author} {\bibfnamefont {S.~S.}\ \bibnamefont
  {Agaev}}, \bibinfo {author} {\bibfnamefont {K.}~\bibnamefont {Azizi}},
  \bibinfo {author} {\bibfnamefont {B.}~\bibnamefont {Barsbay}},\ and\ \bibinfo
  {author} {\bibfnamefont {H.}~\bibnamefont {Sundu}},\ }\href
  {https://doi.org/10.1016/j.physletb.2023.138089} {\bibfield  {journal}
  {\bibinfo  {journal} {Phys. Lett. B}\ }\textbf {\bibinfo {volume} {844}},\
  \bibinfo {pages} {138089} (\bibinfo {year} {2023}{\natexlab{a}})},\ \Eprint
  {https://arxiv.org/abs/2304.03244} {arXiv:2304.03244 [hep-ph]} \BibitemShut
  {NoStop}%
\bibitem [{\citenamefont {Agaev}\ \emph {et~al.}(2024)\citenamefont {Agaev},
  \citenamefont {Azizi}, \citenamefont {Barsbay},\ and\ \citenamefont
  {Sundu}}]{Agaev:2023gaq}%
  \BibitemOpen
  \bibfield  {author} {\bibinfo {author} {\bibfnamefont {S.~S.}\ \bibnamefont
  {Agaev}}, \bibinfo {author} {\bibfnamefont {K.}~\bibnamefont {Azizi}},
  \bibinfo {author} {\bibfnamefont {B.}~\bibnamefont {Barsbay}},\ and\ \bibinfo
  {author} {\bibfnamefont {H.}~\bibnamefont {Sundu}},\ }\href
  {https://doi.org/10.1016/j.nuclphysa.2023.122768} {\bibfield  {journal}
  {\bibinfo  {journal} {Nucl. Phys. A}\ }\textbf {\bibinfo {volume} {1041}},\
  \bibinfo {pages} {122768} (\bibinfo {year} {2024})},\ \Eprint
  {https://arxiv.org/abs/2304.09943} {arXiv:2304.09943 [hep-ph]} \BibitemShut
  {NoStop}%
\bibitem [{\citenamefont {Agaev}\ \emph
  {et~al.}(2023{\natexlab{b}})\citenamefont {Agaev}, \citenamefont {Azizi},
  \citenamefont {Barsbay},\ and\ \citenamefont {Sundu}}]{Agaev:2023rpj}%
  \BibitemOpen
  \bibfield  {author} {\bibinfo {author} {\bibfnamefont {S.~S.}\ \bibnamefont
  {Agaev}}, \bibinfo {author} {\bibfnamefont {K.}~\bibnamefont {Azizi}},
  \bibinfo {author} {\bibfnamefont {B.}~\bibnamefont {Barsbay}},\ and\ \bibinfo
  {author} {\bibfnamefont {H.}~\bibnamefont {Sundu}},\ }\href
  {https://doi.org/10.1140/epjc/s10052-023-12145-y} {\bibfield  {journal}
  {\bibinfo  {journal} {Eur. Phys. J. C}\ }\textbf {\bibinfo {volume} {83}},\
  \bibinfo {pages} {994} (\bibinfo {year} {2023}{\natexlab{b}})},\ \Eprint
  {https://arxiv.org/abs/2307.01857} {arXiv:2307.01857 [hep-ph]} \BibitemShut
  {NoStop}%
\bibitem [{\citenamefont {Wang}\ and\ \citenamefont
  {Yang}(2024)}]{Wang:2023kir}%
  \BibitemOpen
  \bibfield  {author} {\bibinfo {author} {\bibfnamefont {Z.-G.}\ \bibnamefont
  {Wang}}\ and\ \bibinfo {author} {\bibfnamefont {X.-S.}\ \bibnamefont
  {Yang}},\ }\href {https://doi.org/10.1007/s43673-023-00112-4} {\bibfield
  {journal} {\bibinfo  {journal} {AAPPS Bull.}\ }\textbf {\bibinfo {volume}
  {34}},\ \bibinfo {pages} {5} (\bibinfo {year} {2024})},\ \Eprint
  {https://arxiv.org/abs/2310.16583} {arXiv:2310.16583 [hep-ph]} \BibitemShut
  {NoStop}%
\bibitem [{\citenamefont {Anwar}\ and\ \citenamefont
  {Burns}(2024)}]{Anwar:2023fbp}%
  \BibitemOpen
  \bibfield  {author} {\bibinfo {author} {\bibfnamefont {M.~N.}\ \bibnamefont
  {Anwar}}\ and\ \bibinfo {author} {\bibfnamefont {T.~J.}\ \bibnamefont
  {Burns}},\ }\href {https://doi.org/10.1103/PhysRevD.110.034012} {\bibfield
  {journal} {\bibinfo  {journal} {Phys. Rev. D}\ }\textbf {\bibinfo {volume}
  {110}},\ \bibinfo {pages} {034012} (\bibinfo {year} {2024})},\ \Eprint
  {https://arxiv.org/abs/2311.15853} {arXiv:2311.15853 [hep-ph]} \BibitemShut
  {NoStop}%
\bibitem [{\citenamefont {Tang}\ \emph {et~al.}(2024)\citenamefont {Tang},
  \citenamefont {Duan},\ and\ \citenamefont {Tang}}]{Tang:2024zvf}%
  \BibitemOpen
  \bibfield  {author} {\bibinfo {author} {\bibfnamefont {C.-M.}\ \bibnamefont
  {Tang}}, \bibinfo {author} {\bibfnamefont {C.-G.}\ \bibnamefont {Duan}},\
  and\ \bibinfo {author} {\bibfnamefont {L.}~\bibnamefont {Tang}},\ }\href
  {https://doi.org/10.1140/epjc/s10052-024-13102-z} {\bibfield  {journal}
  {\bibinfo  {journal} {Eur. Phys. J. C}\ }\textbf {\bibinfo {volume} {84}},\
  \bibinfo {pages} {743} (\bibinfo {year} {2024})},\ \Eprint
  {https://arxiv.org/abs/2405.05042} {arXiv:2405.05042 [hep-ph]} \BibitemShut
  {NoStop}%
\bibitem [{\citenamefont {Yang}\ and\ \citenamefont
  {Wang}(2025)}]{Yang:2024guo}%
  \BibitemOpen
  \bibfield  {author} {\bibinfo {author} {\bibfnamefont {X.-S.}\ \bibnamefont
  {Yang}}\ and\ \bibinfo {author} {\bibfnamefont {Z.-G.}\ \bibnamefont
  {Wang}},\ }\href {https://doi.org/10.1088/1674-1137/adc081} {\bibfield
  {journal} {\bibinfo  {journal} {Chin. Phys. C}\ }\textbf {\bibinfo {volume}
  {49}},\ \bibinfo {pages} {063108} (\bibinfo {year} {2025})},\ \Eprint
  {https://arxiv.org/abs/2409.05428} {arXiv:2409.05428 [hep-ph]} \BibitemShut
  {NoStop}%
\bibitem [{\citenamefont {Wu}\ and\ \citenamefont {Zhu}(2025)}]{Wu:2024ocq}%
  \BibitemOpen
  \bibfield  {author} {\bibinfo {author} {\bibfnamefont {W.-L.}\ \bibnamefont
  {Wu}}\ and\ \bibinfo {author} {\bibfnamefont {S.-L.}\ \bibnamefont {Zhu}},\
  }\href {https://doi.org/10.1103/PhysRevD.111.034044} {\bibfield  {journal}
  {\bibinfo  {journal} {Phys. Rev. D}\ }\textbf {\bibinfo {volume} {111}},\
  \bibinfo {pages} {034044} (\bibinfo {year} {2025})},\ \Eprint
  {https://arxiv.org/abs/2411.17962} {arXiv:2411.17962 [hep-ph]} \BibitemShut
  {NoStop}%
\bibitem [{\citenamefont {Chen}\ \emph {et~al.}(2024)\citenamefont {Chen},
  \citenamefont {Liu}, \citenamefont {Zhao}, \citenamefont {Zhong},
  \citenamefont {Zhu},\ and\ \citenamefont {Zou}}]{Chen:2024orv}%
  \BibitemOpen
  \bibfield  {author} {\bibinfo {author} {\bibfnamefont {K.}~\bibnamefont
  {Chen}}, \bibinfo {author} {\bibfnamefont {F.-X.}\ \bibnamefont {Liu}},
  \bibinfo {author} {\bibfnamefont {Q.}~\bibnamefont {Zhao}}, \bibinfo {author}
  {\bibfnamefont {X.-H.}\ \bibnamefont {Zhong}}, \bibinfo {author}
  {\bibfnamefont {R.}~\bibnamefont {Zhu}},\ and\ \bibinfo {author}
  {\bibfnamefont {B.-S.}\ \bibnamefont {Zou}},\ }\href@noop {} {\  (\bibinfo
  {year} {2024})},\ \Eprint {https://arxiv.org/abs/2412.13455}
  {arXiv:2412.13455 [hep-ph]} \BibitemShut {NoStop}%
\bibitem [{\citenamefont {Kalamidas}\ and\ \citenamefont
  {Vanderhaeghen}(2025)}]{Kalamidas:2025gen}%
  \BibitemOpen
  \bibfield  {author} {\bibinfo {author} {\bibfnamefont {P.}~\bibnamefont
  {Kalamidas}}\ and\ \bibinfo {author} {\bibfnamefont {M.}~\bibnamefont
  {Vanderhaeghen}},\ }\href {https://doi.org/10.1103/PhysRevD.111.094033}
  {\bibfield  {journal} {\bibinfo  {journal} {Phys. Rev. D}\ }\textbf {\bibinfo
  {volume} {111}},\ \bibinfo {pages} {094033} (\bibinfo {year} {2025})},\
  \Eprint {https://arxiv.org/abs/2501.06034} {arXiv:2501.06034 [hep-ph]}
  \BibitemShut {NoStop}%
\bibitem [{\citenamefont {Ayd{\i}n}\ \emph {et~al.}(2025)\citenamefont
  {Ayd{\i}n}, \citenamefont {Sundu}, \citenamefont {S{\"u}ng{\"u}},\ and\
  \citenamefont {Veliev}}]{Aydin:2025lbl}%
  \BibitemOpen
  \bibfield  {author} {\bibinfo {author} {\bibfnamefont {A.}~\bibnamefont
  {Ayd{\i}n}}, \bibinfo {author} {\bibfnamefont {H.}~\bibnamefont {Sundu}},
  \bibinfo {author} {\bibfnamefont {J.~Y.}\ \bibnamefont {S{\"u}ng{\"u}}},\
  and\ \bibinfo {author} {\bibfnamefont {E.~V.}\ \bibnamefont {Veliev}},\
  }\href {https://doi.org/10.1140/epjc/s10052-025-14090-4} {\bibfield
  {journal} {\bibinfo  {journal} {Eur. Phys. J. C}\ }\textbf {\bibinfo {volume}
  {85}},\ \bibinfo {pages} {567} (\bibinfo {year} {2025})},\ \Eprint
  {https://arxiv.org/abs/2501.10191} {arXiv:2501.10191 [hep-ph]} \BibitemShut
  {NoStop}%
\bibitem [{\citenamefont {Silva}\ \emph {et~al.}(2025)\citenamefont {Silva},
  \citenamefont {Pigozzo},\ and\ \citenamefont {Abreu}}]{Silva:2025bdg}%
  \BibitemOpen
  \bibfield  {author} {\bibinfo {author} {\bibfnamefont {V.~S.}\ \bibnamefont
  {Silva}}, \bibinfo {author} {\bibfnamefont {C.}~\bibnamefont {Pigozzo}},\
  and\ \bibinfo {author} {\bibfnamefont {L.~M.}\ \bibnamefont {Abreu}},\ }\href
  {https://doi.org/10.1140/epjc/s10052-025-14871-x} {\bibfield  {journal}
  {\bibinfo  {journal} {Eur. Phys. J. C}\ }\textbf {\bibinfo {volume} {85}},\
  \bibinfo {pages} {1154} (\bibinfo {year} {2025})},\ \Eprint
  {https://arxiv.org/abs/2503.12160} {arXiv:2503.12160 [hep-ph]} \BibitemShut
  {NoStop}%
\bibitem [{\citenamefont {Wang}\ and\ \citenamefont
  {Zhu}(2025)}]{Wang:2025hex}%
  \BibitemOpen
  \bibfield  {author} {\bibinfo {author} {\bibfnamefont {Y.}~\bibnamefont
  {Wang}}\ and\ \bibinfo {author} {\bibfnamefont {R.}~\bibnamefont {Zhu}},\
  }\href@noop {} {\  (\bibinfo {year} {2025})},\ \Eprint
  {https://arxiv.org/abs/2510.02085} {arXiv:2510.02085 [hep-ph]} \BibitemShut
  {NoStop}%
\bibitem [{\citenamefont {Lu}\ and\ \citenamefont {Jiang}(2026)}]{Lu:2025lyu}%
  \BibitemOpen
  \bibfield  {author} {\bibinfo {author} {\bibfnamefont {D.-D.}\ \bibnamefont
  {Lu}}\ and\ \bibinfo {author} {\bibfnamefont {S.-Z.}\ \bibnamefont {Jiang}},\
  }\href {https://doi.org/10.1103/tnmr-tj6v} {\bibfield  {journal} {\bibinfo
  {journal} {Phys. Rev. D}\ }\textbf {\bibinfo {volume} {113}},\ \bibinfo
  {pages} {094034} (\bibinfo {year} {2026})},\ \Eprint
  {https://arxiv.org/abs/2512.18569} {arXiv:2512.18569 [hep-ph]} \BibitemShut
  {NoStop}%
\bibitem [{\citenamefont {Liu}\ \emph {et~al.}(2026)\citenamefont {Liu},
  \citenamefont {Wang},\ and\ \citenamefont {Zhu}}]{Liu:2025mxv}%
  \BibitemOpen
  \bibfield  {author} {\bibinfo {author} {\bibfnamefont {X.}~\bibnamefont
  {Liu}}, \bibinfo {author} {\bibfnamefont {Y.}~\bibnamefont {Wang}},\ and\
  \bibinfo {author} {\bibfnamefont {R.}~\bibnamefont {Zhu}},\ }\href
  {https://doi.org/10.1103/6fcd-bj74} {\bibfield  {journal} {\bibinfo
  {journal} {Phys. Rev. D}\ }\textbf {\bibinfo {volume} {114}},\ \bibinfo
  {pages} {014044} (\bibinfo {year} {2026})},\ \Eprint
  {https://arxiv.org/abs/2512.22070} {arXiv:2512.22070 [hep-ph]} \BibitemShut
  {NoStop}%
\bibitem [{\citenamefont {Feng}\ and\ \citenamefont
  {Liu}(2026)}]{Feng:2026orq}%
  \BibitemOpen
  \bibfield  {author} {\bibinfo {author} {\bibfnamefont {F.}~\bibnamefont
  {Feng}}\ and\ \bibinfo {author} {\bibfnamefont {M.-M.}\ \bibnamefont {Liu}},\
  }\href {https://doi.org/10.1103/n6n2-d8mv} {\bibfield  {journal} {\bibinfo
  {journal} {Phys. Rev. D}\ }\textbf {\bibinfo {volume} {113}},\ \bibinfo
  {pages} {054010} (\bibinfo {year} {2026})},\ \Eprint
  {https://arxiv.org/abs/2601.07243} {arXiv:2601.07243 [hep-ph]} \BibitemShut
  {NoStop}%
\bibitem [{\citenamefont {Celiberto}(2026)}]{Celiberto:2026kks}%
  \BibitemOpen
  \bibfield  {author} {\bibinfo {author} {\bibfnamefont {F.~G.}\ \bibnamefont
  {Celiberto}},\ }\href {https://doi.org/10.1103/t5j1-84y2} {\bibfield
  {journal} {\bibinfo  {journal} {Phys. Rev. D}\ }\textbf {\bibinfo {volume}
  {113}},\ \bibinfo {pages} {114037} (\bibinfo {year} {2026})},\ \Eprint
  {https://arxiv.org/abs/2604.11646} {arXiv:2604.11646 [hep-ph]} \BibitemShut
  {NoStop}%
\bibitem [{\citenamefont {Wang}\ \emph {et~al.}(2026)\citenamefont {Wang},
  \citenamefont {Li}, \citenamefont {Zhong},\ and\ \citenamefont
  {Wang}}]{Wang:2026kcw}%
  \BibitemOpen
  \bibfield  {author} {\bibinfo {author} {\bibfnamefont {Y.}~\bibnamefont
  {Wang}}, \bibinfo {author} {\bibfnamefont {R.}~\bibnamefont {Li}}, \bibinfo
  {author} {\bibfnamefont {B.}~\bibnamefont {Zhong}},\ and\ \bibinfo {author}
  {\bibfnamefont {Y.-Q.}\ \bibnamefont {Wang}},\ }\href
  {https://doi.org/10.1088/1674-1137/ae643e} {\bibfield  {journal} {\bibinfo
  {journal} {Chin. Phys. C}\ }\textbf {\bibinfo {volume} {50}},\ \bibinfo
  {pages} {073003} (\bibinfo {year} {2026})},\ \Eprint
  {https://arxiv.org/abs/2604.18061} {arXiv:2604.18061 [hep-ex]} \BibitemShut
  {NoStop}%
\bibitem [{\citenamefont {Bernard}\ \emph {et~al.}(1997)\citenamefont {Bernard}
  \emph {et~al.}}]{MILC:1997usn}%
  \BibitemOpen
  \bibfield  {author} {\bibinfo {author} {\bibfnamefont {C.~W.}\ \bibnamefont
  {Bernard}} \emph {et~al.} (\bibinfo {collaboration} {MILC}),\ }\href
  {https://doi.org/10.1103/PhysRevD.56.7039} {\bibfield  {journal} {\bibinfo
  {journal} {Phys. Rev. D}\ }\textbf {\bibinfo {volume} {56}},\ \bibinfo
  {pages} {7039} (\bibinfo {year} {1997})},\ \Eprint
  {https://arxiv.org/abs/hep-lat/9707008} {arXiv:hep-lat/9707008} \BibitemShut
  {NoStop}%
\bibitem [{\citenamefont {Sasaki}(2004)}]{Sasaki:2003gi}%
  \BibitemOpen
  \bibfield  {author} {\bibinfo {author} {\bibfnamefont {S.}~\bibnamefont
  {Sasaki}},\ }\href {https://doi.org/10.1103/PhysRevLett.93.152001} {\bibfield
   {journal} {\bibinfo  {journal} {Phys. Rev. Lett.}\ }\textbf {\bibinfo
  {volume} {93}},\ \bibinfo {pages} {152001} (\bibinfo {year} {2004})},\
  \Eprint {https://arxiv.org/abs/hep-lat/0310014} {arXiv:hep-lat/0310014}
  \BibitemShut {NoStop}%
\bibitem [{\citenamefont {Okiharu}\ \emph {et~al.}(2005)\citenamefont
  {Okiharu}, \citenamefont {Suganuma},\ and\ \citenamefont
  {Takahashi}}]{Okiharu:2004ve}%
  \BibitemOpen
  \bibfield  {author} {\bibinfo {author} {\bibfnamefont {F.}~\bibnamefont
  {Okiharu}}, \bibinfo {author} {\bibfnamefont {H.}~\bibnamefont {Suganuma}},\
  and\ \bibinfo {author} {\bibfnamefont {T.~T.}\ \bibnamefont {Takahashi}},\
  }\href {https://doi.org/10.1103/PhysRevD.72.014505} {\bibfield  {journal}
  {\bibinfo  {journal} {Phys. Rev. D}\ }\textbf {\bibinfo {volume} {72}},\
  \bibinfo {pages} {014505} (\bibinfo {year} {2005})},\ \Eprint
  {https://arxiv.org/abs/hep-lat/0412012} {arXiv:hep-lat/0412012} \BibitemShut
  {NoStop}%
\bibitem [{\citenamefont {Prelovsek}\ and\ \citenamefont
  {Mohler}(2009)}]{Prelovsek:2008rf}%
  \BibitemOpen
  \bibfield  {author} {\bibinfo {author} {\bibfnamefont {S.}~\bibnamefont
  {Prelovsek}}\ and\ \bibinfo {author} {\bibfnamefont {D.}~\bibnamefont
  {Mohler}},\ }\href {https://doi.org/10.1103/PhysRevD.79.014503} {\bibfield
  {journal} {\bibinfo  {journal} {Phys. Rev. D}\ }\textbf {\bibinfo {volume}
  {79}},\ \bibinfo {pages} {014503} (\bibinfo {year} {2009})},\ \Eprint
  {https://arxiv.org/abs/0810.1759} {arXiv:0810.1759 [hep-lat]} \BibitemShut
  {NoStop}%
\bibitem [{\citenamefont {Dudek}\ \emph {et~al.}(2009)\citenamefont {Dudek},
  \citenamefont {Edwards},\ and\ \citenamefont {Thomas}}]{Dudek:2009kk}%
  \BibitemOpen
  \bibfield  {author} {\bibinfo {author} {\bibfnamefont {J.~J.}\ \bibnamefont
  {Dudek}}, \bibinfo {author} {\bibfnamefont {R.}~\bibnamefont {Edwards}},\
  and\ \bibinfo {author} {\bibfnamefont {C.~E.}\ \bibnamefont {Thomas}},\
  }\href {https://doi.org/10.1103/PhysRevD.79.094504} {\bibfield  {journal}
  {\bibinfo  {journal} {Phys. Rev. D}\ }\textbf {\bibinfo {volume} {79}},\
  \bibinfo {pages} {094504} (\bibinfo {year} {2009})},\ \Eprint
  {https://arxiv.org/abs/0902.2241} {arXiv:0902.2241 [hep-ph]} \BibitemShut
  {NoStop}%
\bibitem [{\citenamefont {Prelovsek}\ \emph {et~al.}(2010)\citenamefont
  {Prelovsek}, \citenamefont {Draper}, \citenamefont {Lang}, \citenamefont
  {Limmer}, \citenamefont {Liu}, \citenamefont {Mathur},\ and\ \citenamefont
  {Mohler}}]{Prelovsek:2010kg}%
  \BibitemOpen
  \bibfield  {author} {\bibinfo {author} {\bibfnamefont {S.}~\bibnamefont
  {Prelovsek}}, \bibinfo {author} {\bibfnamefont {T.}~\bibnamefont {Draper}},
  \bibinfo {author} {\bibfnamefont {C.~B.}\ \bibnamefont {Lang}}, \bibinfo
  {author} {\bibfnamefont {M.}~\bibnamefont {Limmer}}, \bibinfo {author}
  {\bibfnamefont {K.-F.}\ \bibnamefont {Liu}}, \bibinfo {author} {\bibfnamefont
  {N.}~\bibnamefont {Mathur}},\ and\ \bibinfo {author} {\bibfnamefont
  {D.}~\bibnamefont {Mohler}},\ }\href
  {https://doi.org/10.1103/PhysRevD.82.094507} {\bibfield  {journal} {\bibinfo
  {journal} {Phys. Rev. D}\ }\textbf {\bibinfo {volume} {82}},\ \bibinfo
  {pages} {094507} (\bibinfo {year} {2010})},\ \Eprint
  {https://arxiv.org/abs/1005.0948} {arXiv:1005.0948 [hep-lat]} \BibitemShut
  {NoStop}%
\bibitem [{\citenamefont {Beane}\ \emph {et~al.}(2012)\citenamefont {Beane},
  \citenamefont {Chang}, \citenamefont {Detmold}, \citenamefont {Lin},
  \citenamefont {Luu}, \citenamefont {Orginos}, \citenamefont {Parreno},
  \citenamefont {Savage}, \citenamefont {Torok},\ and\ \citenamefont
  {Walker-Loud}}]{NPLQCD:2011naw}%
  \BibitemOpen
  \bibfield  {author} {\bibinfo {author} {\bibfnamefont {S.~R.}\ \bibnamefont
  {Beane}}, \bibinfo {author} {\bibfnamefont {E.}~\bibnamefont {Chang}},
  \bibinfo {author} {\bibfnamefont {W.}~\bibnamefont {Detmold}}, \bibinfo
  {author} {\bibfnamefont {H.~W.}\ \bibnamefont {Lin}}, \bibinfo {author}
  {\bibfnamefont {T.~C.}\ \bibnamefont {Luu}}, \bibinfo {author} {\bibfnamefont
  {K.}~\bibnamefont {Orginos}}, \bibinfo {author} {\bibfnamefont
  {A.}~\bibnamefont {Parreno}}, \bibinfo {author} {\bibfnamefont {M.~J.}\
  \bibnamefont {Savage}}, \bibinfo {author} {\bibfnamefont {A.}~\bibnamefont
  {Torok}},\ and\ \bibinfo {author} {\bibfnamefont {A.}~\bibnamefont
  {Walker-Loud}} (\bibinfo {collaboration} {NPLQCD}),\ }\href
  {https://doi.org/10.1103/PhysRevD.85.054511} {\bibfield  {journal} {\bibinfo
  {journal} {Phys. Rev. D}\ }\textbf {\bibinfo {volume} {85}},\ \bibinfo
  {pages} {054511} (\bibinfo {year} {2012})},\ \Eprint
  {https://arxiv.org/abs/1109.2889} {arXiv:1109.2889 [hep-lat]} \BibitemShut
  {NoStop}%
\bibitem [{\citenamefont {Liu}\ \emph {et~al.}(2012)\citenamefont {Liu},
  \citenamefont {Moir}, \citenamefont {Peardon}, \citenamefont {Ryan},
  \citenamefont {Thomas}, \citenamefont {Vilaseca}, \citenamefont {Dudek},
  \citenamefont {Edwards}, \citenamefont {Joo},\ and\ \citenamefont
  {Richards}}]{HadronSpectrum:2012gic}%
  \BibitemOpen
  \bibfield  {author} {\bibinfo {author} {\bibfnamefont {L.}~\bibnamefont
  {Liu}}, \bibinfo {author} {\bibfnamefont {G.}~\bibnamefont {Moir}}, \bibinfo
  {author} {\bibfnamefont {M.}~\bibnamefont {Peardon}}, \bibinfo {author}
  {\bibfnamefont {S.~M.}\ \bibnamefont {Ryan}}, \bibinfo {author}
  {\bibfnamefont {C.~E.}\ \bibnamefont {Thomas}}, \bibinfo {author}
  {\bibfnamefont {P.}~\bibnamefont {Vilaseca}}, \bibinfo {author}
  {\bibfnamefont {J.~J.}\ \bibnamefont {Dudek}}, \bibinfo {author}
  {\bibfnamefont {R.~G.}\ \bibnamefont {Edwards}}, \bibinfo {author}
  {\bibfnamefont {B.}~\bibnamefont {Joo}},\ and\ \bibinfo {author}
  {\bibfnamefont {D.~G.}\ \bibnamefont {Richards}} (\bibinfo {collaboration}
  {Hadron Spectrum}),\ }\href {https://doi.org/10.1007/JHEP07(2012)126}
  {\bibfield  {journal} {\bibinfo  {journal} {JHEP}\ }\textbf {\bibinfo
  {volume} {07}},\ \bibinfo {pages} {126}},\ \Eprint
  {https://arxiv.org/abs/1204.5425} {arXiv:1204.5425 [hep-ph]} \BibitemShut
  {NoStop}%
\bibitem [{\citenamefont {Bicudo}\ and\ \citenamefont
  {Wagner}(2013)}]{Bicudo:2012qt}%
  \BibitemOpen
  \bibfield  {author} {\bibinfo {author} {\bibfnamefont {P.}~\bibnamefont
  {Bicudo}}\ and\ \bibinfo {author} {\bibfnamefont {M.}~\bibnamefont {Wagner}}
  (\bibinfo {collaboration} {European Twisted Mass}),\ }\href
  {https://doi.org/10.1103/PhysRevD.87.114511} {\bibfield  {journal} {\bibinfo
  {journal} {Phys. Rev. D}\ }\textbf {\bibinfo {volume} {87}},\ \bibinfo
  {pages} {114511} (\bibinfo {year} {2013})},\ \Eprint
  {https://arxiv.org/abs/1209.6274} {arXiv:1209.6274 [hep-ph]} \BibitemShut
  {NoStop}%
\bibitem [{\citenamefont {Ikeda}\ \emph {et~al.}(2014)\citenamefont {Ikeda},
  \citenamefont {Charron}, \citenamefont {Aoki}, \citenamefont {Doi},
  \citenamefont {Hatsuda}, \citenamefont {Inoue}, \citenamefont {Ishii},
  \citenamefont {Murano}, \citenamefont {Nemura},\ and\ \citenamefont
  {Sasaki}}]{Ikeda:2013vwa}%
  \BibitemOpen
  \bibfield  {author} {\bibinfo {author} {\bibfnamefont {Y.}~\bibnamefont
  {Ikeda}}, \bibinfo {author} {\bibfnamefont {B.}~\bibnamefont {Charron}},
  \bibinfo {author} {\bibfnamefont {S.}~\bibnamefont {Aoki}}, \bibinfo {author}
  {\bibfnamefont {T.}~\bibnamefont {Doi}}, \bibinfo {author} {\bibfnamefont
  {T.}~\bibnamefont {Hatsuda}}, \bibinfo {author} {\bibfnamefont
  {T.}~\bibnamefont {Inoue}}, \bibinfo {author} {\bibfnamefont
  {N.}~\bibnamefont {Ishii}}, \bibinfo {author} {\bibfnamefont
  {K.}~\bibnamefont {Murano}}, \bibinfo {author} {\bibfnamefont
  {H.}~\bibnamefont {Nemura}},\ and\ \bibinfo {author} {\bibfnamefont
  {K.}~\bibnamefont {Sasaki}},\ }\href
  {https://doi.org/10.1016/j.physletb.2014.01.002} {\bibfield  {journal}
  {\bibinfo  {journal} {Phys. Lett. B}\ }\textbf {\bibinfo {volume} {729}},\
  \bibinfo {pages} {85} (\bibinfo {year} {2014})},\ \Eprint
  {https://arxiv.org/abs/1311.6214} {arXiv:1311.6214 [hep-lat]} \BibitemShut
  {NoStop}%
\bibitem [{\citenamefont {Bicudo}\ \emph {et~al.}(2015)\citenamefont {Bicudo},
  \citenamefont {Cichy}, \citenamefont {Peters}, \citenamefont {Wagenbach},\
  and\ \citenamefont {Wagner}}]{Bicudo:2015vta}%
  \BibitemOpen
  \bibfield  {author} {\bibinfo {author} {\bibfnamefont {P.}~\bibnamefont
  {Bicudo}}, \bibinfo {author} {\bibfnamefont {K.}~\bibnamefont {Cichy}},
  \bibinfo {author} {\bibfnamefont {A.}~\bibnamefont {Peters}}, \bibinfo
  {author} {\bibfnamefont {B.}~\bibnamefont {Wagenbach}},\ and\ \bibinfo
  {author} {\bibfnamefont {M.}~\bibnamefont {Wagner}},\ }\href
  {https://doi.org/10.1103/PhysRevD.92.014507} {\bibfield  {journal} {\bibinfo
  {journal} {Phys. Rev. D}\ }\textbf {\bibinfo {volume} {92}},\ \bibinfo
  {pages} {014507} (\bibinfo {year} {2015})},\ \Eprint
  {https://arxiv.org/abs/1505.00613} {arXiv:1505.00613 [hep-lat]} \BibitemShut
  {NoStop}%
\bibitem [{\citenamefont {Ikeda}\ \emph {et~al.}(2016)\citenamefont {Ikeda},
  \citenamefont {Aoki}, \citenamefont {Doi}, \citenamefont {Gongyo},
  \citenamefont {Hatsuda}, \citenamefont {Inoue}, \citenamefont {Iritani},
  \citenamefont {Ishii}, \citenamefont {Murano},\ and\ \citenamefont
  {Sasaki}}]{HALQCD:2016ofq}%
  \BibitemOpen
  \bibfield  {author} {\bibinfo {author} {\bibfnamefont {Y.}~\bibnamefont
  {Ikeda}}, \bibinfo {author} {\bibfnamefont {S.}~\bibnamefont {Aoki}},
  \bibinfo {author} {\bibfnamefont {T.}~\bibnamefont {Doi}}, \bibinfo {author}
  {\bibfnamefont {S.}~\bibnamefont {Gongyo}}, \bibinfo {author} {\bibfnamefont
  {T.}~\bibnamefont {Hatsuda}}, \bibinfo {author} {\bibfnamefont
  {T.}~\bibnamefont {Inoue}}, \bibinfo {author} {\bibfnamefont
  {T.}~\bibnamefont {Iritani}}, \bibinfo {author} {\bibfnamefont
  {N.}~\bibnamefont {Ishii}}, \bibinfo {author} {\bibfnamefont
  {K.}~\bibnamefont {Murano}},\ and\ \bibinfo {author} {\bibfnamefont
  {K.}~\bibnamefont {Sasaki}} (\bibinfo {collaboration} {HAL QCD}),\ }\href
  {https://doi.org/10.1103/PhysRevLett.117.242001} {\bibfield  {journal}
  {\bibinfo  {journal} {Phys. Rev. Lett.}\ }\textbf {\bibinfo {volume} {117}},\
  \bibinfo {pages} {242001} (\bibinfo {year} {2016})},\ \Eprint
  {https://arxiv.org/abs/1602.03465} {arXiv:1602.03465 [hep-lat]} \BibitemShut
  {NoStop}%
\bibitem [{\citenamefont {Francis}\ \emph {et~al.}(2017)\citenamefont
  {Francis}, \citenamefont {Hudspith}, \citenamefont {Lewis},\ and\
  \citenamefont {Maltman}}]{Francis:2016hui}%
  \BibitemOpen
  \bibfield  {author} {\bibinfo {author} {\bibfnamefont {A.}~\bibnamefont
  {Francis}}, \bibinfo {author} {\bibfnamefont {R.~J.}\ \bibnamefont
  {Hudspith}}, \bibinfo {author} {\bibfnamefont {R.}~\bibnamefont {Lewis}},\
  and\ \bibinfo {author} {\bibfnamefont {K.}~\bibnamefont {Maltman}},\ }\href
  {https://doi.org/10.1103/PhysRevLett.118.142001} {\bibfield  {journal}
  {\bibinfo  {journal} {Phys. Rev. Lett.}\ }\textbf {\bibinfo {volume} {118}},\
  \bibinfo {pages} {142001} (\bibinfo {year} {2017})},\ \Eprint
  {https://arxiv.org/abs/1607.05214} {arXiv:1607.05214 [hep-lat]} \BibitemShut
  {NoStop}%
\bibitem [{\citenamefont {Cheung}\ \emph {et~al.}(2016)\citenamefont {Cheung},
  \citenamefont {O'Hara}, \citenamefont {Moir}, \citenamefont {Peardon},
  \citenamefont {Ryan}, \citenamefont {Thomas},\ and\ \citenamefont
  {Tims}}]{Cheung:2016bym}%
  \BibitemOpen
  \bibfield  {author} {\bibinfo {author} {\bibfnamefont {G.~K.~C.}\
  \bibnamefont {Cheung}}, \bibinfo {author} {\bibfnamefont {C.}~\bibnamefont
  {O'Hara}}, \bibinfo {author} {\bibfnamefont {G.}~\bibnamefont {Moir}},
  \bibinfo {author} {\bibfnamefont {M.}~\bibnamefont {Peardon}}, \bibinfo
  {author} {\bibfnamefont {S.~M.}\ \bibnamefont {Ryan}}, \bibinfo {author}
  {\bibfnamefont {C.~E.}\ \bibnamefont {Thomas}},\ and\ \bibinfo {author}
  {\bibfnamefont {D.}~\bibnamefont {Tims}} (\bibinfo {collaboration} {Hadron
  Spectrum}),\ }\href {https://doi.org/10.1007/JHEP12(2016)089} {\bibfield
  {journal} {\bibinfo  {journal} {JHEP}\ }\textbf {\bibinfo {volume} {12}},\
  \bibinfo {pages} {089}},\ \Eprint {https://arxiv.org/abs/1610.01073}
  {arXiv:1610.01073 [hep-lat]} \BibitemShut {NoStop}%
\bibitem [{\citenamefont {Bicudo}\ \emph {et~al.}(2017)\citenamefont {Bicudo},
  \citenamefont {Scheunert},\ and\ \citenamefont {Wagner}}]{Bicudo:2016ooe}%
  \BibitemOpen
  \bibfield  {author} {\bibinfo {author} {\bibfnamefont {P.}~\bibnamefont
  {Bicudo}}, \bibinfo {author} {\bibfnamefont {J.}~\bibnamefont {Scheunert}},\
  and\ \bibinfo {author} {\bibfnamefont {M.}~\bibnamefont {Wagner}},\ }\href
  {https://doi.org/10.1103/PhysRevD.95.034502} {\bibfield  {journal} {\bibinfo
  {journal} {Phys. Rev. D}\ }\textbf {\bibinfo {volume} {95}},\ \bibinfo
  {pages} {034502} (\bibinfo {year} {2017})},\ \Eprint
  {https://arxiv.org/abs/1612.02758} {arXiv:1612.02758 [hep-lat]} \BibitemShut
  {NoStop}%
\bibitem [{\citenamefont {Cheung}\ \emph {et~al.}(2017)\citenamefont {Cheung},
  \citenamefont {Thomas}, \citenamefont {Dudek},\ and\ \citenamefont
  {Edwards}}]{Cheung:2017tnt}%
  \BibitemOpen
  \bibfield  {author} {\bibinfo {author} {\bibfnamefont {G.~K.~C.}\
  \bibnamefont {Cheung}}, \bibinfo {author} {\bibfnamefont {C.~E.}\
  \bibnamefont {Thomas}}, \bibinfo {author} {\bibfnamefont {J.~J.}\
  \bibnamefont {Dudek}},\ and\ \bibinfo {author} {\bibfnamefont {R.~G.}\
  \bibnamefont {Edwards}} (\bibinfo {collaboration} {Hadron Spectrum}),\ }\href
  {https://doi.org/10.1007/JHEP11(2017)033} {\bibfield  {journal} {\bibinfo
  {journal} {JHEP}\ }\textbf {\bibinfo {volume} {11}},\ \bibinfo {pages}
  {033}},\ \Eprint {https://arxiv.org/abs/1709.01417} {arXiv:1709.01417
  [hep-lat]} \BibitemShut {NoStop}%
\bibitem [{\citenamefont {Hughes}\ \emph {et~al.}(2018)\citenamefont {Hughes},
  \citenamefont {Eichten},\ and\ \citenamefont {Davies}}]{Hughes:2017xie}%
  \BibitemOpen
  \bibfield  {author} {\bibinfo {author} {\bibfnamefont {C.}~\bibnamefont
  {Hughes}}, \bibinfo {author} {\bibfnamefont {E.}~\bibnamefont {Eichten}},\
  and\ \bibinfo {author} {\bibfnamefont {C.~T.~H.}\ \bibnamefont {Davies}},\
  }\href {https://doi.org/10.1103/PhysRevD.97.054505} {\bibfield  {journal}
  {\bibinfo  {journal} {Phys. Rev. D}\ }\textbf {\bibinfo {volume} {97}},\
  \bibinfo {pages} {054505} (\bibinfo {year} {2018})},\ \Eprint
  {https://arxiv.org/abs/1710.03236} {arXiv:1710.03236 [hep-lat]} \BibitemShut
  {NoStop}%
\bibitem [{\citenamefont {Francis}\ \emph {et~al.}(2019)\citenamefont
  {Francis}, \citenamefont {Hudspith}, \citenamefont {Lewis},\ and\
  \citenamefont {Maltman}}]{Francis:2018jyb}%
  \BibitemOpen
  \bibfield  {author} {\bibinfo {author} {\bibfnamefont {A.}~\bibnamefont
  {Francis}}, \bibinfo {author} {\bibfnamefont {R.~J.}\ \bibnamefont
  {Hudspith}}, \bibinfo {author} {\bibfnamefont {R.}~\bibnamefont {Lewis}},\
  and\ \bibinfo {author} {\bibfnamefont {K.}~\bibnamefont {Maltman}},\ }\href
  {https://doi.org/10.1103/PhysRevD.99.054505} {\bibfield  {journal} {\bibinfo
  {journal} {Phys. Rev. D}\ }\textbf {\bibinfo {volume} {99}},\ \bibinfo
  {pages} {054505} (\bibinfo {year} {2019})},\ \Eprint
  {https://arxiv.org/abs/1810.10550} {arXiv:1810.10550 [hep-lat]} \BibitemShut
  {NoStop}%
\bibitem [{\citenamefont {Junnarkar}\ \emph {et~al.}(2019)\citenamefont
  {Junnarkar}, \citenamefont {Mathur},\ and\ \citenamefont
  {Padmanath}}]{Junnarkar:2018twb}%
  \BibitemOpen
  \bibfield  {author} {\bibinfo {author} {\bibfnamefont {P.}~\bibnamefont
  {Junnarkar}}, \bibinfo {author} {\bibfnamefont {N.}~\bibnamefont {Mathur}},\
  and\ \bibinfo {author} {\bibfnamefont {M.}~\bibnamefont {Padmanath}},\ }\href
  {https://doi.org/10.1103/PhysRevD.99.034507} {\bibfield  {journal} {\bibinfo
  {journal} {Phys. Rev. D}\ }\textbf {\bibinfo {volume} {99}},\ \bibinfo
  {pages} {034507} (\bibinfo {year} {2019})},\ \Eprint
  {https://arxiv.org/abs/1810.12285} {arXiv:1810.12285 [hep-lat]} \BibitemShut
  {NoStop}%
\bibitem [{\citenamefont {Leskovec}\ \emph {et~al.}(2019)\citenamefont
  {Leskovec}, \citenamefont {Meinel}, \citenamefont {Pflaumer},\ and\
  \citenamefont {Wagner}}]{Leskovec:2019ioa}%
  \BibitemOpen
  \bibfield  {author} {\bibinfo {author} {\bibfnamefont {L.}~\bibnamefont
  {Leskovec}}, \bibinfo {author} {\bibfnamefont {S.}~\bibnamefont {Meinel}},
  \bibinfo {author} {\bibfnamefont {M.}~\bibnamefont {Pflaumer}},\ and\
  \bibinfo {author} {\bibfnamefont {M.}~\bibnamefont {Wagner}},\ }\href
  {https://doi.org/10.1103/PhysRevD.100.014503} {\bibfield  {journal} {\bibinfo
   {journal} {Phys. Rev. D}\ }\textbf {\bibinfo {volume} {100}},\ \bibinfo
  {pages} {014503} (\bibinfo {year} {2019})},\ \Eprint
  {https://arxiv.org/abs/1904.04197} {arXiv:1904.04197 [hep-lat]} \BibitemShut
  {NoStop}%
\bibitem [{\citenamefont {Chen}\ \emph
  {et~al.}(2019{\natexlab{a}})\citenamefont {Chen}, \citenamefont {Chen},
  \citenamefont {Gong}, \citenamefont {Liu}, \citenamefont {Liu}, \citenamefont
  {Liu}, \citenamefont {Liu}, \citenamefont {Ma}, \citenamefont {Werner},\ and\
  \citenamefont {Zhang}}]{CLQCD:2019npr}%
  \BibitemOpen
  \bibfield  {author} {\bibinfo {author} {\bibfnamefont {T.}~\bibnamefont
  {Chen}}, \bibinfo {author} {\bibfnamefont {Y.}~\bibnamefont {Chen}}, \bibinfo
  {author} {\bibfnamefont {M.}~\bibnamefont {Gong}}, \bibinfo {author}
  {\bibfnamefont {C.}~\bibnamefont {Liu}}, \bibinfo {author} {\bibfnamefont
  {L.}~\bibnamefont {Liu}}, \bibinfo {author} {\bibfnamefont {Y.-B.}\
  \bibnamefont {Liu}}, \bibinfo {author} {\bibfnamefont {Z.}~\bibnamefont
  {Liu}}, \bibinfo {author} {\bibfnamefont {J.-P.}\ \bibnamefont {Ma}},
  \bibinfo {author} {\bibfnamefont {M.}~\bibnamefont {Werner}},\ and\ \bibinfo
  {author} {\bibfnamefont {J.-B.}\ \bibnamefont {Zhang}} (\bibinfo
  {collaboration} {CLQCD}),\ }\href
  {https://doi.org/10.1088/1674-1137/43/10/103103} {\bibfield  {journal}
  {\bibinfo  {journal} {Chin. Phys. C}\ }\textbf {\bibinfo {volume} {43}},\
  \bibinfo {pages} {103103} (\bibinfo {year} {2019}{\natexlab{a}})},\ \Eprint
  {https://arxiv.org/abs/1907.03371} {arXiv:1907.03371 [hep-lat]} \BibitemShut
  {NoStop}%
\bibitem [{\citenamefont {Hudspith}\ \emph {et~al.}(2020)\citenamefont
  {Hudspith}, \citenamefont {Colquhoun}, \citenamefont {Francis}, \citenamefont
  {Lewis},\ and\ \citenamefont {Maltman}}]{Hudspith:2020tdf}%
  \BibitemOpen
  \bibfield  {author} {\bibinfo {author} {\bibfnamefont {R.~J.}\ \bibnamefont
  {Hudspith}}, \bibinfo {author} {\bibfnamefont {B.}~\bibnamefont {Colquhoun}},
  \bibinfo {author} {\bibfnamefont {A.}~\bibnamefont {Francis}}, \bibinfo
  {author} {\bibfnamefont {R.}~\bibnamefont {Lewis}},\ and\ \bibinfo {author}
  {\bibfnamefont {K.}~\bibnamefont {Maltman}},\ }\href
  {https://doi.org/10.1103/PhysRevD.102.114506} {\bibfield  {journal} {\bibinfo
   {journal} {Phys. Rev. D}\ }\textbf {\bibinfo {volume} {102}},\ \bibinfo
  {pages} {114506} (\bibinfo {year} {2020})},\ \Eprint
  {https://arxiv.org/abs/2006.14294} {arXiv:2006.14294 [hep-lat]} \BibitemShut
  {NoStop}%
\bibitem [{\citenamefont {Ryan}\ and\ \citenamefont
  {Wilson}(2021)}]{Ryan:2020iog}%
  \BibitemOpen
  \bibfield  {author} {\bibinfo {author} {\bibfnamefont {S.~M.}\ \bibnamefont
  {Ryan}}\ and\ \bibinfo {author} {\bibfnamefont {D.~J.}\ \bibnamefont
  {Wilson}} (\bibinfo {collaboration} {Hadron Spectrum}),\ }\href
  {https://doi.org/10.1007/JHEP02(2021)214} {\bibfield  {journal} {\bibinfo
  {journal} {JHEP}\ }\textbf {\bibinfo {volume} {02}},\ \bibinfo {pages}
  {214}},\ \Eprint {https://arxiv.org/abs/2008.02656} {arXiv:2008.02656
  [hep-lat]} \BibitemShut {NoStop}%
\bibitem [{\citenamefont {Padmanath}\ and\ \citenamefont
  {Prelovsek}(2022)}]{Padmanath:2022cvl}%
  \BibitemOpen
  \bibfield  {author} {\bibinfo {author} {\bibfnamefont {M.}~\bibnamefont
  {Padmanath}}\ and\ \bibinfo {author} {\bibfnamefont {S.}~\bibnamefont
  {Prelovsek}},\ }\href {https://doi.org/10.1103/PhysRevLett.129.032002}
  {\bibfield  {journal} {\bibinfo  {journal} {Phys. Rev. Lett.}\ }\textbf
  {\bibinfo {volume} {129}},\ \bibinfo {pages} {032002} (\bibinfo {year}
  {2022})},\ \Eprint {https://arxiv.org/abs/2202.10110} {arXiv:2202.10110
  [hep-lat]} \BibitemShut {NoStop}%
\bibitem [{\citenamefont {Bicudo}(2023)}]{Bicudo:2022cqi}%
  \BibitemOpen
  \bibfield  {author} {\bibinfo {author} {\bibfnamefont {P.}~\bibnamefont
  {Bicudo}},\ }\href {https://doi.org/10.1016/j.physrep.2023.10.001} {\bibfield
   {journal} {\bibinfo  {journal} {Phys. Rept.}\ }\textbf {\bibinfo {volume}
  {1039}},\ \bibinfo {pages} {1} (\bibinfo {year} {2023})},\ \Eprint
  {https://arxiv.org/abs/2212.07793} {arXiv:2212.07793 [hep-lat]} \BibitemShut
  {NoStop}%
\bibitem [{\citenamefont {Lyu}\ \emph {et~al.}(2023)\citenamefont {Lyu},
  \citenamefont {Aoki}, \citenamefont {Doi}, \citenamefont {Hatsuda},
  \citenamefont {Ikeda},\ and\ \citenamefont {Meng}}]{Lyu:2023xro}%
  \BibitemOpen
  \bibfield  {author} {\bibinfo {author} {\bibfnamefont {Y.}~\bibnamefont
  {Lyu}}, \bibinfo {author} {\bibfnamefont {S.}~\bibnamefont {Aoki}}, \bibinfo
  {author} {\bibfnamefont {T.}~\bibnamefont {Doi}}, \bibinfo {author}
  {\bibfnamefont {T.}~\bibnamefont {Hatsuda}}, \bibinfo {author} {\bibfnamefont
  {Y.}~\bibnamefont {Ikeda}},\ and\ \bibinfo {author} {\bibfnamefont
  {J.}~\bibnamefont {Meng}},\ }\href
  {https://doi.org/10.1103/PhysRevLett.131.161901} {\bibfield  {journal}
  {\bibinfo  {journal} {Phys. Rev. Lett.}\ }\textbf {\bibinfo {volume} {131}},\
  \bibinfo {pages} {161901} (\bibinfo {year} {2023})},\ \Eprint
  {https://arxiv.org/abs/2302.04505} {arXiv:2302.04505 [hep-lat]} \BibitemShut
  {NoStop}%
\bibitem [{\citenamefont {Wilson}\ \emph {et~al.}(2024)\citenamefont {Wilson},
  \citenamefont {Thomas}, \citenamefont {Dudek},\ and\ \citenamefont
  {Edwards}}]{Wilson:2023anv}%
  \BibitemOpen
  \bibfield  {author} {\bibinfo {author} {\bibfnamefont {D.~J.}\ \bibnamefont
  {Wilson}}, \bibinfo {author} {\bibfnamefont {C.~E.}\ \bibnamefont {Thomas}},
  \bibinfo {author} {\bibfnamefont {J.~J.}\ \bibnamefont {Dudek}},\ and\
  \bibinfo {author} {\bibfnamefont {R.~G.}\ \bibnamefont {Edwards}} (\bibinfo
  {collaboration} {Hadron Spectrum}),\ }\href
  {https://doi.org/10.1103/PhysRevD.109.114503} {\bibfield  {journal} {\bibinfo
   {journal} {Phys. Rev. D}\ }\textbf {\bibinfo {volume} {109}},\ \bibinfo
  {pages} {114503} (\bibinfo {year} {2024})},\ \Eprint
  {https://arxiv.org/abs/2309.14071} {arXiv:2309.14071 [hep-lat]} \BibitemShut
  {NoStop}%
\bibitem [{\citenamefont {Meng}\ \emph {et~al.}(2025)\citenamefont {Meng},
  \citenamefont {Liu}, \citenamefont {Tuo}, \citenamefont {Yan},\ and\
  \citenamefont {Zhang}}]{Meng:2024czd}%
  \BibitemOpen
  \bibfield  {author} {\bibinfo {author} {\bibfnamefont {Y.}~\bibnamefont
  {Meng}}, \bibinfo {author} {\bibfnamefont {C.}~\bibnamefont {Liu}}, \bibinfo
  {author} {\bibfnamefont {X.-Y.}\ \bibnamefont {Tuo}}, \bibinfo {author}
  {\bibfnamefont {H.}~\bibnamefont {Yan}},\ and\ \bibinfo {author}
  {\bibfnamefont {Z.}~\bibnamefont {Zhang}},\ }\href
  {https://doi.org/10.1140/epjc/s10052-025-14192-z} {\bibfield  {journal}
  {\bibinfo  {journal} {Eur. Phys. J. C}\ }\textbf {\bibinfo {volume} {85}},\
  \bibinfo {pages} {458} (\bibinfo {year} {2025})},\ \Eprint
  {https://arxiv.org/abs/2411.11533} {arXiv:2411.11533 [hep-lat]} \BibitemShut
  {NoStop}%
\bibitem [{\citenamefont {Guo}\ and\ \citenamefont
  {Hanhart}(2025)}]{Guo:2025imr}%
  \BibitemOpen
  \bibfield  {author} {\bibinfo {author} {\bibfnamefont {F.-K.}\ \bibnamefont
  {Guo}}\ and\ \bibinfo {author} {\bibfnamefont {C.}~\bibnamefont {Hanhart}},\
  }\href {https://doi.org/10.1088/1572-9494/ade2ea} {\bibfield  {journal}
  {\bibinfo  {journal} {Commun. Theor. Phys.}\ }\textbf {\bibinfo {volume}
  {77}},\ \bibinfo {pages} {125201} (\bibinfo {year} {2025})},\ \Eprint
  {https://arxiv.org/abs/2503.05372} {arXiv:2503.05372 [hep-ph]} \BibitemShut
  {NoStop}%
\bibitem [{\citenamefont {Maiani}\ \emph {et~al.}(2005)\citenamefont {Maiani},
  \citenamefont {Piccinini}, \citenamefont {Polosa},\ and\ \citenamefont
  {Riquer}}]{Maiani:2004vq}%
  \BibitemOpen
  \bibfield  {author} {\bibinfo {author} {\bibfnamefont {L.}~\bibnamefont
  {Maiani}}, \bibinfo {author} {\bibfnamefont {F.}~\bibnamefont {Piccinini}},
  \bibinfo {author} {\bibfnamefont {A.~D.}\ \bibnamefont {Polosa}},\ and\
  \bibinfo {author} {\bibfnamefont {V.}~\bibnamefont {Riquer}},\ }\href
  {https://doi.org/10.1103/PhysRevD.71.014028} {\bibfield  {journal} {\bibinfo
  {journal} {Phys. Rev. D}\ }\textbf {\bibinfo {volume} {71}},\ \bibinfo
  {pages} {014028} (\bibinfo {year} {2005})},\ \Eprint
  {https://arxiv.org/abs/hep-ph/0412098} {arXiv:hep-ph/0412098} \BibitemShut
  {NoStop}%
\bibitem [{\citenamefont {Maiani}\ \emph {et~al.}(2014)\citenamefont {Maiani},
  \citenamefont {Piccinini}, \citenamefont {Polosa},\ and\ \citenamefont
  {Riquer}}]{Maiani:2014aja}%
  \BibitemOpen
  \bibfield  {author} {\bibinfo {author} {\bibfnamefont {L.}~\bibnamefont
  {Maiani}}, \bibinfo {author} {\bibfnamefont {F.}~\bibnamefont {Piccinini}},
  \bibinfo {author} {\bibfnamefont {A.~D.}\ \bibnamefont {Polosa}},\ and\
  \bibinfo {author} {\bibfnamefont {V.}~\bibnamefont {Riquer}},\ }\href
  {https://doi.org/10.1103/PhysRevD.89.114010} {\bibfield  {journal} {\bibinfo
  {journal} {Phys. Rev. D}\ }\textbf {\bibinfo {volume} {89}},\ \bibinfo
  {pages} {114010} (\bibinfo {year} {2014})},\ \Eprint
  {https://arxiv.org/abs/1405.1551} {arXiv:1405.1551 [hep-ph]} \BibitemShut
  {NoStop}%
\bibitem [{\citenamefont {Lebed}(2015)}]{Lebed:2015tna}%
  \BibitemOpen
  \bibfield  {author} {\bibinfo {author} {\bibfnamefont {R.~F.}\ \bibnamefont
  {Lebed}},\ }\href {https://doi.org/10.1016/j.physletb.2015.08.032} {\bibfield
   {journal} {\bibinfo  {journal} {Phys. Lett. B}\ }\textbf {\bibinfo {volume}
  {749}},\ \bibinfo {pages} {454} (\bibinfo {year} {2015})},\ \Eprint
  {https://arxiv.org/abs/1507.05867} {arXiv:1507.05867 [hep-ph]} \BibitemShut
  {NoStop}%
\bibitem [{\citenamefont {Anwar}\ and\ \citenamefont
  {Burns}(2023)}]{Anwar:2023svj}%
  \BibitemOpen
  \bibfield  {author} {\bibinfo {author} {\bibfnamefont {M.~N.}\ \bibnamefont
  {Anwar}}\ and\ \bibinfo {author} {\bibfnamefont {T.~J.}\ \bibnamefont
  {Burns}},\ }\href {https://doi.org/10.1016/j.physletb.2023.138248} {\bibfield
   {journal} {\bibinfo  {journal} {Phys. Lett. B}\ }\textbf {\bibinfo {volume}
  {847}},\ \bibinfo {pages} {138248} (\bibinfo {year} {2023})},\ \Eprint
  {https://arxiv.org/abs/2309.03309} {arXiv:2309.03309 [hep-ph]} \BibitemShut
  {NoStop}%
\bibitem [{\citenamefont {AlFiky}\ \emph {et~al.}(2006)\citenamefont {AlFiky},
  \citenamefont {Gabbiani},\ and\ \citenamefont {Petrov}}]{AlFiky:2005jd}%
  \BibitemOpen
  \bibfield  {author} {\bibinfo {author} {\bibfnamefont {M.~T.}\ \bibnamefont
  {AlFiky}}, \bibinfo {author} {\bibfnamefont {F.}~\bibnamefont {Gabbiani}},\
  and\ \bibinfo {author} {\bibfnamefont {A.~A.}\ \bibnamefont {Petrov}},\
  }\href {https://doi.org/10.1016/j.physletb.2006.07.069} {\bibfield  {journal}
  {\bibinfo  {journal} {Phys. Lett. B}\ }\textbf {\bibinfo {volume} {640}},\
  \bibinfo {pages} {238} (\bibinfo {year} {2006})},\ \Eprint
  {https://arxiv.org/abs/hep-ph/0506141} {arXiv:hep-ph/0506141} \BibitemShut
  {NoStop}%
\bibitem [{\citenamefont {Haidenbauer}\ and\ \citenamefont
  {Meissner}(2012)}]{Haidenbauer:2011za}%
  \BibitemOpen
  \bibfield  {author} {\bibinfo {author} {\bibfnamefont {J.}~\bibnamefont
  {Haidenbauer}}\ and\ \bibinfo {author} {\bibfnamefont {U.~G.}\ \bibnamefont
  {Meissner}},\ }\href {https://doi.org/10.1016/j.nuclphysa.2012.01.021}
  {\bibfield  {journal} {\bibinfo  {journal} {Nucl. Phys. A}\ }\textbf
  {\bibinfo {volume} {881}},\ \bibinfo {pages} {44} (\bibinfo {year} {2012})},\
  \Eprint {https://arxiv.org/abs/1111.4069} {arXiv:1111.4069 [nucl-th]}
  \BibitemShut {NoStop}%
\bibitem [{\citenamefont {Guo}(2017)}]{Guo:2017vcf}%
  \BibitemOpen
  \bibfield  {author} {\bibinfo {author} {\bibfnamefont {Z.-H.}\ \bibnamefont
  {Guo}},\ }\href {https://doi.org/10.1103/PhysRevD.96.074004} {\bibfield
  {journal} {\bibinfo  {journal} {Phys. Rev. D}\ }\textbf {\bibinfo {volume}
  {96}},\ \bibinfo {pages} {074004} (\bibinfo {year} {2017})},\ \Eprint
  {https://arxiv.org/abs/1708.04145} {arXiv:1708.04145 [hep-ph]} \BibitemShut
  {NoStop}%
\bibitem [{\citenamefont {Meng}\ \emph {et~al.}(2023)\citenamefont {Meng},
  \citenamefont {Wang}, \citenamefont {Wang},\ and\ \citenamefont
  {Zhu}}]{Meng:2022ozq}%
  \BibitemOpen
  \bibfield  {author} {\bibinfo {author} {\bibfnamefont {L.}~\bibnamefont
  {Meng}}, \bibinfo {author} {\bibfnamefont {B.}~\bibnamefont {Wang}}, \bibinfo
  {author} {\bibfnamefont {G.-J.}\ \bibnamefont {Wang}},\ and\ \bibinfo
  {author} {\bibfnamefont {S.-L.}\ \bibnamefont {Zhu}},\ }\href
  {https://doi.org/10.1016/j.physrep.2023.04.003} {\bibfield  {journal}
  {\bibinfo  {journal} {Phys. Rept.}\ }\textbf {\bibinfo {volume} {1019}},\
  \bibinfo {pages} {1} (\bibinfo {year} {2023})},\ \Eprint
  {https://arxiv.org/abs/2204.08716} {arXiv:2204.08716 [hep-ph]} \BibitemShut
  {NoStop}%
\bibitem [{\citenamefont {Nielsen}\ \emph {et~al.}(2010)\citenamefont
  {Nielsen}, \citenamefont {Navarra},\ and\ \citenamefont
  {Lee}}]{Nielsen:2009uh}%
  \BibitemOpen
  \bibfield  {author} {\bibinfo {author} {\bibfnamefont {M.}~\bibnamefont
  {Nielsen}}, \bibinfo {author} {\bibfnamefont {F.~S.}\ \bibnamefont
  {Navarra}},\ and\ \bibinfo {author} {\bibfnamefont {S.~H.}\ \bibnamefont
  {Lee}},\ }\href {https://doi.org/10.1016/j.physrep.2010.07.005} {\bibfield
  {journal} {\bibinfo  {journal} {Phys. Rept.}\ }\textbf {\bibinfo {volume}
  {497}},\ \bibinfo {pages} {41} (\bibinfo {year} {2010})},\ \Eprint
  {https://arxiv.org/abs/0911.1958} {arXiv:0911.1958 [hep-ph]} \BibitemShut
  {NoStop}%
\bibitem [{\citenamefont {Albuquerque}\ \emph {et~al.}(2019)\citenamefont
  {Albuquerque}, \citenamefont {Dias}, \citenamefont {Khemchandani},
  \citenamefont {Mart{\'\i}nez~Torres}, \citenamefont {Navarra}, \citenamefont
  {Nielsen},\ and\ \citenamefont {Zanetti}}]{Albuquerque:2018jkn}%
  \BibitemOpen
  \bibfield  {author} {\bibinfo {author} {\bibfnamefont {R.~M.}\ \bibnamefont
  {Albuquerque}}, \bibinfo {author} {\bibfnamefont {J.~M.}\ \bibnamefont
  {Dias}}, \bibinfo {author} {\bibfnamefont {K.~P.}\ \bibnamefont
  {Khemchandani}}, \bibinfo {author} {\bibfnamefont {A.}~\bibnamefont
  {Mart{\'\i}nez~Torres}}, \bibinfo {author} {\bibfnamefont {F.~S.}\
  \bibnamefont {Navarra}}, \bibinfo {author} {\bibfnamefont {M.}~\bibnamefont
  {Nielsen}},\ and\ \bibinfo {author} {\bibfnamefont {C.~M.}\ \bibnamefont
  {Zanetti}},\ }\href {https://doi.org/10.1088/1361-6471/ab2678} {\bibfield
  {journal} {\bibinfo  {journal} {J. Phys. G}\ }\textbf {\bibinfo {volume}
  {46}},\ \bibinfo {pages} {093002} (\bibinfo {year} {2019})},\ \Eprint
  {https://arxiv.org/abs/1812.08207} {arXiv:1812.08207 [hep-ph]} \BibitemShut
  {NoStop}%
\bibitem [{\citenamefont {Wang}(2026)}]{Wang:2025sic}%
  \BibitemOpen
  \bibfield  {author} {\bibinfo {author} {\bibfnamefont {Z.-G.}\ \bibnamefont
  {Wang}},\ }\href {https://doi.org/10.15302/frontphys.2026.016300} {\bibfield
  {journal} {\bibinfo  {journal} {Front. Phys. (Beijing)}\ }\textbf {\bibinfo
  {volume} {21}},\ \bibinfo {pages} {016300} (\bibinfo {year} {2026})},\
  \Eprint {https://arxiv.org/abs/2502.11351} {arXiv:2502.11351 [hep-ph]}
  \BibitemShut {NoStop}%
\bibitem [{\citenamefont {Dong}\ \emph {et~al.}(2017)\citenamefont {Dong},
  \citenamefont {Faessler},\ and\ \citenamefont {Lyubovitskij}}]{Dong:2017gaw}%
  \BibitemOpen
  \bibfield  {author} {\bibinfo {author} {\bibfnamefont {Y.}~\bibnamefont
  {Dong}}, \bibinfo {author} {\bibfnamefont {A.}~\bibnamefont {Faessler}},\
  and\ \bibinfo {author} {\bibfnamefont {V.~E.}\ \bibnamefont {Lyubovitskij}},\
  }\href {https://doi.org/10.1016/j.ppnp.2017.01.002} {\bibfield  {journal}
  {\bibinfo  {journal} {Prog. Part. Nucl. Phys.}\ }\textbf {\bibinfo {volume}
  {94}},\ \bibinfo {pages} {282} (\bibinfo {year} {2017})}\BibitemShut
  {NoStop}%
\bibitem [{\citenamefont {Guo}\ \emph {et~al.}(2018)\citenamefont {Guo},
  \citenamefont {Hanhart}, \citenamefont {Mei{\ss}ner}, \citenamefont {Wang},
  \citenamefont {Zhao},\ and\ \citenamefont {Zou}}]{Guo:2017jvc}%
  \BibitemOpen
  \bibfield  {author} {\bibinfo {author} {\bibfnamefont {F.-K.}\ \bibnamefont
  {Guo}}, \bibinfo {author} {\bibfnamefont {C.}~\bibnamefont {Hanhart}},
  \bibinfo {author} {\bibfnamefont {U.-G.}\ \bibnamefont {Mei{\ss}ner}},
  \bibinfo {author} {\bibfnamefont {Q.}~\bibnamefont {Wang}}, \bibinfo {author}
  {\bibfnamefont {Q.}~\bibnamefont {Zhao}},\ and\ \bibinfo {author}
  {\bibfnamefont {B.-S.}\ \bibnamefont {Zou}},\ }\href
  {https://doi.org/10.1103/RevModPhys.90.015004} {\bibfield  {journal}
  {\bibinfo  {journal} {Rev. Mod. Phys.}\ }\textbf {\bibinfo {volume} {90}},\
  \bibinfo {pages} {015004} (\bibinfo {year} {2018})},\ \bibinfo {note}
  {[Erratum: Rev.Mod.Phys. 94, 029901 (2022)]},\ \Eprint
  {https://arxiv.org/abs/1705.00141} {arXiv:1705.00141 [hep-ph]} \BibitemShut
  {NoStop}%
\bibitem [{\citenamefont {Chen}\ \emph
  {et~al.}(2019{\natexlab{b}})\citenamefont {Chen}, \citenamefont {Sun},
  \citenamefont {Liu},\ and\ \citenamefont {Zhu}}]{Chen:2019asm}%
  \BibitemOpen
  \bibfield  {author} {\bibinfo {author} {\bibfnamefont {R.}~\bibnamefont
  {Chen}}, \bibinfo {author} {\bibfnamefont {Z.-F.}\ \bibnamefont {Sun}},
  \bibinfo {author} {\bibfnamefont {X.}~\bibnamefont {Liu}},\ and\ \bibinfo
  {author} {\bibfnamefont {S.-L.}\ \bibnamefont {Zhu}},\ }\href
  {https://doi.org/10.1103/PhysRevD.100.011502} {\bibfield  {journal} {\bibinfo
   {journal} {Phys. Rev. D}\ }\textbf {\bibinfo {volume} {100}},\ \bibinfo
  {pages} {011502} (\bibinfo {year} {2019}{\natexlab{b}})},\ \Eprint
  {https://arxiv.org/abs/1903.11013} {arXiv:1903.11013 [hep-ph]} \BibitemShut
  {NoStop}%
\bibitem [{\citenamefont {Guo}\ \emph {et~al.}(2019)\citenamefont {Guo},
  \citenamefont {Jing}, \citenamefont {Mei{\ss}ner},\ and\ \citenamefont
  {Sakai}}]{Guo:2019fdo}%
  \BibitemOpen
  \bibfield  {author} {\bibinfo {author} {\bibfnamefont {F.-K.}\ \bibnamefont
  {Guo}}, \bibinfo {author} {\bibfnamefont {H.-J.}\ \bibnamefont {Jing}},
  \bibinfo {author} {\bibfnamefont {U.-G.}\ \bibnamefont {Mei{\ss}ner}},\ and\
  \bibinfo {author} {\bibfnamefont {S.}~\bibnamefont {Sakai}},\ }\href
  {https://doi.org/10.1103/PhysRevD.99.091501} {\bibfield  {journal} {\bibinfo
  {journal} {Phys. Rev. D}\ }\textbf {\bibinfo {volume} {99}},\ \bibinfo
  {pages} {091501} (\bibinfo {year} {2019})},\ \Eprint
  {https://arxiv.org/abs/1903.11503} {arXiv:1903.11503 [hep-ph]} \BibitemShut
  {NoStop}%
\bibitem [{\citenamefont {Yamaguchi}\ \emph {et~al.}(2020)\citenamefont
  {Yamaguchi}, \citenamefont {Hosaka}, \citenamefont {Takeuchi},\ and\
  \citenamefont {Takizawa}}]{Yamaguchi:2019vea}%
  \BibitemOpen
  \bibfield  {author} {\bibinfo {author} {\bibfnamefont {Y.}~\bibnamefont
  {Yamaguchi}}, \bibinfo {author} {\bibfnamefont {A.}~\bibnamefont {Hosaka}},
  \bibinfo {author} {\bibfnamefont {S.}~\bibnamefont {Takeuchi}},\ and\
  \bibinfo {author} {\bibfnamefont {M.}~\bibnamefont {Takizawa}},\ }\href
  {https://doi.org/10.1088/1361-6471/ab72b0} {\bibfield  {journal} {\bibinfo
  {journal} {J. Phys. G}\ }\textbf {\bibinfo {volume} {47}},\ \bibinfo {pages}
  {053001} (\bibinfo {year} {2020})},\ \Eprint
  {https://arxiv.org/abs/1908.08790} {arXiv:1908.08790 [hep-ph]} \BibitemShut
  {NoStop}%
\bibitem [{\citenamefont {Dong}\ \emph
  {et~al.}(2021{\natexlab{b}})\citenamefont {Dong}, \citenamefont {Guo},\ and\
  \citenamefont {Zou}}]{Dong:2021juy}%
  \BibitemOpen
  \bibfield  {author} {\bibinfo {author} {\bibfnamefont {X.-K.}\ \bibnamefont
  {Dong}}, \bibinfo {author} {\bibfnamefont {F.-K.}\ \bibnamefont {Guo}},\ and\
  \bibinfo {author} {\bibfnamefont {B.-S.}\ \bibnamefont {Zou}},\ }\href
  {https://doi.org/10.13725/j.cnki.pip.2021.02.001} {\bibfield  {journal}
  {\bibinfo  {journal} {Progr. Phys.}\ }\textbf {\bibinfo {volume} {41}},\
  \bibinfo {pages} {65} (\bibinfo {year} {2021}{\natexlab{b}})},\ \Eprint
  {https://arxiv.org/abs/2101.01021} {arXiv:2101.01021 [hep-ph]} \BibitemShut
  {NoStop}%
\bibitem [{\citenamefont {Dong}\ \emph
  {et~al.}(2021{\natexlab{c}})\citenamefont {Dong}, \citenamefont {Guo},\ and\
  \citenamefont {Zou}}]{Dong:2021bvy}%
  \BibitemOpen
  \bibfield  {author} {\bibinfo {author} {\bibfnamefont {X.-K.}\ \bibnamefont
  {Dong}}, \bibinfo {author} {\bibfnamefont {F.-K.}\ \bibnamefont {Guo}},\ and\
  \bibinfo {author} {\bibfnamefont {B.-S.}\ \bibnamefont {Zou}},\ }\href
  {https://doi.org/10.1088/1572-9494/ac27a2} {\bibfield  {journal} {\bibinfo
  {journal} {Commun. Theor. Phys.}\ }\textbf {\bibinfo {volume} {73}},\
  \bibinfo {pages} {125201} (\bibinfo {year} {2021}{\natexlab{c}})},\ \Eprint
  {https://arxiv.org/abs/2108.02673} {arXiv:2108.02673 [hep-ph]} \BibitemShut
  {NoStop}%
\bibitem [{\citenamefont {Wu}\ \emph {et~al.}(2022)\citenamefont {Wu},
  \citenamefont {Pan}, \citenamefont {Liu},\ and\ \citenamefont
  {Geng}}]{Wu:2022ftm}%
  \BibitemOpen
  \bibfield  {author} {\bibinfo {author} {\bibfnamefont {T.-W.}\ \bibnamefont
  {Wu}}, \bibinfo {author} {\bibfnamefont {Y.-W.}\ \bibnamefont {Pan}},
  \bibinfo {author} {\bibfnamefont {M.-Z.}\ \bibnamefont {Liu}},\ and\ \bibinfo
  {author} {\bibfnamefont {L.-S.}\ \bibnamefont {Geng}},\ }\href
  {https://doi.org/10.1016/j.scib.2022.08.007} {\bibfield  {journal} {\bibinfo
  {journal} {Sci. Bull.}\ }\textbf {\bibinfo {volume} {67}},\ \bibinfo {pages}
  {1735} (\bibinfo {year} {2022})},\ \Eprint {https://arxiv.org/abs/2208.00882}
  {arXiv:2208.00882 [hep-ph]} \BibitemShut {NoStop}%
\bibitem [{\citenamefont {Su}\ \emph {et~al.}(2025)\citenamefont {Su},
  \citenamefont {Song}, \citenamefont {L{\"u}},\ and\ \citenamefont
  {Zhu}}]{Su:2025toa}%
  \BibitemOpen
  \bibfield  {author} {\bibinfo {author} {\bibfnamefont {J.-C.}\ \bibnamefont
  {Su}}, \bibinfo {author} {\bibfnamefont {Q.-F.}\ \bibnamefont {Song}},
  \bibinfo {author} {\bibfnamefont {Q.-F.}\ \bibnamefont {L{\"u}}},\ and\
  \bibinfo {author} {\bibfnamefont {J.}~\bibnamefont {Zhu}},\ }\href
  {https://doi.org/10.1140/epjc/s10052-025-14905-4} {\bibfield  {journal}
  {\bibinfo  {journal} {Eur. Phys. J. C}\ }\textbf {\bibinfo {volume} {85}},\
  \bibinfo {pages} {1181} (\bibinfo {year} {2025})},\ \Eprint
  {https://arxiv.org/abs/2504.13431} {arXiv:2504.13431 [hep-ph]} \BibitemShut
  {NoStop}%
\bibitem [{\citenamefont {Chen}\ \emph
  {et~al.}(2020{\natexlab{b}})\citenamefont {Chen}, \citenamefont {Chen},\ and\
  \citenamefont {Liu}}]{Chen:2019thk}%
  \BibitemOpen
  \bibfield  {author} {\bibinfo {author} {\bibfnamefont {C.-Y.}\ \bibnamefont
  {Chen}}, \bibinfo {author} {\bibfnamefont {M.}~\bibnamefont {Chen}},\ and\
  \bibinfo {author} {\bibfnamefont {Y.-X.}\ \bibnamefont {Liu}},\ }\href
  {https://doi.org/10.1088/1572-9494/abb7cd} {\bibfield  {journal} {\bibinfo
  {journal} {Commun. Theor. Phys.}\ }\textbf {\bibinfo {volume} {72}},\
  \bibinfo {pages} {125202} (\bibinfo {year} {2020}{\natexlab{b}})},\ \Eprint
  {https://arxiv.org/abs/1912.01931} {arXiv:1912.01931 [hep-ph]} \BibitemShut
  {NoStop}%
\bibitem [{\citenamefont {Edwards}\ and\ \citenamefont
  {Joo}(2005)}]{Edwards:2004sx}%
  \BibitemOpen
  \bibfield  {author} {\bibinfo {author} {\bibfnamefont {R.~G.}\ \bibnamefont
  {Edwards}}\ and\ \bibinfo {author} {\bibfnamefont {B.}~\bibnamefont {Joo}}
  (\bibinfo {collaboration} {SciDAC, LHPC, UKQCD}),\ }\href
  {https://doi.org/10.1016/j.nuclphysbps.2004.11.254} {\bibfield  {journal}
  {\bibinfo  {journal} {Nucl. Phys. B Proc. Suppl.}\ }\textbf {\bibinfo
  {volume} {140}},\ \bibinfo {pages} {832} (\bibinfo {year} {2005})},\ \Eprint
  {https://arxiv.org/abs/hep-lat/0409003} {arXiv:hep-lat/0409003} \BibitemShut
  {NoStop}%
\bibitem [{\citenamefont {Clark}\ \emph {et~al.}(2010)\citenamefont {Clark},
  \citenamefont {Babich}, \citenamefont {Barros}, \citenamefont {Brower},\ and\
  \citenamefont {Rebbi}}]{Clark:2009wm}%
  \BibitemOpen
  \bibfield  {author} {\bibinfo {author} {\bibfnamefont {M.~A.}\ \bibnamefont
  {Clark}}, \bibinfo {author} {\bibfnamefont {R.}~\bibnamefont {Babich}},
  \bibinfo {author} {\bibfnamefont {K.}~\bibnamefont {Barros}}, \bibinfo
  {author} {\bibfnamefont {R.~C.}\ \bibnamefont {Brower}},\ and\ \bibinfo
  {author} {\bibfnamefont {C.}~\bibnamefont {Rebbi}} (\bibinfo {collaboration}
  {QUDA}),\ }\href {https://doi.org/10.1016/j.cpc.2010.05.002} {\bibfield
  {journal} {\bibinfo  {journal} {Comput. Phys. Commun.}\ }\textbf {\bibinfo
  {volume} {181}},\ \bibinfo {pages} {1517} (\bibinfo {year} {2010})},\ \Eprint
  {https://arxiv.org/abs/0911.3191} {arXiv:0911.3191 [hep-lat]} \BibitemShut
  {NoStop}%
\bibitem [{\citenamefont {Jaffe}(1977{\natexlab{a}})}]{Jaffe:1976ig}%
  \BibitemOpen
  \bibfield  {author} {\bibinfo {author} {\bibfnamefont {R.~L.}\ \bibnamefont
  {Jaffe}},\ }\href {https://doi.org/10.1103/PhysRevD.15.267} {\bibfield
  {journal} {\bibinfo  {journal} {Phys. Rev. D}\ }\textbf {\bibinfo {volume}
  {15}},\ \bibinfo {pages} {267} (\bibinfo {year}
  {1977}{\natexlab{a}})}\BibitemShut {NoStop}%
\bibitem [{\citenamefont {Jaffe}(1977{\natexlab{b}})}]{Jaffe:1976ih}%
  \BibitemOpen
  \bibfield  {author} {\bibinfo {author} {\bibfnamefont {R.~L.}\ \bibnamefont
  {Jaffe}},\ }\href {https://doi.org/10.1103/PhysRevD.15.281} {\bibfield
  {journal} {\bibinfo  {journal} {Phys. Rev. D}\ }\textbf {\bibinfo {volume}
  {15}},\ \bibinfo {pages} {281} (\bibinfo {year}
  {1977}{\natexlab{b}})}\BibitemShut {NoStop}%
\bibitem [{\citenamefont {Jaffe}(1977{\natexlab{c}})}]{Jaffe:1976yi}%
  \BibitemOpen
  \bibfield  {author} {\bibinfo {author} {\bibfnamefont {R.~L.}\ \bibnamefont
  {Jaffe}},\ }\href {https://doi.org/10.1103/PhysRevLett.38.195} {\bibfield
  {journal} {\bibinfo  {journal} {Phys. Rev. Lett.}\ }\textbf {\bibinfo
  {volume} {38}},\ \bibinfo {pages} {195} (\bibinfo {year}
  {1977}{\natexlab{c}})},\ \bibinfo {note} {[Erratum: Phys.Rev.Lett. 38, 617
  (1977)]}\BibitemShut {NoStop}%
\bibitem [{\citenamefont {Bao}\ and\ \citenamefont
  {Lin}(1995)}]{PhysRevA.52.3586}%
  \BibitemOpen
  \bibfield  {author} {\bibinfo {author} {\bibfnamefont {C.~G.}\ \bibnamefont
  {Bao}}\ and\ \bibinfo {author} {\bibfnamefont {C.~D.}\ \bibnamefont {Lin}},\
  }\href {https://doi.org/10.1103/PhysRevA.52.3586} {\bibfield  {journal}
  {\bibinfo  {journal} {Phys. Rev. A}\ }\textbf {\bibinfo {volume} {52}},\
  \bibinfo {pages} {3586} (\bibinfo {year} {1995})}\BibitemShut {NoStop}%
\bibitem [{\citenamefont {Bao}(1997)}]{Bao:1997zz}%
  \BibitemOpen
  \bibfield  {author} {\bibinfo {author} {\bibfnamefont {C.~G.}\ \bibnamefont
  {Bao}},\ }\href {https://doi.org/10.1103/PhysRevLett.79.3475} {\bibfield
  {journal} {\bibinfo  {journal} {Phys. Rev. Lett.}\ }\textbf {\bibinfo
  {volume} {79}},\ \bibinfo {pages} {3475} (\bibinfo {year}
  {1997})}\BibitemShut {NoStop}%
\bibitem [{\citenamefont {Bao}(1998)}]{Bao:1998an}%
  \BibitemOpen
  \bibfield  {author} {\bibinfo {author} {\bibfnamefont {C.~G.}\ \bibnamefont
  {Bao}},\ }\href {https://doi.org/10.1016/S0375-9474(98)00239-5} {\bibfield
  {journal} {\bibinfo  {journal} {Nucl. Phys. A}\ }\textbf {\bibinfo {volume}
  {637}},\ \bibinfo {pages} {520} (\bibinfo {year} {1998})},\ \Eprint
  {https://arxiv.org/abs/nucl-th/9805001} {arXiv:nucl-th/9805001} \BibitemShut
  {NoStop}%
\bibitem [{\citenamefont {Cheng-guang}(1999)}]{Bao:1999}%
  \BibitemOpen
  \bibfield  {author} {\bibinfo {author} {\bibfnamefont {B.}~\bibnamefont
  {Cheng-guang}},\ }\href {https://doi.org/10.1088/0256-307X/16/3/009}
  {\bibfield  {journal} {\bibinfo  {journal} {Chin. Phys. Lett.}\ }\textbf
  {\bibinfo {volume} {16}},\ \bibinfo {pages} {178} (\bibinfo {year}
  {1999})}\BibitemShut {NoStop}%
\bibitem [{\citenamefont {Bao}\ and\ \citenamefont {Liu}(1999)}]{Bao:1998kj}%
  \BibitemOpen
  \bibfield  {author} {\bibinfo {author} {\bibfnamefont {C.~G.}\ \bibnamefont
  {Bao}}\ and\ \bibinfo {author} {\bibfnamefont {Y.~X.}\ \bibnamefont {Liu}},\
  }\href {https://doi.org/10.1103/PhysRevLett.82.61} {\bibfield  {journal}
  {\bibinfo  {journal} {Phys. Rev. Lett.}\ }\textbf {\bibinfo {volume} {82}},\
  \bibinfo {pages} {61} (\bibinfo {year} {1999})},\ \Eprint
  {https://arxiv.org/abs/nucl-th/9805013} {arXiv:nucl-th/9805013} \BibitemShut
  {NoStop}%
\bibitem [{\citenamefont {Liu}\ \emph {et~al.}(2002)\citenamefont {Liu},
  \citenamefont {Li},\ and\ \citenamefont {Bao}}]{Liu:2002qm}%
  \BibitemOpen
  \bibfield  {author} {\bibinfo {author} {\bibfnamefont {Y.-X.}\ \bibnamefont
  {Liu}}, \bibinfo {author} {\bibfnamefont {J.-S.}\ \bibnamefont {Li}},\ and\
  \bibinfo {author} {\bibfnamefont {C.-G.}\ \bibnamefont {Bao}},\ }\href
  {https://doi.org/10.1016/S0370-2693(02)02514-5} {\bibfield  {journal}
  {\bibinfo  {journal} {Phys. Lett. B}\ }\textbf {\bibinfo {volume} {544}},\
  \bibinfo {pages} {280} (\bibinfo {year} {2002})}\BibitemShut {NoStop}%
\bibitem [{\citenamefont {Liu}\ \emph {et~al.}(2003{\natexlab{a}})\citenamefont
  {Liu}, \citenamefont {Li},\ and\ \citenamefont {Bao}}]{Liu:2002vi}%
  \BibitemOpen
  \bibfield  {author} {\bibinfo {author} {\bibfnamefont {Y.-x.}\ \bibnamefont
  {Liu}}, \bibinfo {author} {\bibfnamefont {J.-s.}\ \bibnamefont {Li}},\ and\
  \bibinfo {author} {\bibfnamefont {C.-g.}\ \bibnamefont {Bao}},\ }\href
  {https://doi.org/10.1103/PhysRevC.67.055207} {\bibfield  {journal} {\bibinfo
  {journal} {Phys. Rev. C}\ }\textbf {\bibinfo {volume} {67}},\ \bibinfo
  {pages} {055207} (\bibinfo {year} {2003}{\natexlab{a}})},\ \Eprint
  {https://arxiv.org/abs/nucl-th/0212069} {arXiv:nucl-th/0212069} \BibitemShut
  {NoStop}%
\bibitem [{\citenamefont {Liu}\ \emph {et~al.}(2003{\natexlab{b}})\citenamefont
  {Liu}, \citenamefont {Li},\ and\ \citenamefont {Bao}}]{Liu:2003gi}%
  \BibitemOpen
  \bibfield  {author} {\bibinfo {author} {\bibfnamefont {Y.-X.}\ \bibnamefont
  {Liu}}, \bibinfo {author} {\bibfnamefont {J.-S.}\ \bibnamefont {Li}},\ and\
  \bibinfo {author} {\bibfnamefont {C.-G.}\ \bibnamefont {Bao}},\ }\href
  {https://doi.org/10.1142/S0217732303010594} {\bibfield  {journal} {\bibinfo
  {journal} {Mod. Phys. Lett. A}\ }\textbf {\bibinfo {volume} {18}},\ \bibinfo
  {pages} {414} (\bibinfo {year} {2003}{\natexlab{b}})}\BibitemShut {NoStop}%
\bibitem [{\citenamefont {Liu}\ \emph {et~al.}(2004)\citenamefont {Liu},
  \citenamefont {Li},\ and\ \citenamefont {Bao}}]{Liu:2004xyi}%
  \BibitemOpen
  \bibfield  {author} {\bibinfo {author} {\bibfnamefont {Y.-x.}\ \bibnamefont
  {Liu}}, \bibinfo {author} {\bibfnamefont {J.-s.}\ \bibnamefont {Li}},\ and\
  \bibinfo {author} {\bibfnamefont {C.-g.}\ \bibnamefont {Bao}}\ }(\bibinfo
  {year} {2004})\ \Eprint {https://arxiv.org/abs/hep-ph/0401197}
  {arXiv:hep-ph/0401197} \BibitemShut {NoStop}%
\bibitem [{\citenamefont {Wilczek}(2004)}]{Wilczek:2004im}%
  \BibitemOpen
  \bibfield  {author} {\bibinfo {author} {\bibfnamefont {F.}~\bibnamefont
  {Wilczek}},\ }in\ \href {https://doi.org/10.1142/9789812775344_0007} {\emph
  {\bibinfo {booktitle} {{Deserfest: A Celebration of the Life and Works of
  Stanley Deser}}}}\ (\bibinfo {year} {2004})\ pp.\ \bibinfo {pages}
  {322--338},\ \Eprint {https://arxiv.org/abs/hep-ph/0409168}
  {arXiv:hep-ph/0409168} \BibitemShut {NoStop}%
\end{thebibliography}%

\end{document}